\documentclass[11pt,a4paper]{article}
\pdfoutput=1

\usepackage[utf8]{inputenc}
\usepackage{jheppub}
\usepackage{bbm}			%
\usepackage{mathtools}		%
\usepackage{dsfont} 		%
\usepackage{bookmark}		%
\usepackage{braket}

\hypersetup{bookmarksdepth=subsection}
\newcommand{\pheq}{\phantom{{}={}}}

\newcommand{\mc}[1]{\mathcal{#1}}
\newcommand{\undernote}[2]{\underset{#1}{\underbrace{#2}}}

\newcommand{\p}{\partial}
\newcommand{\pd}[2]{\tfrac{\partial #1}{\partial #2}}
\renewcommand{\d}{\mathrm{d}}
\newcommand{\del}[1]{\frac{\partial}{\partial #1}}
\newcommand{\tdel}[1]{\tfrac{\partial}{\partial #1}}
\newcommand{\cpint}[1]{\int \limits_{\mathclap{#1}}}

\renewcommand{\Im}{\operatorname{Im}}
\renewcommand{\Re}{\operatorname{Re}}
\newcommand{\Unit}{\mathds{1}}
\DeclareMathOperator{\diag}{diag}
\DeclareMathOperator{\tr}{tr}

\newcommand{\ga}{\alpha}
\newcommand{\gb}{\beta}

\newcommand{\gd}{\delta}

\newcommand{\eps}{\epsilon}
\newcommand{\veps}{\varepsilon}

\newcommand{\gt}{\theta}
\newcommand{\vt}{\vartheta}

\newcommand{\gS}{\Sigma}

\newcommand{\vPhi}{\varPhi}

\renewcommand{\bar}{\overline}
\newcommand{\vPhib}{\bar{\vPhi}}
\newcommand{\thetab}{\overline \theta}
\newcommand{\tb}{\overline{\theta}} %
\newcommand{\phib}{\bar \phi}
\newcommand{\Phib}{\bar \Phi}
\newcommand{\Db}{\bar D}
\newcommand{\Qb}{\bar Q}

\newcommand{\Fb}{\overline{F}}
\newcommand{\Cb}{\overline{C}}

\newcommand{\Hb}{\bar H}
\newcommand{\Wb}{\bar W}
\newcommand{\Eb}{\bar E}
\newcommand{\Jb}{\bar J}
\newcommand{\chib}{\bar \chi}

\newcommand{\psib}{\bar \psi}
\newcommand{\Psib}{\bar \Psi}
\newcommand{\etab}{\bar \eta}
\newcommand{\omegab}{\bar \omega}

\newcommand{\lambdab}{\bar \lambda}
\newcommand{\Yb}{\overline{Y}}

\newcommand{\xib}{\bar \xi}

\newcommand{\Lag}{\mathcal{L}}
\newcommand{\N}{\mathcal{N}}
\newcommand{\Qc}{\mathcal{Q}}
\newcommand{\Qbc}{{\bar{\mathcal{Q}}}}
\newcommand{\D}{\mathcal{D}}
\newcommand{\Jc}{\mathcal{J}}

\newcommand{\Rc}{\mathcal R}
\newcommand{\Tc}{\mathcal T}
\newcommand{\Y}{\mathcal Y}
\newcommand{\Ac}{\mathcal A}
\newcommand{\X}{\mathcal X}
\newcommand{\Wc}{\mathcal W}

\newcommand{\Sc}{\mathcal S}
\newcommand{\Sb}{\overline{S}}

\newcommand{\epsb}{\bar \eps}

\newcommand{\R}{\mathbb{R}}

\newcommand{\hT}{{\widehat{T}}}

\newcommand{\fQ}{\mathbbm{Q}}			%
\newcommand{\fD}{\mathbbm{D}}			%

\newcommand{\dsym}{\delta_\text{sym}}
\newcommand{\eom}{\operatorname{EoM}}
\newcommand{\hmu}{{\hat{\mu}}}
\newcommand{\hnu}{{\hat{\nu}}}
\newcommand{\hrho}{{\hat{\rho}}}
\newcommand{\proj}{\mc{P}}
\newcommand{\cc}{\text{cc.}}
\newcommand{\full}{\text{full}}
\newcommand{\kin}{\text{kin.}}
\newcommand{\integ}{\text{int.}}

\newcommand{\ph}{\phantom}
\newcommand{\tp}{\theta^+}
\newcommand{\tm}{\theta^-}
\newcommand{\tbp}{\thetab^+}
\newcommand{\tbm}{\thetab^-}
\newcommand{\zerotwo}{{(0,2)}}
\newcommand{\oneone}{{(1,1)}}
\newcommand{\zero}{{(0)}}
\newcommand{\one}{{(1)}}
\newcommand{\onea}{{(1a)}}
\newcommand{\oneb}{{(1b)}}
\newcommand{\two}{{(2)}}

\newcommand{\vtp}{\vt^+}
\newcommand{\vtm}{\vt^-}
\newcommand{\tvt}{\tilde{\vt}}
\newcommand{\tvtp}{\tvt^+}
\newcommand{\tvtm}{\tvt^-}
\newcommand{\nzerotwo}{{\mathcal{N}=(0,2)}}

\newcommand{\pgS}{{\partial \Sigma}}
\newcommand{\bdy}{{\text{bdy}}}
\newcommand{\bulk}{{\text{bulk}}}

\title{Boundaries and supercurrent multiplets in 3D~Landau-Ginzburg~models}
\author{Ilka Brunner,}
\author{Jonathan Schulz}
\author{and Alexander Tabler}
\affiliation{Arnold Sommerfeld Center, Ludwig-Maximilians-Universität,\\ Theresienstraße 37, 80333 München, Germany}
\emailAdd{ilka.brunner@physik.uni-muenchen.de}
\emailAdd{j.schulz@physik.uni-muenchen.de}
\emailAdd{tabler.alexander@physik.uni-muenchen.de}
\abstract{Theories with 3D $\mathcal{N}=2$ bulk supersymmetry may preserve a 2D $\mathcal{N}=(0,2)$ subalgebra when a boundary is introduced, possibly with localized degrees of freedom. We propose generalized supercurrent multiplets with bulk and boundary parts adapted to such setups. Using their structure, we comment on implications for the $\overline{Q}_+$-cohomology. As an example, we apply the developed framework to Landau-Ginzburg models. In these models, we study the role of boundary degrees of freedom and matrix factorizations. We verify our results using quantization.}
\preprint{LMU-ASC 16/19}

\begin{document}

\maketitle 

\section{Introduction}

In this paper we study $\N = 2$ supersymmetric theories in flat $(2+1)$-dimensional spacetime with (spacelike) boundaries. The boundary necessarily breaks translational invariance and hence can only preserve part of the bulk supersymmetry. Explicitly, the $\N = 2$ algebra in 3 dimensions is
\begin{equation}\label{eq:SUSYalg}
\{ Q_\pm , \Qb_\pm \} = -4 P_\pm , \quad \{ Q_+, \Qb_- \} = 2 P_\perp ,
\end{equation}
where spacetime has coordinates $x^\pm$ and $x_\perp$, see appendix \ref{app:conventions} for a summary of our conventions. We want to consider the case with a boundary in $x_\perp$-direction, breaking supersymmetry to a subalgebra of \eqref{eq:SUSYalg} that does not contain $P_\perp $, the generator of translations in that direction. 
As has been analyzed before \cite{Dimofte:2017tpi,Okazaki:2013kaa,Gadde:2013sca}, in the case of $\N=2$ supersymmetry in $3$ dimensions, there are two types of supersymmetric boundary conditions, referred to as A-type and B-type. Each of them is associated to a subalgebra of the initial bulk supersymmetry algebra, containing two momentum operators and two supersymmetry charges. A-type boundary conditions preserve $(1,1)$ supersymmetry, whereas B-type boundary conditions preserve a chiral $\nzerotwo$ subalgebra, generated by $Q_+$ and $\Qb_+$.
We will focus on $(0,2)$ boundary conditions and analyze them from two points of view: On the one hand, for theories defined by a Lagrangian, we employ a Noether procedure. On the other hand, we discuss the structure of the supercurrent multiplets \cite{Komargodski:2010rb,Dumitrescu:2011iu} and formulate boundary multiplets. The two points of view are interrelated, as the (improved) Noether currents form components of the current multiplets.

From a Lagrangian point of view, the supersymmetric bulk Lagrangian transforms under SUSY-variations into a total derivative. In the presence of a boundary, this generically yields a boundary term which must be canceled for the symmetry to be preserved. This can be achieved by choosing boundary conditions on the fields, such that the boundary variation vanishes. Alternatively, and this is the main focus of the present paper, one can cancel the boundary variation by adding a suitable boundary part to the action, such that the action is invariant under symmetry variations without reference to the boundary conditions on the fields. This term can contain extra boundary degrees of freedom that are not inherited from the bulk. The full invariant action thus contains a bulk and boundary term.
\begin{equation}
S=\int_M {\cal L}^B + \int_{\partial M} {\cal L}^\partial.
\end{equation}

Given an action which is invariant under a symmetry, Noether's procedure yields a set of conserved charges and currents. In the case of supersymmetry, this includes the (canonical) energy momentum tensor and the supersymmetry currents. After imposing canonical commutation relations between the fields, the Noether charges provide a representation of the symmetry algebra in terms of the fields.

In the case of pure bulk theories, it is very useful to arrange the supercurrents together with other conserved currents into multiplets. This can be done independently of a Lagrangian definition of a theory. The supercurrents form a representation of the supersymmetry algebra, and their most general form has been discussed recently in \cite{Komargodski:2010rb,Dumitrescu:2011iu}.
Supercurrent multiplets are indecomposable SUSY multiplets that contain the stress energy tensor $T_{\mu \nu}$ and  the supercurrents $S_{\alpha \mu}$  as part of their components. In addition, there are brane currents, whose integrals yield brane charges. The components of the supercurrent multiplets appear in a local version of the supersymmetry algebra, which very schematically takes the form
\begin{align}\label{eq:local-schematic}
\{ \bar{Q}_\alpha , S_{\beta \mu} \} &= 2 \gamma_{\alpha \beta}^\nu T_{\mu \nu}  + \dots, \\
\{ Q_\alpha , S_{\beta \mu} \} &= \dots ,
\end{align}
where the dots indicate various currents that will be explained later in the text.
As is well-known, the stress tensor for a theory is not unique, but can be modified by improvement transformations. Indeed, the same holds for the supercurrents, and the notion of improvement transformations can be lifted to the full multiplet. In any three-dimensional $\N=2$ supersymmetric theory (and also in other dimensions with the same amount of supersymmetry), there exists a so-called $\Sc$-multiplet. Under special conditions, the $\Sc$-multiplet can be decomposable, such that there exist smaller multiplets. Of special interest in the context of the current work is the $\Rc$-multiplet, which exists in theories which exhibit an $R$-symmetry.

The notion of supercurrent multiplets has been extended to theories with defects in \cite{Drukker:2017xrb}, where a new so-called defect multiplet was constructed.  As a consequence of the violation of translation symmetry perpendicular to the defect, the stress tensor is no longer conserved. This violation is encoded in the displacement operator. The defect multiplet contains the displacement operator as one of its components \cite{Gaiotto:2013sma,Drukker:2017xrb}.

In the current paper, we consider supermultiplets in situations with boundary, focusing on the \emph{preserved} symmetries. As mentioned above, to formulate boundary conditions means to specify a subalgebra of the SUSY algebra such that the momentum operator in the direction perpendicular to the boundary is not contained. The supercurrent multiplet is in particular a representation of the larger (bulk) algebra and hence also of the smaller algebra. In the case of the $\N=(0,2)$ subalgebra, we show how the bulk supercurrent multiplets decompose under the smaller algebra. Of course, due to the presence of the boundary, the currents contained in the multiplet are no longer conserved by themselves. To formulate a consistent multiplet for a theory with boundary, we discuss how  to add boundary parts to the $(0,2)$-components of the initial bulk multiplet. Our ansatz for a full $\Rc$-multiplet is
\begin{equation}
	\Rc_\mu^\full= \Rc_\mu^B+\proj_\mu^{\ph\mu\hmu} \delta(x_\perp) \Rc^\p_\hmu,
\end{equation}
where $\Rc_\mu^B$ is the bulk part, $\Rc^\p_\hmu$ is the boundary part and $\proj_\mu^{\ph\mu\hmu}$ denotes an embedding. Both parts decompose into $(0,2)$-components. The boundary part is added to the bulk multiplet in such a way that the initial divergence-freeness of the bulk currents is completed to bulk-boundary conservation laws. We do not discuss possible modifications of the bulk currents corresponding to the \emph{broken} symmetries. 

One important feature of supercurrent multiplets is that they fall into short representations of the supersymmetry algebra. Therefore, they are protected under RG flow and retain their form \cite{Dumitrescu:2011iu}. The supercurrents of the quantum theory can thus be used to constrain the IR behavior of a theory using the UV information. In the case of \emph{two-dimensional} $\nzerotwo$ models, the supercurrent multiplet for theories with an $R$-symmetry was used to study renormalization group invariants in \cite{Dedushenko:2015opz}. In particular, it was shown that an RG invariant  chiral algebra exists, extending earlier works \cite{Witten:1993jg,Silverstein:1994ih,Witten:2005px}. The chiral algebra arises as the cohomology of the supercharge $\Qb_+$. Using only the form of the supercurrent multiplet, \cite{Dedushenko:2015opz} shows that there is a half-twisted stress tensor (the original stress tensor modified by the $R$-current) in the cohomology. As a consequence, conformal symmetry is part of the chiral algebra.

Given these findings, we consider the consequences of the supercurrent multiplets for the case that the $\nzerotwo$ supersymmetry is the symmetry preserved at the \emph{boundary} of a three-dimensional theory. We do not find a stress-tensor in the cohomology following the steps in \cite{Dedushenko:2015opz}, however, there is a weaker statement. For this, one makes the $(0,2)$-structure manifest by regarding the three-dimensional $\N =2$ theory as a two-dimensional theory living on the boundary $\R^{1,1}$. The bosonic fields of this theory are valued in maps from $\R_{\leq 0}$  to the original target. The currents  are obtained from the original three-dimensional ones by integrating over the direction perpendicular to the boundary and are preserved in the boundary theory. In this theory, we then do have a stress energy tensor that is part of the cohomology. Formulated in the initial theory, this cohomology element is obtained by an integration in the perpendicular direction from infinity (or a second boundary, which we do not discuss here) to the boundary. Note that the action of any charge computed from the currents applied to an insertion at the boundary would involve an integration over this direction, as well as all other spatial directions. In this sense we can also interpret the partially  integrated currents in the original theory. 

The integration along the perpendicular direction also gives another perspective on the boundary multiplets. The theory effectively becomes a two-dimensional theory with $\N=(0,2)$ supersymmetry, and the integrated multiplets are genuine $\N=(0,2)$ multiplets.

In the second part of the paper, we study a specific example, namely a theory of three-dimensional chiral multiplets with a superpotential. We restrict our explicit discussion to the case of   a single chiral field with a monomial superpotential. However, a generalization to more than one superfield and an arbitrary superpotential preserving an $R$-symmetry is straightforward and our discussion applies to this case as well. As has already been shown in \cite{Gadde:2013sca,Yoshida:2014ssa}, the condition of preserving $\nzerotwo$ supersymmetry leads to a three-dimensional generalization of matrix factorizations \cite{Kapustin:2002bi,Brunner:2003dc}. In this case, boundary terms are canceled by adding fermionic degrees of freedom and a potential at the boundary. Using Noether's procedure, we compute the conserved currents for different boundary conditions. These currents contain pure bulk as well as boundary pieces and  only the combination of both  is conserved. We also discuss the current multiplets, following \cite{Dumitrescu:2011iu} for the bulk case. Starting from the Noether currents, one needs to apply improvement transformations to symmetrize the stress energy tensor and subsequently organize the currents into supercurrent multiplets. As we consider the case where $R$-symmetry is preserved, the relevant multiplet is the $\Rc$-multiplet. We spell out all components of the bulk-boundary multiplet in the example following the strategy outlined above.

To complete our understanding of the symmetries in the model, we study the algebra of the supercharges and supercurrents \eqref{eq:local-schematic} in the example. We start with the explicit expressions of the currents in terms of fields and impose canonical commutation relations on the fields. We then verify the relations \eqref{eq:local-schematic} as well as the $\N=(0,2)$ superalgebra in the specific representation of the example. In the computation, it is essential to make use of the correct factorization properties of the superpotential to recover the correct form of the algebra.

This paper is organized as follows: In section \ref{sec:currents and charges general} we review and elaborate on various aspects of theories formulated on Minkowski space $\R^{1,N-1}$ with a flat boundary $\R^{1, N-2}$, following and extending \cite{DiPietro:2015zia}. In section \ref{sec:general boundary supercurrent multiplets} we discuss supercurrent multiplets. We review the constraints that have to be satisfied by such a multiplet from \cite{Dumitrescu:2011iu}. As explained there, the most general supercurrent multiplets consists of certain superfields satisfying defining constraint relations. Their solutions are unique up to improvement transformations. In special situations, such improvements can be used to formulate shorter multiplets, in particular the $\Rc$-multiplet. We decompose the multiplets and constraints according to the $(0,2)$-substructure and formulate consistent bulk-boundary multiplets. Here, current conservation of the combined bulk-boundary system is imposed. We then discuss an integrated structure, where we integrate in the direction perpendicular to the boundary. This provides a two-dimensional version of the conservation equations, taking the familiar form of divergence-freeness of the currents. 

We then turn to a discussion of the Landau-Ginzburg example in section \ref{sec:LG model}. First, in \ref{subsec:LG bulk theory}, we introduce the bulk model, then introduce a boundary in \ref{subsec:LG boundary introduction}. We distinguish the cases with and without superpotential and show that without specifying boundary conditions, $\nzerotwo$ SUSY can be preserved by introducing boundary fermions and matrix factorizations. Boundary conditions do however have to be imposed to make the action stationary and we discuss Dirichlet and Neumann conditions. In particular, there is a possibility to make boundary conditions dynamical, imposing them as equations of motion. We then turn to the formulation of currents in \ref{subsec:LG currents}. Here, Noether's procedure is employed to compute the conserved currents in the combined bulk boundary system. We discuss improvements to symmetrize the stress-energy tensor. Section \ref{subsec:LG multiplets} contains a discussion of the supercurrent multiplets in the example. 

In section \ref{sec:Quantization} we study the realization of the symmetry algebra in the Landau-Ginzburg model in terms of the fields. By imposing canonical commutation relations for the bulk and boundary fields, we verify that the supercharges implement the correct symmetry transformations on the fields and their derivatives, and we compute the brackets between supercharges and supercurrents. Special attention is paid to the contributions from the boundary. We do not impose any explicit boundary conditions on the fields in our computations. In \ref{subsec:quantization supercharge} we use the action of the supercharge on the derivative of the scalars to verify how the bulk fermions decompose into boundary fermions --- a decomposition that was seen in earlier sections from the point of view of the action. In the following subsections \ref{subsec:Qbc commutator} and \ref{subsec:Qc commutator} we verify the brackets between supercharges and currents and finally integrate them to the global supersymmetry algebra. The factorization condition of the superpotential arises as a consistency condition on the SUSY algebra. Finally, section \ref{sec:outlook} contains some conclusions and outlook.

We summarize various technical points in a number of appendices. Here we also collect some supplementary material for completeness. In particular, we present some results on $\N = (1,1)$ supersymmetry in appendix \ref{app:branching to oneone}.

\section{Currents and charges in theories with boundaries}
\label{sec:currents and charges general}
In this section we want to study (classical) theories on flat Minkowski half-space with spacelike boundary. In particular, we will consider
\begin{equation}
M= \Set{x^\mu=\big(x^0,\ldots,x^n,\ldots ,x^{N-1}\big)\in \R^{1,N-1} | x^n\leq 0}, \quad \p M=\big\{x^n=0\big\}.
\end{equation}
On the flat boundary we will denote the tangential coordinates by $x^\hmu$ while the normal coordinate is $x^n$, i.e.\ $\hmu$ takes all values except $n$. While our main focus will be $N=3$ later on, we will keep the discussion more general in this section.

\subsection{Bulk and boundary Lagrangians}
We want to study theories that have both bulk and boundary fields with an action of the form
\begin{equation}
S=S^B+S^\p=\int_M \Lag^B+\int_{\p M}\Lag^\p, \quad \Lag^B=\Lag^B[\Phi,\p_\mu\Phi], \quad \Lag^\p=\Lag^\p[X,\p_\hmu X, \Phi|_\p,\p_\hmu \Phi|_\p],\footnote{We assume that the boundary Lagrangian $\Lag^\p$ contains only tangential derivatives $\p_\hmu$. We also assume for simplicity that there are only first order derivatives of any kind.}
\end{equation}
where $\Phi$ and $X$ denotes bulk and boundary fields, respectively. Furthermore, one has to impose \emph{boundary conditions}. We follow the similar discussion from \cite{DiPietro:2015zia}, which we generalize here. The most general boundary conditions will be of the form
\begin{equation}
G(\text{fields}|_\p,\text{derivatives of fields}|_\p)=0.
\end{equation}
Under a (rigid) variation of the fields and an integration by parts, we get 
\begin{equation}\label{variation of action}
	\begin{aligned}
		\delta S&= \int_M \delta \Lag^B +\int_{\p M} \delta \Lag^\p
		\\&=\int_M \Big[\undernote{\text{Bulk } \eom^B[\Phi]}{\pd{\Lag^B}{\Phi}-\p_\mu \pd{\Lag^B}{(\p_\mu\Phi)}}\Big]\delta \Phi
		\\&\hspace{.4cm}
		+\int_{\p M} \Big[\big(\undernote{\text{Boundary } \eom^\p[X]}{
			\pd{\Lag^\p}{X}-\p_\hmu \pd{\Lag^\p}{(\p_\hmu X)}}\big)\delta X
		+\big(\undernote{\eqqcolon \Ac}{\pd{\Lag^\p}{\Phi|_\p}-\p_\hmu \pd{\Lag^\p}{(\p_\hmu\Phi|_\p)} +\pd{\Lag^B}{(\p_n\Phi|_\p)}}\big)\delta \Phi|_\p 
		+\p_\hmu (\cdots)
		\Big]\mathclap{.}
	\end{aligned}
\end{equation}
Stationarity of the action requires that
\begin{equation}\label{stationarity condition}
	\big[\Ac\cdot \delta \Phi|_\p\big]_{G=0}=\p_\hmu A^\hmu,
\end{equation}
for some boundary vector field $A^\hmu$. Note that in any case, the variations $\delta\Phi|_\p$, $\delta X$ we consider must be \emph{compatible} with the chosen boundary condition, i.e.\ we may only consider such variations that satisfy
\begin{equation}\label{compatibility of variations with G}
	\delta G|_{G=0}=0.
\end{equation}
A special kind of boundary condition is the \emph{dynamical boundary condition}, which amounts to requiring
\begin{equation}\label{dynamical BC}
	G\coloneqq \Ac \overset{!}{=}0.
\end{equation}
This is equivalent to the paradigm (e.g.\ found in \cite{Dimofte:2017tpi,Bilal:2011gp}) not to impose static boundary conditions, but instead adopt the boundary conditions that are naturally imposed by the tendency of the system to make the action stationary.

\subsection{Symmetries, currents and charges in boundary theories}\label{sec:generalcharges}

\subsubsection{Symmetries in boundary theories}\label{sec:symmetries in boundary theories}

Let us try to understand symmetries in theories with a boundary. 
If we want the full theory to be invariant under some symmetry transformation of the fields, the boundary condition must be compatible with the symmetry transformation, as mentioned in \cite{DiPietro:2015zia}:
\begin{equation}\label{symmetric boundary condition}
	\dsym G|_{G=0}=0.
\end{equation}
If this is satisfied, $G$ is called a \emph{symmetric} boundary condition with respect to $\dsym$. Conveniently, this requirement is equivalent to demanding that the symmetry variation $\dsym$ is a \emph{permitted variation} in the sense of equation \eqref{compatibility of variations with G}.

The definition of a symmetry in the presence of a boundary is analogous to the case of a pure bulk theory: a symmetry is an off-shell transformation of both bulk and boundary fields that leaves the action invariant, \emph{possibly} after using boundary conditions: 
\begin{equation} \label{action symmetry invariance}
	0=\dsym S|_{G=0}=(\dsym S^B +\dsym S^\p)|_{G=0}=0.
\end{equation}
It is natural to restrict to symmetries that arise from a symmetry of the bulk theory, i.e.\ $\dsym \Lag^B=\p_\mu V^\mu$  holds for some bulk vector fields $V^\mu$. Noether's theorem in the bulk then ensures that $\p_\mu J^\mu_B=0$, where $J_B^\mu=-\pd{\Lag^B}{(\p_\mu \Phi)}\dsym \Phi + V^\mu$ still holds. 
In terms of Lagrangians, we can then write
\begin{equation}
	0 = \dsym S|_{G=0} = \int_{M} \dsym \Lag^B + \int_{\p M} \dsym \Lag^\p = \int_{\p M}( V^n + \dsym \Lag^\p)|_{G=0}.
\end{equation}

If the above condition holds without imposing any boundary condition $G=0$, one says that the symmetry is preserved \emph{without reference to specific boundary conditions}. As we start from a bulk theory $\Lag^B$ with a symmetry, it is interesting to investigate whether a \emph{boundary compensating term} $\Lag^\p$ exists, so that \eqref{action symmetry invariance} holds without referring to a boundary condition \cite{DiPietro:2015zia,Bilal:2011gp}.
However, in general cases, one must impose specific symmetric boundary conditions \emph{and} add boundary terms so that the full action is stationary \eqref{stationarity condition} and symmetric \eqref{action symmetry invariance}.

\subsubsection{Currents and charges}

It is clear that the bulk theory charge $Q_B = \int_\Sigma J_B^0$ of the aforementioned symmetry  is, in general, no longer conserved after introducing a boundary, since the constant-time slice $\Sigma$ now has a boundary $\p \Sigma$. As a physical interpretation, the bulk current ``leaks'' from the boundary, and this ``leakage'' must be compensated by a boundary term. More precisely, what we need in addition to the bulk current $J^\mu_B$ is a \emph{boundary current} $J_\p^\hmu$ which lives on the boundary of the full theory, such that the equations
\begin{equation} \label{full conservation equations}
	\p_\mu J_B^\mu=0, \quad \p_\hmu J_\p^\hmu= J_B^n|_\p
\end{equation}
are satisfied. We also introduce the \emph{total} conserved current\footnote{See appendix \ref{app:deltadistributions} for details on $\gd$-distributions at the boundary.}
\begin{equation} \label{full current definition}
	J^\mu_\full= J_B^\mu+\delta(x^n) \proj^\mu_{\ph\mu\hmu} J^\hmu_\p,
\end{equation}
where $\proj^\mu_{\ph \mu \hmu}$ is a projector/embedding with $\proj^n_{\ph n \hmu}\!=0$, $ \proj^\hnu_{\ph \hnu \hmu}\!=\delta^\hnu_{\ph \hnu \hmu}$. The conservation equations \eqref{full conservation equations} can be expressed as:
\begin{equation}
	\p_\mu J^\mu_\full=\delta(x^n) J^n_\full.
\end{equation}
Note that the (boundary) conservation equation might only hold modulo boundary conditions. 
The full conserved charge of the theory is then given by
\begin{equation} \label{full theory charge}
	Q= \int_\Sigma J^0_B+\int_{\p\Sigma} J^0_\p=\int_\Sigma J_\full^0,
\end{equation}
whose conservation is easy to see from \eqref{full conservation equations}. 

Just as in pure bulk theories, there is more than one current leading to the same conserved charge associated to a given symmetry. Transformations of the currents that preserve the conservation equations and the charges are called the \emph{improvements} of the currents. 

For a pure bulk theory, an improvement locally takes the form
\begin{equation}
	J_B^\mu\mapsto \tilde J_B^\mu=J_B^\mu+\p_\nu M^{[\mu \nu]},
\end{equation}
which preserves the conservation equations and charges: $ \p_\mu \tilde J_B^\mu=0$ and  $\tilde Q_B=\int_\Sigma \tilde J_B^0\equiv Q_B$.
\\
For a theory with boundary, an improvement takes the form
\begin{equation}\label{improvement with boundary}
	\begin{Bmatrix}
		J_B^\mu
		\\J_\p^\hmu
	\end{Bmatrix}
	\mapsto 
	\begin{Bmatrix}
		\tilde J_B^\mu=J_B^\mu+\p_\nu M^{[\mu \nu]}
		\\\tilde J_\p^\hmu=J_\p^\hmu +M^{n\hmu} +\p_\hnu m^{[\hmu\hnu]}
	\end{Bmatrix},
\end{equation}
which preserves the conservation equations \eqref{full conservation equations} and the charge \eqref{full theory charge}. The improvement of the bulk current induces an improvement on the boundary current, and the boundary current may be further improved by a pure boundary improvement. 

Note that it is sometimes possible to completely ``improve away'' the boundary part of the conserved current, in particular if there are no degrees of freedom on the boundary. In that case the bulk charge $Q_B$ is conserved even in the presence of a boundary, but then it is, in general, sensitive to bulk improvements. This is the approach of the authors of \cite{DiPietro:2015zia}.

\subsection{Noether's theorem on manifolds with boundary}\label{sec:Noether's thm with bdry}
Now that we have discussed the properties of conserved currents and charges in boundary theories, let us investigate how we can compute them in a particular model. We present a modification of Noether's theorem that yields bulk and boundary currents in the sense we defined above. The special case of a theory with boundary (and boundary terms) but without boundary dynamics is discussed in detail in \cite{DiPietro:2015zia}.

\subsubsection{Currents and charges without boundary}
For completeness, let us quickly repeat Noether's theorem in pure bulk theories. A symmetry is an off-shell transformation of the fields that leaves the action invariant:
\begin{equation}
\dsym S=0.
\end{equation}
If the transformation is rigid (i.e.\ leaves spacetime invariant) and assuming that fields ``fall off'' at infinity the above condition is equivalent to
\begin{equation}
	\dsym \Lag^B=\p_\mu V^\mu
\end{equation}
for some bulk vector field $V^\mu$. On the other hand, a generic variation of the Lagrangian is also given by
\begin{equation}
	\dsym \Lag^B=\eom^B[\Phi] \dsym \Phi +\p_\mu\big(\pd{\Lag^B}{(\p_\mu \Phi|_\p)}\dsym \Phi\big)
	\overset{\text{\tiny on-shell}}{=}
	\p_\mu\big(\pd{\Lag^B}{(\p_\mu \Phi|_\p)}\dsym \Phi\big).
\end{equation}
We find that on-shell $0=\p_\mu(V^\mu-\pd{\Lag^B}{(\p_\mu\Phi)}\dsym \Phi)$, thus the \emph{bulk Noether current}
\begin{equation} \label{bulk Noether current}
	J_B^\mu= -\pd{\Lag^B}{(\p_\mu \Phi)}\dsym \Phi + V^\mu, \quad \p_\mu J_B^\mu=0,
\end{equation}
is divergence-free (on-shell). Since there is no boundary present, divergence-freeness implies the conservation of the charge \begin{equation}
	Q_B=\int_\Sigma J_B^0,
\end{equation}
where $\Sigma$ is a constant-time slice of $M$, since $
\p_0 Q_B= \int_\Sigma \p_0 J^0=-\int_\Sigma \p_i J^i=0$.

\subsubsection{Currents and charges with boundary}

As we restricted to symmetries of boundary theories that come from a bulk theory, $J_B^\mu$ from \eqref{bulk Noether current} is still a valid divergence-free current by the same argument as above, so we can use it as the bulk part of the full current \eqref{full current definition}. The task at hand is to now find a boundary current which satisfies $\p_\hmu J_\p^\hmu= J_B^n|_\p$. We want to apply a similar strategy as in the pure bulk theory: compare the (off-shell) symmetry variation of the action with an on-shell variation. On the off-shell side we get
\begin{equation}
0=\dsym S|_{G=0}=
\int_M\p_\mu V^\mu +\int_{\p M} \dsym \Lag^\p|_{G=0}
=\int_{\p M} [V^n+\dsym \Lag^\p]_{G=0},
\end{equation}
which implies that 
\begin{equation}\label{off-shell variation boundary}
[V^n+\dsym \Lag^\p]_{G=0}=\p_\hmu K^\hmu
\end{equation}
for some boundary vector field $K^\hmu$. 

On the on-shell side, we now use equations of motion \emph{and} the stationarity condition \eqref{stationarity condition}. By varying the boundary Lagrangian directly and assuming $G[\ldots]|_\p=0$ we get
\begin{equation}
\begin{aligned}
\dsym \Lag^\p&=
\left[\pd{\Lag^\p}{\Phi|_\p}-\p_\hmu \pd{\Lag^\p}{(\p_\hmu\Phi|_\p)}\right]\dsym \Phi|_\p 
+ \eom^\p[X]\dsym X \\
&\pheq +\p_\hmu\big(\pd{\Lag^\p}{(\p_\hmu X)}\dsym X+ \pd{\Lag^\p}{(\p_\hmu \Phi|_\p)}\dsym \Phi|_\p\big) .
\end{aligned}
\end{equation}
To rewrite the first term, let us plug in $\delta = \dsym$ into \eqref{stationarity condition} (still assuming $G[\ldots]=0$) and use the definition of the bulk Noether current \eqref{bulk Noether current} to rewrite it:
\begin{equation}\label{identity on Ac}
\p_\hmu A^\hmu \overset{\text{\tiny on-shell}}{=} \left[\pd{\Lag^\p}{\Phi|_\p}-\p_\hmu \pd{\Lag^\p}{(\p_\hmu\Phi|_\p)}\right] \dsym \Phi|_\p +[V^n-J_B^n]_\p.
\end{equation}
Plugging this into the previous equation and going on-shell, we get
\begin{equation}
\dsym \Lag^\p \overset{\text{\tiny on-shell}}{=}
[J_B^n-V^n]_\p+\p_\hmu \big(A^\hmu + \pd{\Lag^\p}{(\p_\hmu X)}\dsym X+ \pd{\Lag^\p}{(\p_\hmu \Phi|_\p)}\dsym \Phi|_\p\big).
\end{equation}
We can now compare this on-shell variation to the off-shell variation in \eqref{off-shell variation boundary} and see
\begin{equation}
J_B^n|_\p = \p_\hmu \big( K^\hmu - A^\hmu - \pd{\Lag^\p}{(\p_\hmu X)}\dsym X - \pd{\Lag^\p}{(\p_\hmu \Phi|_\p)}\dsym \Phi|_\p).
\end{equation}
Thus, we can read off the boundary Noether current
\begin{equation}\label{boundary conservation}
	J_\p^\hmu= K^\hmu-A^\hmu 
	-\pd{\Lag^\p}{(\p_\hmu X)}\dsym X- \pd{\Lag^\p}{(\p_\hmu \Phi|_\p)}\dsym \Phi|_\p, \quad \p_\hmu J_\p^\hmu=J_B^n|_\p.
\end{equation}
Together with the bulk current \eqref{bulk Noether current}, this forms a conserved boundary theory current in the sense of \eqref{full conservation equations}. We recall that $K^\hmu$ is defined by the symmetry condition \eqref{off-shell variation boundary} and $A^\hmu$ is defined by the stationarity condition \eqref{stationarity condition}. Notice that through the dependency on $A^\hmu$, the boundary Noether current may explicitly depend on the boundary condition, even if the bulk variation is compensated at the boundary in a boundary-condition-independent way (cf.\ section \ref{sec:symmetries in boundary theories}).

\subsubsection{Special case: Energy-momentum tensor}\label{sec:special case EMT}
In the presence of a boundary, Noether's theorem applies to spacetime translations as well: the total energy-momentum tensor is\footnote{Strictly speaking, the index $\nu$ does not take the value $n$: translations in $x^n$-direction are no longer symmetries. However, we may still consider this part of the tensor even though it does not lead to a conserved charge.}
\begin{equation}
T_\nu^{\ph \nu \mu}=(T^B)_\nu^{\ph \nu \mu} +\delta(x^n)\proj^\mu_{\ph\mu \hmu}\proj_\nu^{\ph \nu\hnu}(T^\p)_\hnu^{\ph \hnu \hmu}.
\end{equation}
Here, the bulk contribution is
\begin{equation}
(T^B)_\nu^{\ph \nu \mu}=-\frac{\p \Lag^B}{\p(\p_\mu \Phi)}\p_\nu \Phi +\delta_\nu^{\ph\nu\mu} \Lag^B,
\end{equation}
while the boundary contribution is
\begin{equation}
(T^\p)_\hnu^{\ph \hnu \hmu}=-\frac{\p \Lag^\p}{\p(\p_\hmu X)}\p_\hnu X
-\frac{\p \Lag^\p}{\p(\p_\hmu \Phi)}\p_\hnu \Phi
+\delta_{\ph\hmu\hnu}^\hmu \Lag^\p,
\end{equation}
where summation over fields is implied. 
The conservation equations are given by
\begin{equation}
\begin{aligned}
\p_\mu (T^B)_\nu^{\ph \nu \mu}&=0, \\
\p_\hmu(T^\p)_\hnu^{\ph \hnu \hmu}&=T_\hnu^{\ph \hnu n }|_\p,
\end{aligned}
\end{equation}
and the total tensor satisfies $\p_\mu T_\nu^{\ph \nu \mu}= \delta(x^n) \proj_\nu^{\ph \nu\hnu} T_\hnu^{\ph \hnu n }$. The momenta along the tangential $\hnu$-directions are conserved
\begin{equation}
P_\hnu=\int_\Sigma (T^B)_\hnu^{\ph \hnu 0}+\int_{\p\Sigma} (T^\p)_\hnu^{\ph \hnu 0}, \quad\quad  \p_0 P_\hnu=0,
\end{equation}
while $P_n=\int_\Sigma (T^B)_n^{\ph n 0}$
is clearly not conserved in general: $\p_0 P_n =-\int_\Sigma \p_i(T^B)_n^{\ph n 0}=-T_n^{\ph n n}|_\p$.
This is consistent with a flat boundary: the theory is only invariant under spacetime translations tangential to the boundary.

As far as improvements are concerned, the most general improvement takes the form 
\begin{equation}\label{improvements on EMT}
	\begin{Bmatrix}
		(T^B)_{\mu\nu}\\
		(T^\p)_{\hmu\hnu}
	\end{Bmatrix}
	\mapsto
	\begin{Bmatrix}
		(T^B)_{\mu\nu}+\p^\rho M_{\nu[\mu\rho]}\\
		(T^\p)_{\hmu\hnu}+M_{\hnu n \hmu} +\p^\hrho m_{\hnu[\hmu\hrho]}
	\end{Bmatrix},
\end{equation}
which, as before, leads to the same charges. However, if we restrict to improvements of symmetric tensors containing up to spin 1 components \cite{Dumitrescu:2011iu}, the allowed improvements take the form:
\begin{equation}
\begin{Bmatrix}
(T^B)_{\mu\nu}\\
(T^\p)_{\hmu\hnu}
\end{Bmatrix}
\mapsto
\begin{Bmatrix}
(T^B)_{\mu\nu}+\p_\nu U_\mu -\eta_{\mu\nu}\p^\rho U_\rho \\
(T^\p)_{\hmu\hnu}+\eta_{\hmu\hnu}U_n-\eta_{n\hnu} U_\hmu
+\p_\hnu u_\hmu -\eta_{\hmu\hnu}\p^\hrho u_\hrho
\end{Bmatrix},
\end{equation}
where $U_\mu$ is the bulk improvement and $u_\hmu$ the boundary improvement.

\section{Boundary supercurrent multiplets in 3D}
\label{sec:general boundary supercurrent multiplets}
We want to study supercurrent multiplets of theories on manifolds with boundary. In particular, we consider the special case of bulk theories with 3D $\mc N=2$ supersymmetry, broken to 2D $\nzerotwo$ due to the boundary. While our discussion is limited to this particular case, the strategy is expected to work in greater generality.
We start by recalling the definitions and some facts about supercurrent multiplets, following \cite{Dumitrescu:2011iu} (see also \cite{Magro:2001aj} for a connection to a superspace Noether procedure).

The defining properties of a supercurrent multiplet are:
{ %
\renewcommand{\theenumi}{\alph{enumi}}
\begin{enumerate}
	\item \label{condition a}The energy-momentum tensor $(T^B)^{\mu\nu}$ should be a component of the multiplet. It is also the only component with spin 2.
	\item \label{condition b}The supercurrents, i.e.\ conserved currents associated to supersymmetry, are components of the multiplet. They are the only components with spin $3/2$. No component other than the supercurrents and the energy-momentum tensor are allowed to have spin larger than $1$.
	\item \label{condition c}The supercurrent multiplet is not unique: it allows for (supersymmetrically complete) improvements of its components.
	\item \label{condition d}The multiplet is \emph{indecomposable}, so it may have non-trivial submultiplets, but it may not be decomposed into two independent decoupled multiplets. 
\end{enumerate}}
The components of a supercurrent multiplet (in particular, the conserved currents) are only unique up to improvements. However, improving one component and not the others breaks the structure of the multiplet. Hence, to consistently improve the supercurrent multiplet, we must restrict to improvements of all components which are related in a certain way (specifically, the improvement terms have to form a supersymmetry multiplet themselves; details are in \cite{Dumitrescu:2011iu}). In other words, if one is given two components (e.g.\ a supercurrent and an energy-momentum tensor), one may have to improve one of them such that they can be part of the same supercurrent multiplet. We say two conserved currents which have been improved such that they are part of a consistent supercurrent multiplet are in the same \emph{improvement frame}. \label{introduction of 'improvement frame'}

For some theories the supercurrent multiplet may be improved into a smaller multiplet (e.g.\ to obtain an $\Rc$-multiplet or a Ferrara-Zumino multiplet). There are still improvements that preserve this smaller multiplet \cite{Dumitrescu:2011iu,Komargodski:2010rb,Dedushenko:2015opz}. We will recall the case of 3D $\mc N=2$ theories in more detail.

\subsection{In bulk theories}
We will focus on three-dimensional theories with two-dimensional boundaries. In this section we recall the defining relations and properties of supercurrent multiplets in three-dimensional bulk theories with $\mc N=2 $ supersymmetry from \cite{Dumitrescu:2011iu}. 

The most general supercurrent multiplet satisfying the conditions (\ref{condition a})--(\ref{condition d}) (called the \emph{$\Sc$-multiplet}) consists of three superfields, $\Sc_{\alpha\beta}, \chi_\alpha,\Y_\alpha$ with $\Sc_{\alpha\beta}$ real, $\chi_\alpha,\Y_\alpha$ fermionic, and a complex constant $C$. They must satisfy the defining relations:
\begin{equation}\label{constraints 3d n=2}
	\begin{aligned}
		\Db^\beta \Sc_{\alpha\beta}&=\chi_\alpha+\Y_\alpha,
		\\\Db_\alpha\chi_\beta&=\tfrac 1 2 C\eps_{\alpha\beta},
		\\D^\alpha\chi_\alpha+\Db^\alpha\chib_\alpha&=0,
		\\D_\alpha\Y_\beta+D_\beta\Y_\alpha&=0,
		\\\Db^\alpha \Y_\alpha+C&=0.
	\end{aligned}
\end{equation}
These defining relations are solved by the following expansions (using bispinor relations \eqref{bispinor relations}):

\begin{subequations}\label{s,x,y expansions}
	\begin{align}
		\begin{split}
			\Sc_\mu&=
			j_\mu -i \theta(S_\mu +\tfrac i {\sqrt 2} \gamma_\mu \omegab) -i \thetab(\Sb_\mu -\tfrac i {\sqrt 2} \gamma_\mu \omega)+\tfrac i 2 \theta^2 \Yb_\mu +\tfrac i 2 \thetab^2 Y_\mu \\
			&\hspace{.4cm}-(\theta\gamma^\nu \thetab) \big(2T_{\nu\mu} -\eta_{\mu\nu} A-\tfrac 1 4 \eps_{\mu\nu\rho}H^\rho\big)
			-i\theta\thetab\big(\tfrac 1 4 \eps_{\mu\nu\rho}F^{\nu\rho}+\eps_{\mu\nu\rho}\p^\nu j^\rho\big)
			\\&\hspace{.4cm}
			+\tfrac 1 2 \theta^2\thetab\big(\gamma^\nu\p_\nu S_\mu-\tfrac i{\sqrt 2}\gamma_\mu \gamma_\nu\p^\nu \omegab\big)
			+\tfrac 1 2 \thetab^2\theta\big(\gamma^\nu\p_\nu \Sb_\mu+\tfrac i{\sqrt 2}\gamma_\mu \gamma_\nu\p^\nu \omega\big)
			\\&\hspace{.4cm} 
			-\tfrac 1 2 \theta^2\thetab^2\big(\p_\mu\p^\nu j_\nu-\tfrac 1 2 \p^2 j_\mu \big),
		\end{split}
		\\
		\begin{split}
			\chi_\alpha&=-i\lambda_\alpha(y)+\theta_\beta\big[\delta_\alpha^{\ph\alpha\beta} D(y)
			-(\gamma^\mu)_\alpha^{\ph \alpha\beta}\big(H_\mu(y)-\tfrac i 2 \eps_{\mu\nu\rho}F^{\nu\rho}(y)\big)\big]
			\\&\hspace{.4cm}
			+\tfrac 12\thetab_\alpha C -\theta^2(\gamma^\mu\p_\mu\lambdab)_\alpha(y),
		\end{split}
		\\\begin{split}
			\Y_\alpha&=\sqrt 2 \omega_\alpha +2\theta_\alpha B +2i \gamma^\mu_{\alpha\beta}\thetab^\beta Y_\mu 
			+\sqrt 2 i(\theta\gamma^\mu\thetab)\eps_{\mu\nu\rho} (\gamma^\nu\p^\rho\omega)_\alpha
			\\&\hspace{.4cm}
			+\sqrt 2 i\theta\thetab (\gamma^\mu\p_\mu\omega)_\alpha
			+i\theta^2 \gamma^\mu_{\alpha\beta}\thetab^\beta \p_\mu B
			-\thetab^2\theta_\alpha \p_\mu Y^\mu 
			+\tfrac{1}{2\sqrt 2} \theta^2\thetab^2\p^2\omega_\alpha,
		\end{split}
	\end{align}
\end{subequations}
where $(S^\mu)_\alpha,(\Sb^\mu)_\alpha$ are conserved supercurrents, $T_{\mu\nu}$ is a symmetric energy-momentum tensor, and 
\begin{equation}\label{definitions of auxiliaries}
	\begin{aligned}
		\lambda_\alpha&=-2 (\gamma^\mu \Sb_\mu)_\alpha +2\sqrt 2 i \omega_\alpha,
		\\D&= -4 T^\mu_{\ph\mu\mu}+4A,
		\\B&= A+i\p_\mu j^\mu,
		\\ \d H&=0,\quad \d Y=0,\quad \d F=0,
	\end{aligned}
\end{equation}
where  $H, F, Y$ are forms with components $H_\mu, F_{\mu\nu}, Y_\mu$. 
Additionally, $y$ is the ``chiral'' coordinate $y^\mu=x^\mu-i\theta\gamma^\mu\thetab$.
If the forms $Y$ or $H$ are exact, the superfields $\Y_\alpha$ or $\chi_\alpha$ may be written as covariant derivatives: If $Y_\mu=\p_\mu x$, then $\Y_\alpha=D_\alpha X$ where $X|_{\theta^0}=x$, and if $H_\mu=\p_\mu g$, then $\chi_\alpha=i\Db_\alpha G$ where $G|_{\theta^0}=g$.

\subsubsection{Improvements}

The expansions \eqref{s,x,y expansions} together with the relations \eqref{definitions of auxiliaries} and the conservation of currents $\p_\mu (S^\mu)_\alpha=0$, $\p^\mu T_{\mu\nu}=0$ do not form the only solution of the constraints \eqref{constraints 3d n=2}. We may improve without violating the constraints
\begin{equation}\label{bulk improvements of sucumu}
	\begin{aligned}	
		\Sc_\mu &\mapsto \Sc_\mu+\tfrac 1 4 \gamma_\mu^{\alpha\beta} [D_\alpha,\Db_\beta] U,
		\\\chi_\alpha&\mapsto \chi_\alpha-\Db^2 D_\alpha U,
		\\ \Y_\alpha& \mapsto \Y_\alpha -\tfrac 1 2 D_\alpha \Db^2 U,
	\end{aligned}
\end{equation}
where $U=u+\theta\eta-\thetab\etab+\theta^2 N-\thetab^2 \bar N+(\theta\gamma^\mu\thetab) V_\mu -i\theta\thetab K +\ldots$ is a real superfield. The improvement transforms
\begin{equation}\label{improvements explicitly}
	\begin{aligned}
		(S_\mu)_\alpha&\mapsto (S_\mu)_\alpha+\eps_{\mu\nu\rho}(\gamma^\nu\p^\rho\eta)_\alpha, \\
		T_{\mu\nu} & \mapsto T_{\mu\nu}+\tfrac 1 2 (\p_\mu\p_\nu -\eta_{\mu\nu} \p^2)u, 
		\\H_\mu& \mapsto  H_\mu -4\p_\mu K, 
		\\ F_{\mu\nu} &\mapsto F_{\mu\nu}-4 (\p_\mu V_\nu- \p_\nu V_\mu),
		\\ Y_\mu &\mapsto Y_\mu -2\p_\mu \bar N .
	\end{aligned}
\end{equation}
The multiplet $\Sc_\mu$ may be improved into smaller multiplets. In particular:
\begin{enumerate}
	\item If $C=0$, $\chi_\alpha=i \Db_\alpha G$ (i.e.\ $H$ is exact) and there exists a well-defined improvement $U$ such that $G=2i\Db^\alpha D_\alpha U$, then it sends $\chi_\alpha$ to zero and we obtain a \emph{Ferrara-Zumino multiplet} \cite{Ferrara:1974pz}:
	\begin{equation}
		\begin{aligned}
			\Db^\beta \mc J_{\alpha\beta}&=\Y_\alpha,
			\\D_\alpha\Y_\beta+D_\beta \Y_\alpha&=0, \quad \Db^\alpha \Y_\alpha=0.
		\end{aligned}
	\end{equation}
	\item If $C=0$, $\Y_\alpha=D_\alpha X$ (i.e.\ $Y$ is exact) and there exists a well-defined improvement $U$ such that $X=\tfrac 1 2 \Db^2 U$, then it sends $\Y_\alpha$ to zero and we obtain an \emph{$\Rc$-multiplet} \cite{Gates:1983nr}:
	\begin{equation}\label{constraints 3d n=2 R-multiplet}
		\begin{aligned}
			\Db^\beta\Rc_{\alpha\beta}&=\chi_\alpha, \\
			\Db_\alpha \chi_\beta=0, \quad&  D^\alpha\chi_\alpha+\Db^\alpha \chib_\beta=0.
		\end{aligned}
	\end{equation}
	In this case, the lowest component $j_\mu$ of the multiplet $\Rc_\mu$ (we relabel $\Sc_\mu$ to $\Rc_\mu$) is a conserved $R$-current (in the general $\Sc_\mu$-multiplet, $j^\mu$ is not conserved; however, we still call it a ``non-conserved $R$-current'').
	The $\Rc$-multiplet will be the primary focus of our example in section \ref{sec:LG model}.
	\item If $C=0$ and the improvements from (1) and (2) coincide, we can improve both superfields $\chi_\alpha,\Y_\alpha$ to zero \emph{simultaneously}. In that case we obtain a \emph{superconformal multiplet}
	\begin{equation}
		\Db^\beta \Sc_{\alpha\beta}=0.
	\end{equation}
\end{enumerate}
Note that even if smaller multiplets exist, they are still not unique: we may further improve the smaller multiplets without violating the respective additional constraints. For example, in the case of the $\Rc$-multiplet, the improvements that preserve the defining constraints are transformations
\begin{equation}\label{bulk improvements of R}
	\begin{aligned}
		\Rc_\mu &\mapsto \Rc_\mu+\tfrac 1 4 \gamma_\mu^{\alpha\beta} [D_\alpha,\Db_\beta] U,
		\\\chi_\alpha&\mapsto \chi_\alpha-\Db^2 D_\alpha U,
		\\  &D_\alpha \Db^2 U=0.
	\end{aligned}
\end{equation}

\subsubsection{Brane currents}

We may associate to the closed forms $F, H, Y, C$ the \emph{brane currents} defined by taking their Hodge dual:
\begin{equation}\label{brane currents}
	C_\mu\sim \eps_{\mu\nu\rho}F^{\nu\rho}, \quad 
	C_{\mu\nu}\sim \eps_{\mu\nu\rho} H^\rho, \quad
	C'_{\mu\nu}\sim \eps_{\mu\nu\rho}\Yb^\rho,\quad
	C_{\mu\nu\rho}\sim \eps_{\mu\nu\rho} \Cb.
\end{equation}
Note that these are conserved by construction, $\p_\mu C^\mu_{\ph \mu \mu_1\ldots \mu_k}=0$, since the forms are closed.\footnote{In coordinate-free notation this is written as $\d * C=0$ which follows trivially if $C=*A$, with $\d A=0$.} Then, the \emph{brane charges} defined by $Z_{\mu_1\ldots\mu_k}=\int_{\Sigma} C^0_{\ph 0 \mu_1\ldots \mu_k}$ are conserved as well. In addition, they are also invariant under the improvements \eqref{improvements explicitly}. The brane charges, if they are non-trivial, are central charges of the supersymmetry algebra (but not of the Poincaré algebra). This is motivated by studying the explicit commutators \cite{Dumitrescu:2011iu} that follow from the multiplet structure of $\Sc_\mu$:\footnote{Recall, the action of physical supercharges via commutators is related to the action via (super)-differential operators by
\begin{equation*}
	[\xi^\alpha \mc{Q}_\alpha- \xib^\alpha \bar{\mc{Q}}_\alpha, X]=i(\xi^\alpha Q_\alpha -\xib^\alpha \Qb_\alpha) X\eqqcolon i \dsym^{\xi,\xib} X.
\end{equation*}}
\begin{equation}
	\begin{aligned}
		\{\Qbc_\alpha, (S_\mu)_\beta\}&=\gamma^\nu_{\alpha\beta}(2T_{\nu\mu}-\tfrac 1 4 \eps_{\mu\nu\rho}H^\rho)
		+i\eps_{\alpha\beta} \tfrac 1 4 \eps_{\mu\nu\rho}F^{\nu\rho}+\text{total derivatives,}
		\\\{\Qc_\alpha, (S_\mu)_\beta\}&=\tfrac 1 4 (\gamma_\mu)_{\alpha\beta} \Cb 
		+i\eps_{\mu\nu\rho}\gamma^\nu_{\alpha\beta} \Yb^\rho,
	\end{aligned}
\end{equation}
where we may find non-trivial central charges in the supersymmetry algebra upon integration.
Each current $C_{\mu\mu_1\ldots\mu_k}$ and the corresponding charge $Z_{\mu_1\ldots\mu_k}$ is associated to a $k$-brane.
Hence, the brane charges form a physical obstruction to improvements into smaller multiplets. 
In particular, a non-zero charge associated to $F$ or $H$ obstructs the existence of a Ferrara-Zumino multiplet, and a non-zero charge associated to $Y$ obstructs the existence of an $\Rc$-multiplet. 

\subsection{In theories with boundary}\label{sec:in theories with boundary}
The introduction of a boundary affects supercurrent multiplets in two obvious ways. 

First, supersymmetry is broken to a subalgebra, where our main focus will be 2D $\nzerotwo$. The bulk supercurrent multiplets, previously 3D $\mc N=2$ superfields, now decompose under the subalgebra to $\zerotwo$-superfields. We spell out this decomposition in appendix \ref{app: S decomposition to zerotwo}. Similarly, the constraints \eqref{constraints 3d n=2} now decompose into constraints of the $\zerotwo$-superfields. We will spell out this decomposition in the next subsection. 

Second, a boundary changes the conserved currents of \emph{remaining} symmetries by supplementing the bulk currents \eqref{bulk Noether current} with boundary currents \eqref{boundary conservation} satisfying appropriate conservation equations. The conservation equations must follow from the constraints that define the supercurrent multiplets, as in bulk theories, and the \emph{full} supercurrent multiplets will now consist of bulk and boundary pieces. The schematic form of full supercurrent multiplets reads
\begin{equation}\label{form of full SuCuMu}
	\begin{aligned}
		\Sc_\mu^\full&= \Sc_\mu^B+\delta(x^n) \proj^\hmu_{\ph\hmu \mu} \Sc_\hmu^\p,
		\\\chi_\alpha^\full&= \chi_\alpha^B+\delta(x^n) \chi_\alpha^\p,
		\\\Y_\alpha^\full&= \Y_\alpha^B+\delta(x^n)  \Y_\alpha^\p,
	\end{aligned}
\end{equation}
where once again $\proj^\mu_{\ph\mu \hmu}$ is an embedding. 

Let us briefly discuss how the conditions (\ref{condition a})--(\ref{condition d}) are modified. It is clear that the new superfields should contain the \emph{full} conserved currents of unbroken symmetries, in the sense of section \ref{sec:Noether's thm with bdry} (conditions (\ref{condition a}), (\ref{condition b})). Furthermore, improvements of the full conserved currents in the sense of \eqref{improvement with boundary} that form consistent multiplets under the smaller subalgebra are improvements of the $\nzerotwo$ supercurrent multiplets (condition (\ref{condition c})). However, under the smaller symmetry algebra, the previously indecomposable (bulk) multiplet decomposes into possibly several indecomposable multiplets of the remaining symmetry subalgebra. Therefore condition (\ref{condition d}) is not preserved in general.

\subsection{Bulk and boundary constraints by decomposition} \label{subsec:bulk and boundary constraints}

Let us recall the structure of subalgebras of the 3D $\N=2$ algebra which may be preserved after the introduction of a boundary. The (unbroken) symmetry algebra is generated by tangential translations $P_\hmu$, Lorentz transformations $M_{\hmu\hnu}$ in the unbroken directions, and one of the following:
\begin{enumerate}
	\item supercharges $Q_+,\Qb_+$ corresponding to a 2D $\zerotwo$-subalgebra  satisfying
	\begin{equation} \label{eq:remaining zerotwo algebra}
		(Q_+)^2 = 0, \quad \{Q_+,\Qb_+\}=-4i\p_+,
	\end{equation}
	\item their left-moving $(2,0)$ counterparts $Q_-,\Qb_-$,
	\item (real) supercharges $\fQ_-,\fQ_+$ corresponding to a 2D $\oneone$-subalgebra satisfying
	\begin{equation}
		(\fQ_\pm)^2 = - i \p_\pm, \quad \{\fQ_-,\fQ_+\} = 0.
	\end{equation}
\end{enumerate}
In this work we consider the only the first case. We want to determine constraint equations that define supercurrent multiplets in a 3D theory with boundary and 2D $\nzerotwo$ supersymmetry. To do so, we first decompose the 3D $\mc N=2$ bulk constraints into $\nzerotwo$ bulk constraints, and then investigate possible $\nzerotwo$ boundary constraints.

We supplement the superspace operators $Q_+,\Qb_+$ with covariant derivatives $D_+^\zerotwo$, $\Db_+^\zerotwo$ defined in appendix \ref{app:branching to zerotwo}. Let us emphasize that these are \emph{not} the operators $D_\ga$, $\Db_\ga$ acting on 3D $\N=2$ superspace, but are related to them by \eqref{D 3d vs D zerotwo}. We will omit the label $\zerotwo$ from now on; it should be clear from context whether a 3D $\N=2$ or a $\nzerotwo$ covariant derivative is meant.

\subsubsection{Bulk constraints}
Since the bulk conservation equations remain unchanged, the constraints on the bulk pieces in our ansatz \eqref{form of full SuCuMu} will remain the same component-wise. We merely have to decompose the multiplets and their constraints into constraints of $\zerotwo$-submultiplets. For simplicity we choose to do so for the case where the supermultiplet is an $\Rc$-multiplet:\footnote{The more general case of the $\Sc$-multiplet is quite similar and is given in the appendix, see \eqref{branching of S - chi1}--\eqref{branching of S - S}.} We will decompose the superfields $(\Rc_\mu, \chi_\alpha)$ and the constraints \eqref{constraints 3d n=2 R-multiplet}. We can achieve this using the \emph{branching coordinate} $\xi^\mu$. It has the defining property that in the coordinates $(\xi^\mu, \gt^+, \gt^-)$, the preserved supercharges $Q_+$ and $\Qb_+$ commute with $\gt^-$ and $\tb^-$; see appendix \ref{app:branching to zerotwo} for details. Another property is that $Q_+$, $\Qb_+$ do not involve a derivative in $\perp$-direction. In terms of $\xi$ we can decompose 
\begin{equation}\label{R chi decomp ansatz}
	\begin{aligned}
		\Rc_\mu^B(x,\theta,\thetab)&=\Rc^{B\zero}_\mu
		+\tm \Rc^{B\one}_\mu
		-\tbm \bar{\Rc^{B\one}_\mu}
		+\tm\tbm \Rc^{B\two}_\mu ,
		\\
		\chi^B_\alpha(x,\theta,\thetab)&=\chi^{B\zero}_\alpha
		+\tm \chi^{B\onea}_\alpha
		+\tbm \chi^{B\oneb}_\alpha
		+\tm\tbm \chi^{B\two}_\alpha,
	\end{aligned}
\end{equation}
where we now denote bulk fields by a superscript $B$, and boundary fields (to appear later) with a superscript $\p$. The bracketed number subscripts refer to the order in $\tm,\tbm$ we have expanded in. 
Here, each field on each right-hand side is a (super)function of $(\xi,\tp,\tbp)$. Furthermore, because $Q_+$, $\Qb_+$ commute with $\tm$, $\tbm$, the coefficient at each order in $\tm,\tbm$ is a $\zerotwo$-submultiplet --- the remaining supersymmetry group acts independently on each of them. This is a constructive way to decompose 3D $\mc N=2$ superfields with respect to the 2D $\nzerotwo$ subalgebra. In the appendix, we write the above decomposition explicitly for the $\Sc$-multiplet \eqref{+direction}--\eqref{perp-direction}, from which the $\Rc$-multiplet follows by setting appropriate terms to zero. 

In terms of the $\zerotwo$-submultiplets, the constraints \eqref{constraints 3d n=2 R-multiplet} are then written as the following collection of equations, where we use coordinates $\xi^+=\xi^0+\xi^1$, $\xi^-=\xi^0-\xi^1$ and $\xi^\perp=x^\perp+i(\tp\tbm-\tm\tbp)$:\footnote{Note that covariant derivatives acting on $\zerotwo$-superfields are $\zerotwo$-covariant derivatives.}
\\
From $\Db_- \chi_\alpha = 0$:
\begin{subequations}\label{1}
	\begin{align}
	\label{1a}
	\chi^{B\oneb}_\alpha&=0 , 
	\\ \label{1b}
	\chi^{B\two}_\alpha +2i \p_- \chi^{B\zero}_\alpha&=0.
	\end{align}
\end{subequations}
From $\Db_+ \chi_\alpha = 0$:
\begin{subequations}\label{2}
	\begin{align}\label{2a}
	\Db_+  \chi^{B\zero}_\alpha &= 0, 
	\\ \label{2b}
	\Db_+  \chi^{B\onea}_\alpha + 2i\p_\perp \chi^{B\zero}_\alpha&=0, 
	\\ \label{2c}
	\Db_+  \chi^{B\two}_\alpha &= 0.
	\end{align}
\end{subequations}
From $\Im D^\alpha \chi_\alpha = 0$:
\begin{subequations}\label{3}
	\begin{align}
	\label{3a}
	\Im(D_+\chi_-^{B\zero}-\chi_+^{B\onea})&=0,
	\\
	\label{3b}
	\Db_+\bar{\chi_-^{B\onea}}+\chi_+^{B\two} 
	-2i\p_-\chi_+^{B\zero }
	-2i\p_\perp \chi_-^{B\zero}&=0,
	\\
	\label{3c}
	\Im(D_+\chi_-^{B\two}-2i\p_-\chi_+^{B\onea}-2i\p_\perp\chi_-^{B\onea})&=0.
	\end{align}
\end{subequations}
Finally, the relation $\Db^\beta \Rc_{\alpha \beta}=\chi_\alpha$ yields:\footnote{These have already been simplified by some relations we found, e.g.\ \eqref{1a}.}
\begin{subequations}\label{4}
	\begin{align}\label{4a}
	 \chi^{B\zero}_\alpha&=\Db_+ \Rc^{B\zero}_{\alpha-} - \bar{\Rc^{B\one}_{\alpha+}}, 
	\\\label{4b}
	-\chi^{B\onea}_\alpha&=\Db_+ \Rc^{B\one}_{\alpha-} + \Rc^{B\two}_{\alpha+} + 2i\p_\perp\Rc^{B\zero}_{\alpha-} + 2i\p_-\Rc^{B\zero}_{\alpha+} , 
	\\\label{4c}
	0&=\Db_+ \bar\Rc^{B\one}_{\alpha\beta}, 
	\\\label{4d}
	\chi^{B\two}_\alpha&=\Db_+ \Rc^{B\two}_{\alpha-} + 2i\p_\perp\bar{\Rc^{B\one}_{\alpha-}} + 2i\p_-\bar{\Rc^{B\one}_{\alpha+}}.
	\end{align}
\end{subequations}
Note that we have not introduced \emph{any} new structure here: component-wise, equations \eqref{constraints 3d n=2 R-multiplet} have identical content as  \eqref{1a}--\eqref{4d}. In particular, the bulk conservation equations follow from these constraints. Let us explicitly verify this in the example of the conservation of the $R$-current $j_\mu^B$.  We start with equation \eqref{4b} setting $\alpha=+$ and taking the imaginary part. Using the reality of $\Rc_{\ga\gb}$ (which implies the reality of $\Rc^{B\zero}_{\alpha\beta}$ and $\Rc^{B\two}_{\alpha\beta}$), we arrive at
\begin{equation}\label{for conservation of R zero bulk 1}
	-\Im(\chi^{B\onea}_+)=\Im(\Db_+ \Rc^{B\one}_{+-}) + 2\p_\perp\Rc^{B\zero}_{+-} + 2\p_-\Rc^{B\zero}_{++}.
\end{equation}
Now consider equation \eqref{4a}; setting $\alpha=-$, conjugating, applying $\Db_+$ on both sides and finally taking the imaginary part we obtain
\begin{equation}\label{for conservation of R zero bulk 2}
	\Im(\Db_+\bar{\chi^{B\zero}_-})=\Im(\Db_+D_+ \Rc^{B\zero}_{--}) - \Im(\Db_+\Rc^{B\one}_{-+}).
\end{equation}
From the reality of $\Rc^{B\zero}_{\alpha\beta}$ we get $\Im(\Db_+D_+ \Rc^{B\zero}_{--})=2\p_+\Rc^{B\zero}_{--}$. Finally, we use \eqref{3a} to combine \eqref{for conservation of R zero bulk 1} and \eqref{for conservation of R zero bulk 2} into the bulk conservation equation for the $R$-current:
\begin{equation}
	2\p_+\Rc^{B\zero}_{--}+ 2\p_\perp\Rc^{B\zero}_{+-} + 2\p_-\Rc^{B\zero}_{++}=0.
\end{equation}
This equation also implies the bulk conservation of $(S_\mu^B)_+, (\Sb_\mu^B)_+$ and $T_{\mu+}^B$, as can be verified by the expansions \eqref{+direction}--\eqref{perp-direction}.

In an analogous fashion, we may derive the bulk conservation for $\Rc_{\alpha\beta}^{B\one}$ and $\Rc_{\alpha\beta}^{B\two}$. 
The conservation of $\Rc_{\alpha\beta}^{B\one}$ follows from \eqref{4a}--\eqref{4d} together with \eqref{3b}, and implies the conservation of  bulk supercurrents  $(S_\mu^B)_-, (\Sb_\mu^B)_-$ and the tensor $T_{\mu\perp}^B$. 
The conservation of $\Rc_{\alpha\beta}^{B\two}$ follows from \eqref{3c}, \eqref{4b} and \eqref{4d}, and implies the conservation of the bulk tensor  $T_{\mu-}^B$.

\subsubsection{Boundary constraints}
We now want to discuss constraints that we need to impose on the boundary parts $\Rc_\mu^\p$ and $\chi_\alpha^\p$ (and $\Y_\alpha^\p$ in the case of the $\Sc_\mu$-multiplet). Our guiding principle is of course the fact that the constraints need to impose boundary conservation \eqref{boundary conservation} on the components of the boundary multiplets.
The constraints can, in part, be deduced from the bulk constraints \eqref{1}--\eqref{4}. 
Let us elaborate on this point: bulk and boundary superfields are combined to our \emph{total} supercurrent multiplet
\begin{equation}\label{full R multiplet}
	\Rc_\mu^\full= \Rc_\mu^B+\delta(\xi^\perp)\proj_\mu^{\ph\mu\hmu} \Rc^\p_\hmu,
\end{equation}
where both bulk and boundary pieces can be decomposed into $\zerotwo$-multiplets:\footnote{This is essentially an embedding into 3D $\mc N=2$ superspace, see \cite{Drukker:2017xrb,Drukker:2017dgn}.}
\begin{equation}\label{R mult decomp in bulk and boundary}
	\begin{aligned}
		\Rc_\mu^B(x,\theta,\thetab)&=\Rc^{B\zero}_\mu
		+\tm \Rc^{B\one}_\mu
		-\tbm \bar{\Rc^{B\one}_\mu}
		+\tm\tbm \Rc^{B\two}_\mu,
		\\\Rc_\hmu^\p(x,\theta,\thetab)&=\Rc_\hmu^{\p\zero}+\tm\tbm \Rc_\hmu^{\p\two},
	\end{aligned}
\end{equation}
and similarly for auxiliary fields
\begin{equation}
	\begin{aligned}\label{chi multip decomp in bulk and boundary}
		\chi^B_\alpha(x,\theta,\thetab)&=\chi^{B\zero}_\alpha
		+\tm \chi^{B\onea}_\alpha
		+\tbm \chi^{B\oneb}_\alpha
		+\tm\tbm \chi^{B\two}_\alpha,
		\\ 
		\chi^\p_\alpha(x,\theta,\thetab)&=\chi^{\p\zero}_\alpha
		+\tm \chi^{\p\onea}_\alpha
		+\tbm \chi^{\p\oneb}_\alpha
		+\tm\tbm \chi^{\p\two}_\alpha.
	\end{aligned}
\end{equation}
As mentioned before, the bulk part is unchanged compared to the full 3D $\mc N=2$ current multiplet; it is simply decomposed into its $\zerotwo$-submultiplets. The ansatz for the boundary part is directly motivated by the expansions \eqref{+direction}--\eqref{perp-direction} of the bulk multiplet, which show that
\begin{equation}
	\begin{aligned}
		\Rc_\mu^{B\zero}&=j^B_\mu+\ldots,
		\\\Rc_\mu^{B\one}&=-i(S^B_\mu)_-+\ldots,
		\\\Rc_\mu^{B\two}&=-2K_{\mu-}+\ldots,
	\end{aligned}
\end{equation}
where for the $\Rc$-multiplet, $K_{\mu\nu}=2T_{\nu\mu}-\tfrac 1 4 \eps_{\mu\nu\rho}H^\rho$. We can conclude
\begin{equation}
	\begin{aligned}
		\Rc_\hmu^{\p\zero}&=j^\p_\hmu+\ldots,
		\\\Rc_\hmu^{\p\two}&=-2K_{\hmu-}^\p+\ldots,
	\end{aligned}
\end{equation}
as the bulk conserved currents have to be paired with their respective boundary currents. Note that we do not consider a boundary contribution to the ``broken'' $(S_\mu^B)_-$ currents, as we have no guiding principle in this framework. Then, due to the form of the full supercurrent multiplet \eqref{full R multiplet}, we postulate that the constraints applied to the boundary piece must be of similar form as \eqref{1a}--\eqref{4d}, but instead of imposing divergence-freeness \eqref{bulk Noether current}, they should impose \eqref{boundary conservation} on the \emph{remaining, conserved} boundary currents. 
 
We postulate the following adjustments on the constraints obtained from the bulk \eqref{1}--\eqref{4}, now applied to boundary multiplets $\Rc^{\p(*)}_{\alpha\alpha}$, in order to obtain correct conservation equations:
\begin{enumerate}
	\item \label{bulk-to-boundary 1} Firstly, since the boundary is two-dimensional with directions $x_{++}, x_{--}$ (in bispinor notation, cf.\ \eqref{bispinor relations}) we only have superfields $\Rc^{\p(*)}_{++}, \Rc^{\p(*)}_{--}$, and no superfield $\Rc^{\p(*)}_{+-}$.
	
	\item \label{bulk-to-boundary 2} Secondly, we do not consider  boundary contributions to the ``broken'' $(S_\mu^B)_-$ currents, and hence no $\Rc^{\p\one}_{\alpha\alpha}$ should appear. 
	
	\item \label{bulk-to-boundary 3} Lastly, to impose the correct conservation equation on the boundary, we must replace terms of the form $\p_\perp A^\p$ with $-A^B|_\p$ whenever such terms appear. This transformation precisely maps divergence-free equations \eqref{bulk Noether current} into boundary conservation equations \eqref{boundary conservation}. In addition, this replacement parses well with the fact that derivatives in the perpendicular direction make little sense when they act on boundary currents, in particular when the boundary currents are functions of purely boundary fields.
\end{enumerate}
The preliminary constraints on the boundary pieces then read:
\\
Analogs to \eqref{1}:
\begin{subequations}\label{1Bcor}
	\begin{align}
	\chi^{\p\oneb}_\alpha &= 0, 
	\\ \label{1bBcor}
	\chi^{\p\two}_-+2i \p_- \chi^{\p\zero}_- &= 0.
	\end{align}
\end{subequations}
Analogs to \eqref{2}:
\begin{subequations}
	\begin{align} 
	\Db_+\chi_-^{\p\zero}&=0,
	\\\label{2bBcor}
	\Db_+  \chi^{\p\onea}_+ -2i \chi^{B\zero}_+|_\p &=0, 
	\\ 
	\Db_+  \chi^{\p\two}_\alpha &= 0.
	\end{align}
\end{subequations}
Analogs to \eqref{3}:
\begin{subequations}
	\begin{align}
	\label{3aBcor}
	\Im(D_+\chi_-^{\p\zero}-\chi_+^{\p\onea})&=0,
	\\\label{3cBcor}
	\Im(D_+\chi_-^{\p\two}-2i\p_-\chi_+^{\p\onea}
	+2i\chi_-^{B\onea}|_\p)&=0.
	\end{align}
\end{subequations}
Lastly, the analogs to \eqref{4}:
\begin{subequations}\label{4Bcor}
	\begin{align}
	\label{4aBcor}
	\chi^{\p\zero}_+&=0, 
	\\\label{4bBcor}
	\chi^{\p\zero}_-&=\Db_+ \Rc^{\p\zero}_{--},
	\\\label{4cBcor}
	\chi^{\p\onea}_+&= -\Rc^{\p\two}_{++} +2i\Rc^{B\zero}_{+-}|_\p - 2i\p_-\Rc^{\p\zero}_{++} , 
	\\\label{4dBcor}
	\chi^{\p\onea}_-&= 2i\Rc^{B\zero}_{--}|_\p, 
	\\\label{4eBcor}
	\chi^{\p\two}_+&=-2i\bar{\Rc_{+-}^{B\one}}|_\p,
	\\\label{4fBcor}
	\chi^{\p\two}_-&=\Db_+ \Rc^{\p\two}_{--}  -2i\bar{\Rc^{B\one}_{--}}|_\p .
	\end{align}
\end{subequations}
Note that the naive application of adjustments (\ref{bulk-to-boundary 1})--(\ref{bulk-to-boundary 3}) leads to three further relations, which we have intentionally omitted above. These are: the analog of \eqref{1b} for $\alpha=+$, which reads
\begin{equation}\label{discarded eqn 1}
	\chi^{\p\two}_++2i \p_- \chi^{\p\zero}_+=0,
\end{equation}
the analog of \eqref{2b} for $\alpha=-$, which reads
\begin{equation}\label{discarded eqn 2}
	\Db_+  \chi^{\p\onea}_- -2i \chi^{B\zero}_-|_\p,
\end{equation}
and lastly, the analog of \eqref{3b} reads
\begin{equation}\label{discarded eqn 3}
	\Db_+\bar{\chi_-^{\p\onea}}+\chi_+^{\p\two} 
	-2i\p_-\chi_+^{\p\zero} 
	+2i \chi_-^{B\zero}|_\p=0.
\end{equation}
We argue that these relations must be discarded from the set of constraints of boundary multiplets. To see this, note that the first relation \eqref{discarded eqn 1} is compatible with equations \eqref{4aBcor} and \eqref{4eBcor} \emph{only if} $\Rc_{+-}^{B\one}|_\p=0$. Similarly, the second relation \eqref{discarded eqn 2} in agreement with equations \eqref{4dBcor} and \eqref{4a} again \emph{only if} $\Rc_{+-}^{B\one}|_\p=0$. Lastly, the third relation \eqref{discarded eqn 3} is consistent with \eqref{4aBcor}, \eqref{4dBcor}, \eqref{4eBcor} and \eqref{4a}, once more \emph{only if} $\Rc_{+-}^{B\one}|_\p=0$. 
Hence, including any of the three relations in the constraints of boundary $\zerotwo$-supermultiplets would impose
\begin{equation}
	\Rc_{+-}^{B\one}|_\p=-i(S^B_\perp)_-|_\p+\ldots=0.
\end{equation}
However, imposing this condition would imply the conservation of the ``broken'' charge $Q_-$, as the current $(S^B_\mu)_-$ would fulfill \eqref{full conservation equations} with a trivial boundary part. Therefore, if we did not omit relations \eqref{discarded eqn 1}--\eqref{discarded eqn 3}, we would impose the conservation of the ``broken'' charges $Q_-,\Qb_-$, which is inconsistent with the ``breaking'' of $P_\perp$ and the explicit conservation of the charges $Q_+,\Qb_+$. As a last argument, we note that the three omitted relations are \emph{not} required to obtain the boundary conservation equations for the remaining symmetries ($\nzerotwo$ supersymmetry, $R$-symmetry, 2D Poincaré symmetry). We will verify this in the remainder of this subsection by explicitly checking that the boundary conservation equations indeed follow from constraints \eqref{1Bcor}--\eqref{4Bcor}.

We spell out the derivation of the boundary conservation equation in the example of $\Rc_{\alpha\alpha}^{\p\zero}$: Taking the imaginary part of \eqref{4cBcor} and using the reality of the multiplet, we obtain
\begin{equation}\label{for conservation of R zero bdy 1}
	\Im(\chi^{\p\onea}_+)=  2\Rc^{B\zero}_{+-}|_\p - 2\p_-\Rc^{\p\zero}_{++}.
\end{equation}
Now we take equation \eqref{4bBcor}, conjugate it, apply $\Db_+$ on both sides and finally take imaginary part again to obtain:
\begin{equation}\label{for conservation of R zero bdy 2}
	\Im(\Db_+\bar{\chi^{\p\zero}_-})=\Im(\Db_+D_+ \Rc^{\p\zero}_{--}).
\end{equation}
Again, due to the reality of $\Rc^{\p\zero}_{--}$, we have that $\Im(\Db_+D_+ \Rc^{\p\zero}_{--})=2\p_+\Rc^{\p\zero}_{--}$. Finally, we can combine equations \eqref{for conservation of R zero bdy 1} and \eqref{for conservation of R zero bdy 2} using \eqref{3aBcor} into the  conservation equation for the boundary $R$-current:
\begin{equation}
	2\Rc^{B\zero}_{+-}|_\p - 2\p_-\Rc^{\p\zero}_{++}-2\p_+\Rc^{\p\zero}_{--}=0.
\end{equation}
Like its bulk counterpart, this superfield equation also implies the boundary conservation of $(S_\hmu^\p)_+, (\Sb_\hmu^\p)_+$ and $T_{\hmu+}^\p$.

In a similar fashion, we may derive boundary conservation of $\Rc_{\alpha\alpha}^{\p\two}$ using equations \eqref{4fBcor}, \eqref{3cBcor}, \eqref{4cBcor} and \eqref{4dBcor}. Component-wise it implies the conservation of the boundary tensor $T_{\hmu_-}^\p$. No boundary analog to bulk conservation of $(S_\mu^B)_-$ follows from the boundary constraints, which is what we expect.

\subsection{Integrated supercurrent multiplets}\label{sec: integrated sucumus abstract}

\subsubsection{Comparison to pure 2D theories}
The supercurrent multiplets that satisfy the bulk \eqref{1}--\eqref{4} and boundary \eqref{1Bcor}--\eqref{4Bcor} constraints are $\zerotwo$-multiplets of a three-dimensional theory with boundary, with supersymmetry algebra isomorphic to that of a 2D $\zerotwo$-theory. It is interesting to compare and contrast the structure of this theory to a pure bulk 2D theory with $\zerotwo$-supersymmetry. We review some generic aspects of such bulk theories following \cite{Dedushenko:2015opz} (see also \cite{Melnikov:2019tpl} for a comprehensive review on 2D $\nzerotwo$ models).

In the case of a 2D theory with $\nzerotwo$ supersymmetry, the most general $\Sc$-multiplet is given by superfields $(\Sc_{++}^\zerotwo, \Wc_-^\zerotwo, \Tc_{----}^\zerotwo, C)$ and corresponding defining constraints in 2D $\zerotwo$-theories, such that conditions (\ref{condition a})--(\ref{condition d}) are satisfied. For details on their structure see \cite{Dumitrescu:2011iu,Dedushenko:2015opz}. If the $\zerotwo$-model we consider has an $R$-symmetry, there exists a corresponding smaller $\Rc$-multiplet $(\Rc_\hmu^\zerotwo, \Tc_{----}^\zerotwo)$ containing an improved energy momentum tensor $T_{\mu\nu}$. Furthermore, the structure of the multiplet guarantees that we can define the \emph{half-twisted} energy-momentum tensor $\widetilde{T}_{\mu\nu}$
\begin{equation}
	\begin{aligned}
		\widetilde{T}_{++}&=T_{++}+\tfrac i 2 \p_+j_+,
		\\\widetilde{T}_{+-}&=T_{+-}-\tfrac i 2 \p_-j_+,
		\\\widetilde{T}_{--}&=T_{--}-\tfrac i 2 \p_-j_-,
	\end{aligned}
\end{equation}
which satisfies
\begin{equation}
	\begin{aligned}
		\{\Qb_+,\cdots\}&=\widetilde{T}_{+\hmu},
		\\ \{\Qb_+,\widetilde{T}_{--}\}&=0, \quad \text{but } \{\Qb_+,\cdots\}\neq \widetilde{T}_{--}.
	\end{aligned}
\end{equation}
In other words, the components of the twisted energy-momentum tensor are $\Qb_+$-co\-homo\-logy elements, and $\widetilde{T}_{--}$ is a non-trivial element. Starting from these identities, one can show that the $\Qb_+$-cohomology of observables is invariant under renormalization group flow and thus carries information about possible IR fixed points of the model under consideration \cite{Dedushenko:2015opz}. In particular, there is an emergent conformal symmetry on the level of cohomology.

Secondly, it is well known for $\zerotwo$-theories that the cohomology of $\Qb_+$ as an operator on fields is isomorphic to the cohomology of $\Db_+$ as an operator on superfields. In this language, the non-trivial components of the twisted energy-momentum tensor are given by an appropriate $\Db_+$-closed linear combination of superfields from the supercurrent multiplet:
\begin{equation}\label{cohomology element in 2D}
	\Db_+ (\Tc_{----}^\zerotwo -\tfrac i 2  \p_{--}\Rc_{--}^\zerotwo)=0.
\end{equation}

\subsubsection{An energy-momentum tensor (not) in the cohomology}
\label{sec: Dedushenko style}

Since our three-dimensional theory with boundary has the same supersymmetry algebra, we might expect a similar structure as far as $\Qb_+$-cohomology is concerned. Indeed, we can identify the analog of \eqref{cohomology element in 2D} in 3D: we combine bulk equations \eqref{1b},\eqref{4a} and \eqref{4d} for $\alpha=-$ to:\footnote{The factor discrepancy on the left-hand side of \eqref{cohomology element in 2D} and the above equations is merely due to switching between spacetime and bispinor notation (cf.\ \eqref{explicit bispinor relations}).}
\begin{equation}
	\Db_+( \Rc^{B\two}_{--} +2i\p_-\Rc^{B\zero}_{--})=-2i\p_\perp\bar{\Rc^{B\one}_{--}},
\end{equation}
as well as boundary equations \eqref{1bBcor}, \eqref{4bBcor} and \eqref{4fBcor} to
\begin{equation}\label{boundary dedushenko-like relation}
	\Db_+(\Rc^{\p\two}_{--}+2i\p_-\Rc^{\p\zero}_{--})=2i \bar{\Rc^{B\one}_{--}}|_\p.
\end{equation}
Furthermore, we can also combine \eqref{2b}, \eqref{4a} and \eqref{4b} for $\alpha=+$ into 
\begin{equation}
	\Db_+(\Rc_{++}^{B\two}+2i \Rc_{++}^{B\zero})=-2i\p_\perp\bar{\Rc_{++}^{B\one}}.
\end{equation}
Similarly, we combine boundary equations \eqref{2bBcor}, \eqref{3cBcor} and \eqref{4a} into 
\begin{equation}\label{boundary dedushenko-like relation 2}
	\Db_+(\Rc^{\p\two}_{++}+2i\p_-\Rc^{\p\zero}_{++})=2i \bar{\Rc^{B\one}_{++}}|_\p .
\end{equation} 
We can rewrite these relations using the full $\Rc$-multiplet $\Rc_{\alpha\beta}^{\full (*)}=\Rc_{\alpha\beta}^{B (*)}+\delta(\xi^\perp)\Rc_{\alpha\beta}^{\p (*)}$:
\begin{equation}\label{full dedushenko-like relation}
	\begin{aligned}
		\Db_+(\Rc^{\full\two}_{--}+2i\p_-\Rc^{\full\zero}_{--})&=-2i \p_\perp\bar{\Rc^{B\one}_{--}}+2i \delta(\xi^\perp)\bar{\Rc^{B\one}_{--}}|_\p,
		\\
		\Db_+(\Rc^{\full\two}_{++}+2i\p_-\Rc^{\full\zero}_{++})&=-2i \p_\perp\bar{\Rc^{B\one}_{++}}+2i \delta(\xi^\perp)\bar{\Rc^{B\one}_{++}}|_\p.
	\end{aligned}
\end{equation}
The first equation is the analog of \eqref{cohomology element in 2D} in 3D, as we already stated. The second equation is the analog of the $\Qb$-closedness (equivalent to $\Db_+$-closedness) of the half-twisted tensor $T_{+-}-\tfrac i 2 \p_-j_+$, which in 2D follows from $\Qb$-exactness.

\subsubsection{Integrated currents and multiplets}

Equation \eqref{full dedushenko-like relation} shows that one cannot, in general, repeat the pure 2D argument to produce a local energy-momentum tensor twisted by the $R$-symmetry such that it is a $\Qb_+$-cohomology element. However, there is a different point of view which is helpful here: A three-dimensional quantum field theory with a finite number of fields can instead be regarded as a two-dimensional quantum field theory with an infinite number of fields. More precisely, instead of viewing bulk fields, loosely speaking, as maps $\p M \times \R_{\leq 0}\to T$, we view them as maps $\p M\to \{\text{maps: }\R_{\leq 0}\to T\}$ \cite{Dimofte:2017tpi,Bullimore:2016nji}. Now, instead of considering separate bulk and boundary actions, we can write a single Lagrangian for the full theory:
\begin{equation}
\Lag^\integ \coloneqq \Lag^\p + \int_{\R_{\leq 0}} \d x^n \Lag^B; 
\quad S = \int_{\p M} \d x^{N-1} \Lag^\integ .
\end{equation}
We see that the action is the same as before, but integration along $x^n$ is now a conceptually different operation: before, it used to be an integral on the spacetime on which the field theory is defined; now it is an operation on the new target space (i.e.\ a functional). The integration along $x^n$ also translates to conserved currents: As the theory is now formally two-dimensional, applying Noether's theorem to the above Lagrangian yields a two-dimensional current of the form
\begin{equation}
	J_\hmu^\integ=J_\hmu^\p+\int_{\R_{\leq 0}}\hspace{-.3cm} \d x^n  J_\hmu^B.
\end{equation}
We see that its conserved charge
\begin{equation}
	Q=\int_{\p \Sigma} J_\integ^0
\end{equation}
is identical to the one belonging to the local current \eqref{full theory charge}. This also provides an argument why the integrated currents are ``natural'' from the point of view of the three-dimensional QFT: To find the conserved charge of a current, one has to integrate all spatial directions, and the integrated current is an ``intermediate step'' of this integration.
The conservation equations \eqref{boundary conservation} now take the familiar form 
\begin{equation}
	\p^\hmu J_\hmu^\integ=0,
\end{equation}
where boundary conditions are possibly used. 
Extending supersymmetrically, we introduce the \emph{integrated supercurrent multiplets}, in our conventions (recall, mixed indices correspond to the $\perp$-direction, which does not appear here, see \eqref{bispinor relations})
\begin{equation}\label{integr multiplet definition}
	\Rc_{\alpha\alpha}^\integ=\Rc_{\alpha\alpha}^\p+\int_{\R_{\leq 0}}\hspace{-.3cm} \d x^\perp  \Rc_{\alpha\alpha}^B.
\end{equation}
In terms of integrated currents, the right-hand side of \eqref{full dedushenko-like relation} cancels exactly due to the integral:
\begin{equation}\label{dedushenko-like relation integrated}
	\begin{aligned}
		\Db_+(\Rc^{\integ\two}_{--}+2i\p_-\Rc^{\integ\zero}_{--})&=0,
		\\\Db_+(\Rc^{\integ\two}_{++}+2i\p_-\Rc^{\integ\zero}_{++})&=0.
	\end{aligned}
\end{equation}
The general arguments from \cite{Dedushenko:2015opz} presented in section \ref{sec: Dedushenko style} then imply that the lowest component  $-16(T_{--}^\integ -\tfrac i 2\p_- j_-^\integ)$ of the first equation is a non-trivial $\Qb_+$-cohomology element and the lowest component  $-16(T_{-+}^\integ -\tfrac i 2\p_- j_+^\integ)$ of the second equation is a trivial $\Qb_+$-cohomology element in the integrated 2D theory.
In fact, a stronger statement holds: the integrated multiplets are genuine 2D $\nzerotwo$ supersymmetry multiplets. Setting
\begin{equation}
	\begin{aligned}
		\Rc_{\alpha\alpha}^\zerotwo &\coloneqq \Rc_{\alpha\alpha}^{\integ \zero},
		\\ \Tc_{----}^\zerotwo&\coloneqq -\Rc_{--}^{\integ\two},
	\end{aligned}
\end{equation}
the constraints \eqref{1}--\eqref{4} and \eqref{1Bcor}--\eqref{4Bcor} imply that these are indeed 2D $\zerotwo$-supercurrent multiplets in the sense of \cite{Dumitrescu:2011iu,Dedushenko:2015opz}.

\subsection{Summarizing the results on boundary multiplets}

We look at three-dimensional theories with $\N=2$ supersymmetry, broken to a 2D $\zerotwo$-subalgebra due to a boundary. Currents associated to (remaining) symmetries now consist of bulk and boundary pieces. 

First we study 3D bulk supercurrent multiplets, in particular the $\Rc$-multiplet. Its structure remains unchanged, as the bulk parts of conserved currents are still divergence-free; we merely decompose the bulk multiplets into their $\zerotwo$-submultiplets
\begin{equation}
	\begin{aligned}
	\Rc_\mu^B(x,\theta,\thetab)&=\Rc^{B\zero}_\mu
	+\tm \Rc^{B\one}_\mu
	-\tbm \bar{\Rc^{B\one}_\mu}
	+\tm\tbm \Rc^{B\two}_\mu ,
	\\
	\chi^B_\alpha(x,\theta,\thetab)&=\chi^{B\zero}_\alpha
	+\tm \chi^{B\onea}_\alpha
	+\tbm \chi^{B\oneb}_\alpha
	+\tm\tbm \chi^{B\two}_\alpha.
	\end{aligned}
\end{equation}
The defining constraints \eqref{constraints 3d n=2 R-multiplet} now decompose under the $\nzerotwo$-subalgebra into equations \eqref{1}--\eqref{4}.

We investigate possible defining constraints for the boundary parts, using as guiding principles that
\begin{itemize}
	\item bulk and boundary pieces combine into a full multiplet $\Rc_\mu^\full= \Rc_\mu^B+\delta(\xi^\perp)\proj_\mu^{\ph\mu\hmu} \Rc^\p_\hmu$, where the boundary pieces are also decomposed as in \eqref{R mult decomp in bulk and boundary} and \eqref{chi multip decomp in bulk and boundary}, hence the boundary constraints must be of the same form as the bulk constraints \eqref{1}--\eqref{4}, 
	\item boundary constraints must impose boundary conservation \eqref{boundary conservation} on the \emph{remaining, conserved} boundary currents. 
\end{itemize}
We obtain the following list of constraints:
\begin{subequations}
\begin{align}
	\label{constraint used 1}
	0 &= \chi^{\p\two}_-+2i \p_- \chi^{\p\zero}_-, \\
	0&=\Db_+  \chi^{\p\onea}_+ -2i \chi^{B\zero}_+|_\p, 
	\\ 
	0 &= \Im(D_+\chi_-^{\p\zero}-\chi_+^{\p\onea}),
	\\
	0 &= \Im(D_+\chi_-^{\p\two}-2i\p_-\chi_+^{\p\onea}
	+2i\chi_-^{B\onea}|_\p), \\
	\chi^{\p\zero}_-&=\Db_+ \Rc^{\p\zero}_{--},
	\\
	\chi^{\p\onea}_+&= -\Rc^{\p\two}_{++} +2i\Rc^{B\zero}_{+-}|_\p - 2i\p_-\Rc^{\p\zero}_{++} , 
	\\
	\chi^{\p\onea}_-&= 2i\Rc^{B\zero}_{--}|_\p, 
	\\
	\label{constraint used last}
	\chi^{\p\two}_-&=\Db_+ \Rc^{\p\two}_{--}  -2i\bar{\Rc^{B\one}_{--}}|_\p,
	\\
	\label{constraint not used 1}
	\chi^{\p\two}_+&=-2i\bar{\Rc_{+-}^{B\one}}|_\p ,
	\\ \label{constraint not used 2}
	\chi^{\p\zero}_+&=0 , 
	\\ \label{constraint not used 3}
	\chi^{\p\oneb}_\alpha&=0 .
	\end{align}
\end{subequations}
The constraints \eqref{constraint used 1}--\eqref{constraint used last} are necessary to derive boundary conservation equations and equations \eqref{boundary dedushenko-like relation}. The last three relations \eqref{constraint not used 1}--\eqref{constraint not used 3} are not used in any conservation equation and are independent of the rest of the constraints. 

These constraints imply, in particular, equations \eqref{full dedushenko-like relation} on full currents:
\begin{equation} \label{full dedushenko-like relation repeated}
	\Db_+(\Rc^{\full\two}_{\pm\pm}+2i\p_-\Rc^{\full\zero}_{\pm\pm})=-2i \p_\perp\bar{\Rc^{B\one}_{\pm\pm}}+2i \delta(\xi^\perp)\bar{\Rc^{B\one}_{\pm\pm}}|_\p,
\end{equation}
which motivates the introduction of integrated current multiplets
\begin{equation}
	\Rc_{\alpha\alpha}^\integ=\Rc_{\alpha\alpha}^\p+\int \d x^\perp  \Rc_{\alpha\alpha}^B.
\end{equation}
These multiplets are genuine 2D $\nzerotwo$ supercurrent multiplets in the usual sense. In particular, the integration sets the right-hand side of relation \eqref{full dedushenko-like relation repeated} to zero. This implies that $T_{--}^\integ -\tfrac i 2\p_- j_-^\integ$ is a non-trivial $\Qb_+$-cohomology element and $T_{-+}^\integ -\tfrac i 2\p_- j_+^\integ$ is a trivial $\Qb_+$-cohomology element in the effective (integrated) 2D theory.

\section{Landau-Ginzburg model}\label{sec:LG model}

\subsection{Bulk theory} \label{subsec:LG bulk theory}

We now study a particular model where the framework we developed above can be applied. Our bulk theory should be a 3D $\N = 2$ Landau-Ginzburg model which lives on three-dimensional Minkowski space. At a later point, we will introduce a boundary and restrict the theory to the half-space
\begin{equation}
	M = \Set{x \in \R^{1,2} | x^{\smash{\perp}} \coloneqq x^1 \leq 0} .
\end{equation}
We will formulate the bulk theory in 3D $\N=2$ superspace. A generic chiral field is given by
\begin{equation}\label{chiral field}
	\Phi_{3D}(x, \gt, \tb) = \phi(y) + \sqrt{2} \gt \psi(y) + \gt \gt F(y), \qquad y^\mu = x^\mu - i \gt \gamma^\mu \tb,
\end{equation}
where, as usual, $\phi$ is a complex scalar field, $\psi_\ga$ is a complex fermion, and $F$ is a complex auxiliary field. Under the bulk supersymmetry, the components transform as follows:
\begin{equation}
	\label{bulk component variation}
	\begin{aligned}
		\dsym \phi &= \sqrt{2} \eps \psi, \\
		\dsym \psi_\ga &= \sqrt{2} \eps_\ga F - \sqrt 2 i (\gamma^\mu \epsb)_\ga \p_\mu \phi, \\
		\dsym F &= -\sqrt 2i \epsb \gamma^\mu \p_\mu \psi .
	\end{aligned}
\end{equation}
Let us now consider the simplest non-trivial theory: a Landau-Ginzburg model of a single chiral superfield. Its kinetic (Kähler) term is given by
\begin{equation}\label{bulk Lag kinetic}
	\Lag_\kin = \int \d^4 \gt\, \Phi_{3D} \Phib_{3D} 
	= -\p_\mu \phib \p^\mu \phi + \tfrac{i}{2}(\psib\gamma^\mu\p_\mu\psi) - \tfrac{i}{2}(\p_\mu \psib\gamma^\mu\psi) + \bar{F} F + \tfrac{1}{4} \p^2 (\phib \phi).
\end{equation}
At a later point, the term $\tfrac{1}{4} \p^2 (\phib \phi)$ will be removed; as it is a total derivative, it does not influence the bulk theory, but will be relevant once a boundary is introduced.
\\
The superpotential is of the well-known form
\begin{equation}\label{bulk Lag SuPo}
	\Lag_W = \int \d^2 \gt\, W(\Phi_{3D}) + \cc = W'(\phi) F - \tfrac{1}{2} W''(\phi) \psi \psi + \cc
\end{equation}
The bulk equations of motion are given by
\begin{equation}\label{EoM Bulk 3D}
	\Db^2 \Phib = - 4 W'(\Phi) \Leftrightarrow \left\{
	\begin{aligned}
		0 &= \bar{F} + W'(\phi) \\
		0 &= \p_\mu \p^\mu \phib + W''(\phi) F - \tfrac{1}{2} W'''(\phi) \psi \psi \\
		0 &= i (\gamma^\mu \p_\mu \psib)_\ga - W''(\phi) \psi_\ga
	\end{aligned} \right\} .
\end{equation}

\subsection{Introducing a boundary and breaking to \texorpdfstring{$\zerotwo$}{(0,2)}} \label{subsec:LG boundary introduction}

As we already saw in section \ref{subsec:bulk and boundary constraints}, the supersymmetry algebra breaks at least to a 2D $\nzerotwo$ (or 2D $\N=\oneone$, which we do not consider here) subalgebra when we introduce a boundary. Let us study sufficient conditions to preserve exactly $\nzerotwo$.

\subsubsection{Decomposition of the bulk fields}
Under the $\nzerotwo$ subalgebra, the chiral field $\Phi_{3D}$ decomposes into a $\zerotwo$ chiral multiplet and a Fermi multiplet. More details of this decomposition are written in appendix \ref{app:branching to zerotwo}. The resulting $\zerotwo$-superfields are
\begin{equation}
	\begin{aligned}
		\Phi&=\phi+\sqrt 2 \tp\psi_+-2i \tp\tbp \p_+\phi, \\
		\Psi&=\psi_--\sqrt 2 \tp F - 2i \tp\tbp\p_+\psi_- 
		+\sqrt 2 i \tbp \p_\perp \phi -2i \tp\tbp \p_\perp \psi_+.
	\end{aligned}
\end{equation}
The Fermi superfield satisfies 
\begin{equation}
	\Db_+ \Psi= \sqrt 2 E_\Psi, \quad E_\Psi=-i \p_\perp \Phi,
\end{equation}
and the chirality condition reads
\begin{equation}
	\Db_+ \Phi=0.
\end{equation}
The supersymmetry variation in the smaller algebra is given by $\dsym \coloneqq \eps Q_+ - \epsb\Qb_+$ (cf.\ equation \eqref{branched superoperators} for the definition of the superspace operators). The component fields now transform as
\begin{equation}
	\begin{aligned}
		\dsym\phi &= \sqrt 2\eps \psi_+ ,& \dsym \psi_+&=-2\sqrt 2 i\epsb \p_+\phi, \\
		\dsym F& = 2\sqrt 2 i \epsb \p_+\psi_- +\sqrt 2 i \epsb \p_\perp \psi_+, & \dsym \psi_-&= -\sqrt 2\eps  F +\sqrt 2 i\epsb  \p_\perp \phi ,
	\end{aligned}
\end{equation}
which is precisely the restriction of \eqref{bulk component variation} to the $\zerotwo$-subalgebra, given by choosing $\eps^\alpha=\left(\begin{smallmatrix} 0\\ \eps \end{smallmatrix}\right)$.
We may rewrite the Lagrangian in terms of $\zerotwo$-superspace:
\begin{equation} \label{branched bulk Lagrangians}
	\begin{aligned}
		\Lag_\kin &= \frac 1 2\int \d^2 \tp \Big[ i \Phib \p_-\Phi -i\p_- \Phib \Phi+\Psib \Psi
		\\& \hspace{2cm} + \p_\perp \left(\tfrac 1 2
		\tp\tbp  \p_\perp (\Phib \Phi)
		+\tfrac{i}{\sqrt 2} \tp \Phib \Psi+ \tfrac{i}{\sqrt 2}\tbp \Psib \Phi \right)
		\Big], \\
		\Lag_W &= -\tfrac{1}{\sqrt 2}\int \d \tp \Psi W'(\Phi) + \cc
	\end{aligned}
\end{equation}
Note that $\Lag_\kin$ consists of two parts. The first part is invariant under $\zerotwo$-supersymmetry even in the presence of a boundary, as its $\zerotwo$-variation is just a total $x^+$-derivative. The second term (trivially) transforms into an $x^\perp$-derivative, so it breaks $\zerotwo$-supersymmetry in the presence of a boundary, and hence dictates part of the ``boundary compensating term''  (cf.\ discussion at the end of section \ref{sec:symmetries in boundary theories}).

The equations of motion may again be written as superfield equations, now in $\zerotwo$-superspace:
\begin{equation}
	\begin{aligned}
		0&=2i\p_- \Db_+\Phib + \sqrt 2 i \p_\perp \Psib   - \sqrt 2 W''(\Phi) \Psi, \\
		0&=\Db_+\Psib +\sqrt 2  W'(\Phi) .
	\end{aligned}
\end{equation}

\subsubsection{Recovering partial supersymmetry}
As it stands, the pure bulk action \eqref{branched bulk Lagrangians} is not even $\zerotwo$-supersymmetric in the presence of a boundary:
\begin{equation}
	\dsym S = \int_M \dsym (\Lag_\kin + \Lag_W) = \int_M \p_\mu (V^\mu_\kin + V^\mu_W) = \int_{\p M} (V^\perp_\kin + V^\perp_W),
\end{equation}
which, in general, does not vanish. 
To recover at least $\nzerotwo$ supersymmetry, we must compensate these bulk variations.

For the kinetic term, we can add a boundary compensating term $\widetilde \Delta_\kin$ to the boundary Lagrangian (in a boundary-condition-independent way) in the spirit of  \cite{Brunner:2003dc,DiPietro:2015zia}. The boundary term is precisely minus the total $\perp$-derivative from the bulk Lagrangian in $\zerotwo$-superspace \eqref{branched bulk Lagrangians}:
\begin{equation}
	\widetilde \Delta_\kin \coloneqq -\tfrac 1 4 \p_\perp (\phib \phi) - \tfrac{i}{2} \psib_+ \psi_- + \tfrac{i}{2}  \psib_- \psi_+.
\end{equation}
We see that the $-\tfrac 1 4 \p_\perp (\phib \phi)$ cancels the bulk total derivative in $x^\perp$ direction when pulled into the bulk. This means that we can just drop $\tfrac 1 4 \p^2 (\phib \phi)$ from the bulk and $-\tfrac 1 4 \p_\perp (\phib \phi)$ from the boundary simultaneously, leaving us with bulk and boundary Lagrangians:
\begin{equation}\label{Kahler Lagrangian compensated}
	\begin{aligned}
		\Lag^B &= -\p_\mu \phib \p^\mu \phi + \tfrac{i}{2} (\psi \gamma^\mu \p_\mu \psib) - \tfrac{i}{2} (\p_\mu \psi \gamma^\mu \psib) 
		\\&\hspace{.4cm}+ \bar{F} F + W'(\phi) F + \Wb'(\phib) \Fb - \tfrac{1}{2} W''(\phi) \psi \psi + \tfrac{1}{2} \Wb''(\phib) \psib \psib, \\
		\Delta_\kin &= -\tfrac{i}{2} \psib_+ \psi_- + \tfrac{i}{2}  \psib_- \psi_+ = -\tfrac{i}{2} \psib \psi.
	\end{aligned}
\end{equation}
For the bulk superpotential term, the supersymmetry variation yields
\begin{equation} \label{superpotential bulk variation}
	\dsym \Lag_W =\p_\perp\left(-i\int \d \tp\epsb W(\Phi) + \cc \right) + \p_+(\dots) =\p_\perp(-i  \epsb \psi_+ W'(\phi) + \cc) + \p_+(\dots),
\end{equation}
where the right-hand side needs to be compensated. To do this in a boundary-condition-independent way, one can  use a bulk R-symmetry (see \cite{DiPietro:2015zia}), or one can add boundary degrees of freedom, which we will discuss in detail.

\subsubsection{Boundary Fermi multiplet and factorization} \label{sec:Factorization}
To compensate the superpotential term variation \eqref{superpotential bulk variation}, we introduce a 2D boundary Fermi multiplet with $E$- and $J$-potential terms, analogously to \cite{Kapustin:2003ga,Brunner:2003dc, Herbst:2008jq, Lazaroiu:2003zi}, where a 1D Fermi multiplet was used to compensate bulk 2D superpotential terms (see also \cite{Yoshida:2014ssa,Gadde:2013sca} for equivalent, three-dimensional examples). The general superspace expansion of a 2D Fermi multiplet is given by
\begin{equation}
	H=\eta -\sqrt 2 \tp G - 2i \tp \tb^+ \p_+\eta  
	-\sqrt 2  \tb^+ E(\phi) +2 \tp \tb^+ E'(\phi)\psi_+,
\end{equation}
so it has an $E$-potential of
\begin{equation}
	\Db_+ H=\sqrt 2 E(\Phi).
\end{equation}
The $\zerotwo$-supersymmetry variation of the components is given by
\begin{equation}
	\begin{aligned}
		\dsym \eta&=-\sqrt 2(\eps G +\epsb E),\\
		\dsym G&=\sqrt 2\epsb(2 i \p_+\eta -E'\psi_+).
	\end{aligned}
\end{equation}
Its (boundary) Lagrangian is
\begin{equation}
	\begin{aligned}
		\Lag_H&= \int \d^2 \tp \tfrac{1}{2}\Hb H - \int \d \tp \tfrac{i}{\sqrt{2}}  J(\Phi) H+ \int \d \tb^+ \tfrac{i}{\sqrt{2}} \Jb(\Phib) \Hb \\
		&= i\etab \p_+\eta-i\p_+\etab\eta -E'\etab \psi_+ -\Eb'\psib_+\eta+iJ'\eta\psi_+ -i\Jb' \psib_+\etab -|E|^2-|J|^2.
	\end{aligned}
\end{equation}
It consists of a kinetic term, two boundary potentials $E$ and $J$ of the \emph{bulk} chiral field $\phi$, and interactions between the boundary and bulk fermions. The boundary equations of motion are
\begin{equation}\label{EoM boundary}
	\Db_+\Hb+\sqrt 2 i J(\Phi)=0 \Leftrightarrow 
	\left\{
	\begin{aligned}
		G&=i\Jb
		\\2i\p_+\eta&=E'(\phi)\psi_+-i\Jb'(\phib)\psib_+
	\end{aligned}
	\right\}.
\end{equation}
The supersymmetry variation is
\begin{equation} \label{eq:bdy Fermi Lagr variation}
	\dsym\Lag_H = i\int \d\tp \epsb J(\Phi) E(\Phi)
	+ \cc + \p_+ (\dots).
\end{equation}
 We thus find that in case of a \emph{matrix factorization}
\begin{equation} \label{factorization condition}
	W(\Phi)|_\p=E(\Phi)J(\Phi)|_\p,
\end{equation}
the bulk term  from \eqref{superpotential bulk variation} will be compensated precisely, and $\zerotwo$-supersymmetry is preserved. As stated before \cite{Witten:1993yc} and can be seen from \eqref{eq:bdy Fermi Lagr variation}, a pure 2D $\nzerotwo$ theory must fulfill $ E \cdot J = 0$ in order to preserve supersymmetry. However, in our case, the ``failure'' of the boundary Fermi multiplet to meet this condition cancels the failure of the bulk theory to preserve $\nzerotwo$-supersymmetry at the boundary.

The total action of the factorized Landau-Ginzburg model then reads
\begin{equation}
	\begin{aligned}
		S&=\int_M \Lag^B +\int_{\p M}\Lag^\p
		\\&= \frac 1 2 \int_M \bigg\{
		\int \d^2 \tp \big[i\Phib\p_-\Phi-i\p_-\Phib\Phi+\Psib\Psi +\p_\perp \Delta\big]
		-\sqrt 2 \int \d \tp \Psi W(\Phi)+\cc
		\bigg\}
		\\&\hspace{.4cm}
		+\frac 1 2 \int_{\p M} \bigg\{
		\int \d^2\tp \big[-\Delta|_\p+\Hb H\big]
		-\sqrt 2 i \int \d \tp J(\Phi) H+\cc 
		\bigg\},
	\end{aligned}
\end{equation}
where $\tfrac 1 2  \int\d^2\tp \Delta= \tfrac{i}{2\sqrt 2}\int\d^2\tp(\tp\Phib \Psi+\tbp  \Psib \Phi)= \tfrac{i}{2}( \psib_+ \psi_- -  \psib_- \psi_+)$ (cf.\ \eqref{Kahler Lagrangian compensated}). 
After using the algebraic equations of motion, we get the following component expansions:
\begin{equation}\label{Lagrangians in components}
	\begin{aligned}
		\Lag^B&= -\p_\mu \phib \p^\mu \phi + \tfrac{i}{2} (\psi \gamma^\mu \p_\mu \psib) - \tfrac{i}{2} (\p_\mu \psi \gamma^\mu \psib) 
		\\&\hspace{.4cm}
		-|W(\phi)|^2 - \tfrac{1}{2} W''(\phi) \psi \psi + \tfrac{1}{2} \Wb''(\phib) \psib \psib,
		\\\Lag^\p&=i \etab \p_+\eta -i \p_+\etab \eta  -|J|^2	-|E|^2
		-\Eb'\psib_+\eta -E'\etab\psi_+	
		\\&\hspace{.4cm}
		-iJ'\psi_+\eta 	-i\Jb' \psib_+\etab
		-\tfrac i 2 (\psib_+\psi_--\psib_-\psi_+)|_\p.
	\end{aligned}
\end{equation}
The $\zerotwo$-variation of the total action is zero, hence $\nzerotwo$ supersymmetry is preserved in a boundary-condition-independent way. 

\subsubsection{Symmetric boundary conditions}
In any theory with a boundary it is necessary to introduce boundary conditions such that the action can be made stationary. Requiring the boundary condition to be compatible with the $\nzerotwo$ subalgebra in the sense of \eqref{symmetric boundary condition} further restricts the number of options. We now discuss some explicit boundary conditions for our LG model. We consider boundary conditions without superpotential (previously discussed in \cite{Dimofte:2017tpi}) and with superpotential separately. 
\paragraph{Without superpotential}
\begin{itemize}
	\item \emph{(generalized) Dirichlet}: $\Phi=0$ or more generally $\Phi=c$ (in components $\phi=c$ and $\psi_+=0$) is a symmetric boundary condition (the action may require the addition of some boundary terms to be symmetric).
	\item \emph{Neumann}: $\Psi=0$ (in components $\p_\perp\phi=0$ and $\psi_-=0$) is also symmetric. It is also the dynamical boundary condition in the sense of \eqref{dynamical BC} for the action \eqref{Kahler Lagrangian compensated} without superpotential. Note that one can also obtain the (generalized) Dirichlet as a dynamical boundary condition by adding appropriate boundary terms \cite{Dimofte:2017tpi}.
	\item \emph{Mixed conditions}: In models with more than one 3D chiral superfield, we may assign Dirichlet conditions to some and Neumann conditions to others \cite{Dimofte:2017tpi}.
\end{itemize}
\paragraph{With superpotential}
\begin{itemize}
	\item \emph{(generalized) Dirichlet}: Setting $\Phi=c$ is symmetric and also statically cancels the supervariation of the potential \eqref{superpotential bulk variation} (albeit in a boundary-condition-\emph{dependent} way). However, if $W'(c)|_\p \neq 0$, supersymmetry is broken spontaneously, as the vacuum expectation value of $\psi_-$ then transforms non-trivially under supersymmetry.\footnote{We note that if there is a bulk $R$-symmetry, one can also compensate the superpotential variation boundary-condition-independently, see \cite{DiPietro:2015zia}. However, in the case of one chiral field, one can then only impose Dirichlet boundary conditions, as the Neumann condition with $W\neq0$ is not symmetric.}
	\item \emph{Mixed conditions}: Setting $\Psi=0$ (Neumann) is only symmetric if $W'(\phi)|_\p = 0$. For one bosonic field, this holds only if $W=0$, as $\phi$ is unconstrained on the boundary. If $W \neq 0$ and the theory has more than one chiral superfield, one can assign Dirichlet conditions to some and Neumann conditions to others while maintaining supersymmetry (a requirement the authors in \cite{Dimofte:2017tpi} call ``sufficiently Dirichlet'').
	\item \emph{Factorized Neumann}: If we introduce additional degrees of freedom on the boundary as in section \ref{sec:Factorization}, we may again choose dynamical boundary conditions. In the case without superpotential this lead to the Neumann boundary condition. For the action \eqref{Lagrangians in components} the dynamical boundary condition is the analog of the Neumann boundary condition, now with superpotential:
	\begin{equation} \label{factorized neumann BC}
		\Psib = -i\Hb E'(\Phi) - H J'(\Phi) \Leftrightarrow \left\{ \begin{aligned}
		\psib_- &= - i \etab E' - \eta J', \\
		\p_\perp \phib &= -\Eb E' - \Jb J' - (\etab E'' -i \eta J'') \psi_+ \end{aligned} \right\}.
	\end{equation}
	One can check that it is indeed symmetric if the factorization condition \eqref{factorization condition} is met. We use this boundary condition in our computations for currents and current multiplets. This choice of boundary condition in fact encodes a collection of boundary conditions labeled by the choices of \emph{matrix factorizations} of $W$ (since the boundary condition depends explicitly on $E$ and $J$). 
\end{itemize}

\subsection{Currents} \label{subsec:LG currents}
Here we present conserved currents associated to the symmetries of the Landau-Ginzburg theory with one chiral field in the bulk, a Fermi multiplet on the boundary, and factorized Neumann boundary conditions. We will compute the currents in various improvement frames in order to place them into consistent multiplets in the following subsection.

\subsubsection{\boldmath $R$-current}\label{sec:R-current}
If the superpotentials $W,E,J$ are (quasi-)homogeneous functions of $\Phi$ --- in the case of one chiral field, monomials ---, then the action is invariant under the $R$-symmetry transformation\footnote{For multiple chiral fields $\Phi_i$, the condition for quasi-homogeneity reads $W(\Phi_1,\ldots, \Phi_k)=\sum_i \alpha_i\Phi_i\p_{\Phi_i} W $ for some choice of $R$-charges $\ga_i$.}
\begin{equation}\label{R-transformation}
	\begin{aligned}
		\tp&\mapsto e^{-i\tau}\tp, &
		\Phi&\mapsto e^{-2i\tau \alpha}\Phi, \\
		\Psi&\mapsto e^{-i\tau(2\alpha-1)}\Psi, &
		H&\mapsto e^{-i\tau(\ell_E -\ell_J)\alpha}H,
	\end{aligned}
\end{equation}
where $\tau$ is the symmetry variation parameter and we have defined
\begin{equation}
		\alpha\coloneqq(\deg W)^{-1}, \quad
		\ell_E\coloneqq\deg E, \quad
		\ell_J\coloneqq \deg J.
\end{equation}
Note that factorization implies 
\begin{equation}
\alpha(\ell_E+\ell_J)=1.
\end{equation}
The bulk contribution to the $R$-current is given by
\begin{equation}\label{r-current bulk}
	j^B_\mu=2i \alpha (\phib\p_\mu \phi-\p_\mu\phib\phi)+(1-2\alpha)\psib\gamma_\mu\psi,
\end{equation}
while the boundary contribution is given by
\begin{equation}\label{r-current boundary}
	j^\p_\hmu=
	\begin{pmatrix}
		j^\p_+ \\ j^\p_-
	\end{pmatrix}
	=
	\begin{pmatrix}
		0\\
		\alpha(\ell_E-\ell_J)\etab\eta
	\end{pmatrix}.
\end{equation}

\subsubsection{Supercurrents}
After introducing the boundary (with the aforementioned choices), only $\zerotwo$-super\-sym\-metry is preserved. We may however still discuss the full 3D $\mc N=2$ supersymmetry in the bulk, as the $\zerotwo$-restrictions of the bulk currents remain identical (and covariant notation can be conveniently used). 

\paragraph{Noether frame (\boldmath $\Sc$-frame)}
The bulk supercurrent induced by $\dsym=\eps Q$ (full supersymmetry is $\dsym =\eps Q-\epsb\Qb$) in the Noether frame is given by:\footnote{One finds this supercurrent by applying Noether's theorem to \eqref{Kahler Lagrangian compensated} and improving the boundary part to zero.}
\begin{equation}\label{eq:bulk supercurrent S frame}
	(S^B_\mu)_\alpha=\sqrt 2 (\gamma^\nu \gamma_\mu  \psi)_\alpha \p_\nu \phib 
	-\sqrt 2 i(\gamma_\mu \psib)_\alpha \Wb' .
\end{equation}
Its $\zerotwo$-restriction $\dsym=\eps^+ Q_+$ is given by setting $\alpha=+$:
\begin{equation}
	(S^B_\mu)_+=
	\begin{pmatrix}
	(S^B_+)_+\\(S^B_-)_+\\(S^B_\perp)_+
	\end{pmatrix}
	=
	\begin{pmatrix}
		2\sqrt 2 \psi_+\p_+\phib \\
		-\sqrt 2 (\psi_-\p_\perp \phib +i \psib_- \Wb')\\
		\sqrt 2(\psi_+\p_\perp \phib-2\psi_-\p_+\phib +i\psib_+\Wb')
	\end{pmatrix}.
\end{equation}
The boundary contribution is induced by $\dsym=\eps^+ Q_+$ and reads in the Noether frame:
\begin{equation}\label{eq:boundary supercurrent S frame}
	(S_\hmu^\p)_+=
	\begin{pmatrix}
		(S_+^\p)_+\\	(S_-^\p)_+
	\end{pmatrix}
	=
	\begin{pmatrix}
	0\\
	-\sqrt 2 (\Jb \etab -i \Eb \eta)
	\end{pmatrix} .
\end{equation}

\paragraph{\boldmath $\Rc$-frame}
If the Lagrangian has an $R$-symmetry \eqref{R-transformation}, we may improve the above supercurrent to a supercurrent which is part of the $\Rc$-multiplet. We call this improvement frame the $\Rc$-frame.
The bulk components are:
\begin{equation}
	\begin{aligned}
		(S_\mu^B)^\Rc_\alpha
		&=(S_\mu^B)^\Sc_\alpha-2\sqrt 2 \alpha \eps_{\mu\nu\rho}(\gamma^\nu \p^\rho(\phib\psi))_\alpha
		\\&=
		\sqrt 2(1-2\alpha)\big( (\gamma^\nu \gamma_\mu  \psi)_\alpha \p_\nu \phib 
		+i(\gamma_\mu \psib)_\alpha \Wb'\big)
		+2\sqrt 2 \alpha (\p_\mu\phib\psi_\alpha -\phib \p_\mu\psi_\alpha),
	\end{aligned}
\end{equation}
where $(S_\mu^B)^\Sc_\alpha$ denotes the supercurrent in the Noether frame, $\alpha=(\deg W)^{-1}$ and the last equality uses equations of motion \eqref{EoM Bulk 3D} and homogeneity of $W$.

The boundary components are
\begin{equation}
	(S_\hmu^\p)^\Rc_+=(S_\hmu^\p)^\Sc_++ 2\sqrt 2 \alpha \eps_{\hmu\nu n}(\gamma^\nu \psi)_+\phib
	=
	\begin{pmatrix}
	0\\
	\sqrt 2\alpha (\ell_J-\ell_E) (\Jb \etab +i \Eb \eta)
	\end{pmatrix},
\end{equation}
where the last equality uses boundary conditions \eqref{factorized neumann BC}.

\subsubsection{Energy-momentum tensor}
Similarly to the case of supercurrents, we stick to covariant notation for the bulk pieces, even though certain directions are no longer symmetries. Let us start by simplifying the Lagrangians \eqref{Lagrangians in components} on-shell:\footnote{Note that the second equation also uses boundary conditions \eqref{factorized neumann BC}. Without using them, we get
\begin{equation*}
	\Lag^\p\overset{\text{\tiny on-shell}}{=}
	-|E|^2-|J|^2
	-\tfrac 1 2(\Eb'\psib_+\eta 
	+E'\etab\psi_+	
	+iJ'\psi_+\eta 	
	+i\Jb' \psib_+\etab)
	-\tfrac i 2 (\psib_+\psi_--\psib_-\psi_+)|_\p.
\end{equation*}}
\begin{align}
		\Lag^B&\overset{\text{\tiny on-shell}}{=}-\p_\rho\phib\p^\rho\phi-|W'|^2,
		\\\Lag^\p&\overset{\text{\tiny on-shell}}{=}-|E|^2-|J|^2.
\end{align}

\paragraph{Noether frame (\boldmath $\Sc$-frame)}
Using the Noether procedure, in the bulk we find the non-symmetric energy-momentum tensor
\begin{equation} \label{eq:bulk EM tensor nonsym}
	\begin{aligned}
		\hT_{\mu\nu}^B&=
		\p_\mu\phib\p_\nu\phi+\p_\nu\phib\p_\mu\phi 
		+\tfrac i2 \p_\mu\psib \gamma_\nu\psi-\tfrac i2 \psib\gamma_\nu\p_\mu \psi 
		-\eta_{\mu\nu}(\p_\rho\phib\p^\rho\phi+|W'|^2),
	\end{aligned}
\end{equation}
and in the boundary we find (using equations of motion but \emph{not} boundary conditions)
\begin{equation}
\label{eq:bdy EM tensor nonsym no BC}
\begin{aligned}
	\hT^\p_{++} &= 0, \\
	\hT^\p_{--} &= \tfrac{i}{2}\etab \p_-\eta -\tfrac{i}{2}\p_-\etab\eta ,\\
	\hT^\p_{+-} & = \tfrac{i}{4}\psib_+\psi_- - \tfrac{i}{4}\psib_-\psi_+ +\tfrac{1}{2}|E|^2 + \tfrac{1}{2}|J|^2 \\
		& \qquad + \tfrac{1}{2}\big(E'\etab \psi_+ + \Eb'\psib_+\eta - iJ'\eta\psi_+ + i\Jb' \psib_+\etab\big), \\
	\hT^\p_{-+} &= \tfrac{i}{4} \psib_+ \psi_- - \tfrac{i}{4} \psib_- \psi_+ +\tfrac{1}{2}|E|^2 + \tfrac{1}{2}|J|^2 \\
		&\qquad + \tfrac{1}{4} \big( E'\etab \psi_+ + \Eb'\psib_+\eta - iJ'\eta\psi_+ +i\Jb'\psib_+\etab \big).
\end{aligned}
\end{equation}
If we utilize the boundary conditions \eqref{factorized neumann BC}, the expressions simplify to
\begin{equation}
	\begin{aligned}
		\hT_{++}^\p&=0,
		\\\hT_{--}^\p&=\tfrac i 2 \etab\p_-\eta -\tfrac i 2 \p_-\etab\eta,
		\\\hT_{+-}^\p&=\tfrac i 2 \etab\p_+\eta -\tfrac i 2 \p_+\etab\eta+\tfrac 1 2(|E|^2+|J|^2) ,
		\\\hT_{-+}^\p&=\tfrac 1 2(|E|^2+|J|^2).
	\end{aligned}
\end{equation}

\paragraph{Symmetrization}
These can by made symmetric using an improvement. In the bulk we find
\begin{equation}\label{symmetrization}
	\begin{aligned}
		T_{\mu\nu}^B&=\widehat T_{\mu\nu}^B-\tfrac 1 8 \eps_{\mu\nu\rho }H^\rho
		\\&=
		(\p_\mu\phib \p_\nu \phi +\p_\nu \phib \p_\mu \phi)
		-\eta_{\mu\nu} (|\p\phi|^2\!+ |W'|^2)
		+\tfrac i 2 (\p_{(\mu} \psib \gamma_{\nu)} \psi)
		-\tfrac i 2(\psib\gamma_{(\nu} \p_{\mu)} \psi),
	\end{aligned}
\end{equation}
where $H^\rho=-2i\p^\rho(\psib\psi)$.\footnote{This is precisely the brane current from the supercurrent multiplet, see appendix \ref{app:bulk components S multiplet}. To obtain the desired form for $T_{\mu\nu}^B$ we use equations of motion and the Clifford algebra.} The induced boundary improvement \label{induced boundary improvement} is $T_{\hmu\hnu}^\p=\widehat{T}_{\hmu\hnu}^\p-\tfrac i 4 \eps_{\hmu \hnu n} \psib\psi|_\p$, so
\begin{equation}
	\begin{aligned}
		T_{++}^\p&=0,
		\\T_{--}^\p&=\tfrac i 2 \etab\p_-\eta -\tfrac i 2 \p_-\etab\eta,
		\\T_{+-}^\p&=\tfrac i 2 \etab\p_+\eta -\tfrac i 2 \p_+\etab\eta+\tfrac 1 2(|E|^2+|J|^2)-\tfrac i 8 (\psib_-\psi_+-\psib_+\psi_-)|_\p,
		\\T_{-+}^\p&=\tfrac 1 2(|E|^2+|J|^2)+\tfrac i 8 (\psib_-\psi_+-\psib_+\psi_-)|_\p.
	\end{aligned}
\end{equation}
Note that using boundary conditions \eqref{factorized neumann BC} and equations of motion for $\eta$ \eqref{EoM boundary} we find that
\begin{equation}
	\tfrac i 2 (\psib_-\psi_+-\psib_+\psi_-)|_\p=i\etab\p_+\eta-i\p_+\etab\eta,
\end{equation}
which shows that the boundary components are symmetric modulo boundary conditions in this frame as well.

\paragraph{\boldmath $\Rc$-frame}
Again, as in the case of the supercurrent, there is an improved energy-mo\-men\-tum tensor in the $\Rc$-frame. We find
\begin{equation}\label{EMT bulk R-frame}
	\begin{aligned}
		(T_{\mu\nu}^B)^\Rc&=(T_{\mu\nu}^B)^\Sc+\tfrac 12 [\p_\mu\p_\nu-\eta_{\mu\nu}\p^2](-2\alpha \phib\phi)
		\\&=
		(1-\alpha)(\p_\nu\phi \p_\mu \phib +\p_\mu \phi \p_\nu \phib) 
		- \alpha ( \p_\mu\p_\nu\phib\phi + \phib\p_\mu\p_\nu\phi ) 
		+\tfrac i 2 (\p_{(\nu} \psib \gamma_{\mu)} \psi) \\
		&\pheq -\tfrac i 2(\psib\gamma_{(\mu} \p_{\nu)} \psi) 
		- (1-2\alpha)\eta_{\mu\nu} ( |\p\phi|^2 -  |W'|^2)
		+ \alpha \eta_{\mu\nu}(i \psi \gamma^\rho\p_\rho\psib - i\p_\rho\psi\gamma^\rho\psib) \big),
	\end{aligned}
\end{equation}
where for the last equality we have used equations of motion. 
The boundary contributions are given by $(T^\p)_{\hmu\hnu}^R=(T^\p)_{\hmu\hnu}^S+\tfrac 1 2 \eta_{\hmu\hnu} \p_\perp (-2\alpha \phib\phi)$, hence
\begin{equation}
	\begin{aligned}
		(T^\p_{++})^\Rc&=0,
		\\
		(T^\p_{--})^\Rc&=\tfrac i 2 \etab \p_-\eta-\tfrac i 2 \p_-\etab\eta,
		\\
		(T^\p_{+-})^\Rc&=\tfrac i 2 \etab\p_+\eta -\tfrac i 2 \p_+\etab\eta+\tfrac 1 2(|E|^2+|J|^2)
		+\tfrac \alpha {2} \p_\perp(\phib\phi)|_\p
		 -\tfrac i 8 (\psib_-\psi_+-\psib_+\psi_-)|_\p,
		\\
		(T^\p_{-+})^\Rc&=\tfrac 1 2(|E|^2+|J|^2)+\tfrac \alpha {2} \p_\perp(\phib\phi)|_\p+\tfrac i 8 (\psib_-\psi_+-\psib_+\psi_-)|_\p.
	\end{aligned}
\end{equation}
Note that the symmetry of the boundary stress tensor (modulo boundary conditions) was preserved by the improvement.

\subsection{Supercurrent multiplets of the LG model} \label{subsec:LG multiplets}
Let us now assemble the conserved currents of the Landau-Ginzburg model from the previous subsection into supercurrent multiplets. We first recall the supercurrent multiplets of a pure bulk theory, as well as its possible smaller multiplets. After that we will present a valid supercurrent multiplet in the Landau-Ginzburg model with boundary, and also discuss integrated supercurrent multiplets. 
\subsubsection{Bulk theory}
Here we study a pure bulk theory with Lagrangian $\Lag=\Lag_\kin+\Lag_W$ as  in \eqref{bulk Lag kinetic}, \eqref{bulk Lag SuPo}. In such a theory a valid $\Sc$-multiplet is given by
\begin{equation}
\label{bulk S multiplet}
\Sc_{\ga \gb} = D_\ga \Phi_{3D} \Db_\gb \Phib_{3D} + D_\gb \Phi_{3D} \Db_\ga \Phib_{3D}.
\end{equation}
It contains the supercurrent and energy-momentum tensor (in the $\Sc$-frame) in its components. We explicitly compute the components to verify this in the appendix (cf.\ \eqref{explicit components in LG}). The multiplet satisfies
\begin{equation}
\Db^\ga \Sc_{\ga \gb} = \underbrace{-D_\gb\Phi_{3D}\Db^2\Phib_{3D}}_{=\Y_\gb} + \underbrace{(- \tfrac{1}{2})\Db^2 D_\gb(\Phi_{3D}\Phib_{3D})}_{=\chi_\gb}.
\end{equation}
Using the equations of motion \eqref{EoM Bulk 3D}, one may rewrite $\Y_\gb = 4 D_\gb W(\Phi_{3D})$. The defining equations in \eqref{constraints 3d n=2} can be verified easily, proving that this is indeed an $\Sc$-multiplet. The central charge $C$ is zero. This $\Sc$-multiplet can be improved to a Ferrara-Zumino multiplet using the improvement $U_{\text{FZ}} = -\tfrac{1}{2} \Phib_{3D} \Phi_{3D}$ \eqref{bulk improvements of sucumu}, as this implies
\begin{equation}
\chi_\ga' = -\tfrac{1}{2} \Db^2 D_\ga(\Phib_{3D}\Phi_{3D}) - \Db^2 D_\ga U = 0.
\end{equation}
The multiplet is then given by
\begin{equation}
\Jc_{\ga\gb} = \tfrac{1}{2}(D_\ga \Phi_{3D} \Db_\gb \Phib_{3D} + D_\gb \Phi_{3D} \Db_\ga \Phib_{3D} ) + \tfrac{1}{2}(i \Phib_{3D} \p_{\ga \gb} \Phi_{3D} - i \p_{\ga \gb} \Phib_{3D} \Phi_{3D}).
\end{equation}
If the $R$-symmetry (cf.\ section \ref{sec:R-current}) is present, one can instead apply the improvement $U_\Rc = -2\alpha\Phib_{3D} \Phi_{3D}$ ($\alpha=(\deg W)^{-1}$) to the $\Sc$-multiplet, which sets $\Y_\ga$ to zero modulo equations of motion:
\begin{equation}
	\Y_\ga' = 4 D_\ga W(\Phi_{3D}) - \tfrac{1}{2} D_\ga \Db^2 U = 4 D_\ga W(\Phi_{3D}) -4\alpha D_\ga (\Phi_{3D} W'(\Phi_{3D})) = 0.
\end{equation}
Now $\Sc_{\ga \gb}$ transforms to
\begin{equation}\label{r-current-multiplet bulk}
\Rc_{\ga \gb} = (1-2\alpha)(D_\ga \Phi_{3D} \Db_\gb \Phib_{3D} + D_\gb \Phi_{3D} \Db_\ga \Phib_{3D} ) + 2\alpha(i \Phib_{3D} \p_{\ga \gb} \Phi_{3D} - i \p_{\ga \gb} \Phib_{3D} \Phi_{3D}).
\end{equation}
We see that the lowest component of this multiplet is exactly the $R$-current \eqref{r-current bulk}, and one can check that the remaining currents in the $\Rc$-multiplet are in the $\Rc$-frame.

\subsubsection{Adding a boundary}
Now that we have studied the bulk, let us go back to our Landau-Ginzburg theory with a boundary and a boundary Fermi multiplets whose potentials factorizes the superpotential \eqref{factorization condition}. We want to extend the above bulk supercurrent multiplet to a full (bulk and boundary) supercurrent multiplet as described in section \ref{subsec:bulk and boundary constraints}.

 We already computed the bulk and boundary conserved currents in the sense of \eqref{full conservation equations} in various improvements frames in the previous subsection, and now have to organize the components into admissible $\zerotwo$-multiplets. We choose to do so in the case of the $\Rc$-multiplet.

We consider the embedding \cite{Drukker:2017xrb} into 3D $\mc N=2$ superspace:
\begin{equation}
	\begin{aligned}
		\Rc^B_{\alpha\beta}&=\Rc^{B\zero}_{\alpha\beta}
		+\tm \Rc^{B\one}_{\alpha\beta}
		-\tbm\bar{\Rc^{B\one}_{\alpha\beta}}
		+\tm\tbm\Rc^{B\two}_{\alpha\beta},
		\\
		\Rc^\p_{\alpha\alpha}&=\Rc^{\p\zero}_{\alpha\alpha}
		+\tm\undernote{=0}{\Rc^{\p\one}_{\alpha\alpha}}
		-\tbm\undernote{=0}{\bar{\Rc^{\p\one}_{\alpha\alpha}}}
		+\tm\tbm\Rc^{\p\two}_{\alpha\alpha}.
	\end{aligned}
\end{equation}
First, we decompose the bulk contribution to the $\Rc$-multiplet into its $\zerotwo$-submultiplets.
The zeroth-order bulk $\zerotwo$-superfields are
\begin{equation}
	\begin{aligned}
		\Rc^{B\zero}_{++}&=
		8\alpha(i\Phib \p_+ \Phi-i\p_+ \Phib\Phi)-2(1-2\alpha) \Db_+\Phib D_+\Phi
		\\&=4j_+^B+\ldots , 
		\\
		\Rc^{B\zero}_{--}&=
		8\alpha (i\Phib \p_- \Phi-i\p_- \Phib\Phi)-4(1-2\alpha) \Psib \Psi
		\\&=4j_-^B+\ldots , 
		\\
		\Rc^{B\zero}_{+-}&=
		-4\alpha (i\Phib \p_\perp \Phi-i\p_\perp \Phib\Phi)-\sqrt 2(1-2\alpha)(\Db_+ \Phib \Psi +\Psib D_+ \Phi)
		\\&=-2j_\perp^B+\ldots .
	\end{aligned}
\end{equation}
The first-order bulk $\zerotwo$-superfields are
\begin{equation}
	\begin{aligned}
		\Rc_{++}^{B\one }&=
		4(1-2\alpha)\big( i\p_\perp \Phib
		D_+\Phi  
		- \Db_+\Phib \Wb'(\Phib)\big)
		-8i\sqrt 2 \alpha (\p_+\Phib\Psi -\Phib \p_+\Psi)
		\\&=-4i(S_+^B)^\Rc_-+\ldots ,
		\\
		\Rc_{--}^{B\one }&=
		-8i\sqrt 2(1-\alpha)\Psi \p_- \Phib 
		+8i\sqrt 2 \alpha \Phib \p_-\Psi
		\\&=-4i(S_-^B)^\Rc_-+\ldots ,
		\\\Rc_{+-}^{B\one }&=
		2\sqrt 2 i\big(\p_\perp\Phib \Psi
		- \Phib \p_\perp\Psi\big)
		+2(1-2\alpha)\big( 
		iD_+\Phi \p_-\Phib 
		-\sqrt 2 \Psib \Wb'(\Phib)
		+ \sqrt 2i\Phib \p_\perp\Psi\big)
		\\&=2i(S_\perp^B)^\Rc_-+\ldots .
	\end{aligned}
\end{equation}
The second-order bulk $\zerotwo$-superfields are lengthy, but are a straightforward $\zerotwo$-com\-ple\-tion of their lowest components:\footnote{Recall the general expansions \eqref{+direction two component},\eqref{-direction two component} and \eqref{perp-direction two component}, in particular the definition of $K_{\mu\nu}$ \eqref{defs of K and L}.} 
\begin{equation}
	\begin{aligned}
		\Rc_{++}^{B\two }&=-16\big(
		\p_+\Phib\p_-\Phi
		+\p_-\Phib\p_+\Phi
		+\alpha \p_+\p_-(\Phib\Phi)
		-\tfrac \alpha 2 \p_\perp^2(\Phib\Phi)
		\\&\hspace{1cm}
		-\tfrac i 4\p_-\Db_+\Phib D_+\Phi
		+\tfrac i 4\Db_+\Phib \p_-D_+\Phi
		-\tfrac 1 2 \widetilde{\Lag^B}
		\big)
		\\&=-16\big(
		\alpha \p_+\p_-(\Phib\Phi)
		+\tfrac 1 2 \p_\perp\Phib\p_\perp\Phi 
		-\tfrac \alpha 2 \p_\perp^2(\Phib\Phi)
		+\tfrac 1 2 |W'(\Phi)|^2
		\\&\hspace{1cm}
		-\tfrac i 4\p_-\Db_+\Phib D_+\Phi
		+\tfrac i 4\Db_+\Phib \p_-D_+\Phi
		\big)
		\\&=-16 (T_{-+}^B)^\Rc+2 (C^B_{+-})^\Rc+\ldots,
		\\\Rc_{--}^{B\two }&= -16\big(
		2\p_-\Phib\p_-\Phi 
		-\alpha \p_-^2(\Phib\Phi)
		-\tfrac i 2 \p_-\Psib \Psi
		+\tfrac i 2 \Psib\p_-\Psi
		\big)
		\\&=-16 (T_{--}^B)^\Rc+\ldots,
		\\\Rc_{+-}^{B\two }&=
		8\big(
		\p_-\Phib \p_\perp \Phi
		+\p_\perp\Phib\p_-\Phi
		-\alpha\p_-\p_\perp(\Phib\Phi)
		\\&\hspace{.4cm}
		+\tfrac i {2\sqrt 2}(
		\p_-\Db_+\Phib \Psi
		 +\p_- \Psi D_+\Phi 
		 -\Db_+\Phib\p_\Psi
		 -\Psib\p_-D_+\Phi)
		\big)
		\\&=8 (T_{-\perp}^B)^\Rc-(C^B_{\perp-})^\Rc\ldots,
	\end{aligned}
\end{equation}
where the lowest components are given by the energy-momentum tensor  \eqref{EMT bulk R-frame}. The brane current $(C^B_{\mu\nu})^\Rc= \eps_{\mu\nu\rho}(H^{\rho})^\Rc$ in the $\Rc$-frame is given by  $H^\Rc_\mu=-2i(1-4\alpha)\p_\mu(\psib\psi)$ where we have used \eqref{improvements explicitly} and the explicit improvement $U_\Rc$. We have also $\zerotwo$-completed the bulk Lagrangian on-shell
\begin{equation}
	 \widetilde{\Lag^B}=2\p_+\Phib\p_-\Phi+2\p_-\Phib\p_+\Phi-\p_\perp \Phib\p_\perp \Phi-|W'(\Phi)|^2.
\end{equation}
Note that one may also interpret the sum of the tensor and the brane current as a non-symmetric energy-momentum tensor $\hT_{\mu\nu}^B$ (cf.\ \eqref{symmetrization}).

For the zeroth component $\Rc^{\p\zero}_\hmu$, we simply $\zerotwo$-supersymmetrically complete the boundary $R$-current \eqref{r-current boundary}, where again $\alpha=(\deg W)^{-1}, \ell_E=\deg E$ and $\ell_J=\deg J$:
\begin{equation}
\begin{aligned}
\Rc^{\p\zero}_\hmu&=\alpha(\ell_E-\ell_J)\delta_\hmu^{-}\Hb H,
\end{aligned}
\end{equation}
or, in bispinor notation,
\begin{equation}
\begin{aligned}
\Rc^{\p\zero}_{--}&=4\alpha(\ell_E-\ell_J)\Hb H,
\\\Rc^{\p\zero}_{++}&=0.
\end{aligned}
\end{equation}
Note that the $\zerotwo$-completion $(\Rc^{\p\zero})_\hmu$ of $(j^\p)_\hmu$ does not contain \emph{all} the boundary contributions necessary: we need the boundary corrections $T^\p_{--}$ to the energy-momentum tensor, which are not contained in our boundary multiplet $(\Rc^{\p\zero})_{++}$, as can be checked.\footnote{This can be directly verified by the explicit expansions \eqref{+direction}--\eqref{perp-direction}: $T_{+-}$ and $T_{++}$ are contained in the $\zero$-pieces, while $T_{--}$ is contained in the $\two$-piece.} Hence, we must also compute the correction for the second-order terms $(\Rc^{\p\two})_{\alpha\alpha}$:
\begin{equation}
	\begin{aligned}
		\Rc^{\p\two}_{++}&=
		8\widetilde{\Lag^\p}
		-8\alpha \p_\perp(\Phib\Phi)
		+4\sqrt 2 i\alpha 
		(\Db_+\Phib \Psi-\Psib D_+\Phi)|_\p 
		\\&
		=-8|J(\Phi)|^2-8|E(\Phi)|^2
		-8\alpha \p_\perp(\Phib\Phi)
		+4\sqrt 2 i\alpha 
		(\Db_+\Phib \Psi-\Psib D_+\Phi)|_\p 
		\\&=-16 (T^\p_{-+})^\Rc +2C^\p_{+-}+\ldots,
		\\
		\Rc^{\p\two}_{--}&=8i \p_-\Hb H-8i\Hb \p_-H
		\\&=-16(T^\p_{--})^\Rc +\ldots,
	\end{aligned}
\end{equation}
where the boundary contribution $(C^\p_{\hmu\hnu})^\Rc$ to the brane current $(C^B_{\mu\nu})^\Rc$ in the $\Rc$-frame can be found to be $(C^\p_{+-})^\Rc=-i(1-4\alpha)\psib\psi$. It is essentially the induced boundary improvement corresponding to symmetrization of the energy-momentum tensor (cf.\ page \pageref{induced boundary improvement}), now in the $\Rc$-frame. We have also $\zerotwo$-completed the on-shell boundary Lagrangian:
\begin{equation}
	\begin{aligned}
		\widetilde{\Lag^\p}&= -|J(\Phi)|^2	-|E(\Phi)|^2 .
	\end{aligned}
\end{equation}

\subsubsection{Integrated supercurrent multiplets}
We now discuss integrated supercurrent multiplets as in section \ref{sec: integrated sucumus abstract}. The integration along $x^\perp$ will make our Landau-Ginzburg model effectively two-dimensional and we will recover genuine 2D $\nzerotwo$ (integrated) supercurrent multiplets. 

We thus find, according to \eqref{integr multiplet definition}:
\begin{equation}
	\begin{aligned}
		\Rc^{\integ\zero}_{++}&=\int \d x^\perp \;\big[
		8\alpha(i\Phib \p_+ \Phi-i\p_+ \Phib\Phi)-2(1-2\alpha) \Db_+\Phib D_+\Phi\big],
		\\\Rc^{\integ\zero}_{--}&=
		4\alpha(\ell_E-\ell_J)\Hb H
		+\int \d x^\perp \;\big[
		8\alpha (i\Phib \p_- \Phi-i\p_- \Phib\Phi)-4(1-2\alpha) \Psib\,
		\big],
	\end{aligned}
\end{equation}
as well as
\begin{equation}
	\begin{aligned}
		\Rc^{\integ\two}_{++}&=-8|J(\Phi)|^2-8|E(\Phi)|^2
		+4\sqrt 2 i\alpha 
		(\Db_+\Phib \Psi-\Psib D_+\Phi)|_\p 
		\\&\hspace{.4cm}
		-16\int \d x^\perp \; 
		\big[
		\alpha \p_+\p_-(\Phib\Phi)
		+\tfrac 1 2 \p_\perp\Phib\p_\perp\Phi 
		+\tfrac 1 2 |W'(\Phi)|^2
		\\&\hspace{2cm}
		-\tfrac i 4\p_-\Db_+\Phib D_+\Phi
		+\tfrac i 4\Db_+\Phib \p_-D_+\Phi
		\big],
		\\\Rc^{\integ\two}_{--}&=
		8i \p_-\Hb H-8i\Hb \p_-H
		\\&\hspace{.4cm}
		-16\int \d x^\perp \;
		\big[
		2\p_-\Phib\p_-\Phi 
		-\alpha \p_-^2(\Phib\Phi)
		-\tfrac i 2 \p_-\Psib \Psi
		+\tfrac i 2 \Psib\p_-\Psi
		\big].
	\end{aligned}
\end{equation}
Note that from a 2D perspective, the superfields $\Rc^{\integ\zero}_{++},\Rc^{\integ\zero}_{--}$ and $\Rc^{\integ\two}_{--}$ are enough to form a 2D supercurrent multiplet. 

After using equations of motion \eqref{EoM Bulk 3D}, \eqref{EoM boundary}, boundary conditions \eqref{factorized neumann BC}, factorization condition \eqref{factorization condition} and homogeneity of superpotential terms, we find  that these integrated current multiplets indeed satisfy the relations
\begin{equation}
	\begin{aligned}
		\Db_+\big(\Rc^{\integ\two}_{--}+2i\p_-\Rc^{\integ\zero}_{--}\big)&=0,
		\\\Db_+(\Rc^{\integ\two}_{++}+2i\p_-\Rc^{\integ\zero}_{++})&=0.
	\end{aligned}
\end{equation}
which shows that the respective lowest components are  $\Qb_+$-cohomology elements (cf.\ section \ref{sec: integrated sucumus abstract}).

\section{Quantization}
\label{sec:Quantization}

Similarly to \cite{DiPietro:2015zia}, we can follow a canonical quantization approach in this model and verify the supersymmetry conservation, the boundary conditions, and the factorization condition in an independent way. We impose the following canonical quantization conditions:
\begin{subequations} \label{eq:canonical quantization}
\begin{align}
[\p_0 \phi(x), \phib(y) ] &= -i \delta^{(2)} (x - y),\\
\{ \psib_\ga (x), \psi_\gb(y) \} &= - \gamma^0_{\ga \gb}\delta^{(2)} (x - y), \\
\{\etab(x), \eta(y)\} &= \delta (x^2 - y^2).
\end{align}
\end{subequations}
In general, the commutation relations are modified by the introduction of a boundary. Here, however, we follow the point of view of the authors in \cite{DiPietro:2015zia} and use the ``naive'' commutators even after introducing the boundary. This can be justified by considering a full bulk theory first, quantizing it, then introducing a boundary and studying the effect of the boundary on the old bulk fields. Using this method, some properties of the model can be verified independently of the approach in section \ref{sec:LG model}. If we used static boundary conditions and the respective modified commutators instead, these properties would hold trivially. Notice that there are no singularities when moving component fields of chiral multiplets to the boundary as a consequence of supersymmetry.

The following relations hold in the bulk \cite{Dumitrescu:2011iu}:
\begin{align}
\{\Qbc_\ga, S_{\beta \mu} \} &= \gamma^\nu_{\ga \gb} \big(2 T_{\nu \mu} + \tfrac{1}{4} \eps_{\nu \mu \rho} H^\rho + i \p_\nu j_\mu - i \eta_{\mu \nu} \p_\rho j^\rho \big) + i \eps_{\ga \gb} \eps_{\mu \nu \rho} \big( \tfrac{1}{4}  F^{\nu \rho} + \p^\nu j^\rho \big) , \label{eq:bulk QbS commutator} \\
\{\Qc_\ga, S_{\beta \mu} \} &= \tfrac{1}{4} \bar{C} (\gamma_\mu)_{\ga \gb} + i \eps_{\mu \nu \rho} \gamma^\nu_{\ga \gb} \bar{Y}^\rho . \label{eq:bulk QS commutator}
\end{align}
In the absence of a boundary, integration of these relations yields the expected supersymmetry algebra. However, this changes under the introduction of a boundary for two reasons: First, there are additional degrees of freedom at the boundary which appear in $\Qc$ and $S$, and second, there are boundary contributions from pure \emph{bulk} terms as well.

Notice that the half-integrated commutators like $\{\Qc_\ga, S_{\ga\mu}\}$ or $\{\Qbc_\ga, S_{\gb\mu}\}$ are affected by improvements, but the fully integrated commutators like $\{\Qc_+, \Qc_+\}$ and $\{\Qbc_+, \Qc_+\}$ must be invariant under them. This is easy to see in pure bulk theories, but with a boundary, it holds as well. A generic commutator of a charge $\Qc$ and a current $J^\mu$ improves as follows:
\begin{equation}
\{\Qc, J'^\mu(x)\} = \{\Qc, J^\mu(x)\} + \{\Qc, \p_\nu M^{[\mu\nu]}(x)\} + \gd(x^n) \{\Qc, M^{n\hmu}(x)\}.
\end{equation}
Notice that $\Qc$ is a charge and thus invariant under improvements. If $\Qc$ commutes with $\p_\nu$, we find that the integrated algebra $\int \d x \{\Qc, J'^\mu(x)\}$ is unchanged.

\subsection{General properties of the supercharge} \label{subsec:quantization supercharge}

Let us now restrict to the case $\ga=+$, $\gb=+$ as the other supercharge and -current will be broken by the introduction of the boundary. The supercurrent from \eqref{eq:bulk supercurrent S frame}, \eqref{eq:boundary supercurrent S frame} (which we repeat for convenience)
\begin{equation}
S_{+ \mu}(x) = -\sqrt{2} i (\gamma_\mu \psib)_+ \Wb'(\phib) + \sqrt{2} (\psi \gamma_\mu \gamma^\nu)_+ \p_\nu \phib - \sqrt{2} \, \gd(x^\perp) \gd_\mu^- (\Jb \etab - i \Eb \eta) 
\end{equation}
integrates to the full supercharge
\begin{equation}
\Qc_+ =\int_\gS \sqrt{2}\big( \psi_- \p_\perp \phib + i \psib_- \Wb'(\phib) - 2 \psi_+ \p_+\phib \big) + \int_{\p \gS} \sqrt{2}(\Jb \etab - i \Eb \eta),
\end{equation}
where $\int_\gS = \int_{\R \times (-\infty,0]}\d x^2 \d x^\perp$ and $\int_\pgS = \int_{\R} \d x^2$. Let us start by studying the action of the bulk part of $\Qc_+$ on component fields. Analogous to \cite{DiPietro:2015zia}, the action of $[\Qc_{+,\bulk}, \boldsymbol{\cdot}]$ on bulk component fields ($\phi$, $\phib$, $\psi_{\pm}$, $\psib_{\pm}$) is the same as in the pure bulk theory with one exception: The commutator with $\p_0 \phi$ receives an extra boundary term
\begin{equation} \label{eq:change in d0phi bulk}
 [\Qc_{+,\bulk}, \p_0 \phi(x)] = \sqrt{2}i\p_0\psi_+(x) + \sqrt{2}i \gd(x^\perp) \psi_-(x) .
\end{equation}
This result can be derived using the quantization conditions \eqref{eq:canonical quantization} and the delta distribution rule in appendix~\ref{app:deltadistributions}. Notice that for $\nu \neq 0$, the identity $\del{x^\nu}[\Qc_+, \phi(x)] = [\Qc_+, \p_\nu \phi(x)]$ is only true if the correct delta distribution rules derived in appendix \ref{app:deltadistributions} are applied.

The boundary part $\Qc_{+,\bdy}$ has a trivial action on the bulk component fields and on $\p_\nu\phi$, $\nu\neq0$. The action on the boundary fermions $\eta$, $\etab$ is as expected. Again, we get extra terms for $\p_0\phi$. Overall, the full charge acts as follows on $\p_0\phi$:
\begin{equation} \label{eq:change in d0phi full}
[\Qc_{+}, \p_0 \phi(x)] = \sqrt{2} i \p_0 \psi +  \sqrt{2} (i \etab \Jb' + \eta \Eb' + i \psi_-) \gd (x^\perp) .
\end{equation}
We see that the boundary contribution vanishes under the symmetric boundary condition \eqref{factorized neumann BC}, which is an independent way of verifying this boundary condition. However, for reasons outlined above, we will not impose this condition statically and thus treat this extra term like a genuine new contribution.

\subsection{\texorpdfstring{The $\{\Qbc,S\}$ commutator}{The \{Q,S\} commutator}} \label{subsec:Qbc commutator}

We now would like to compute $\{\Qbc_+, S_{+ \mu}(x) \}$ in the presence of a boundary and verify that it integrates to the expected preserved algebra $\{\Qbc_+, \Qc_+\} = -4 P_+$ (see section \ref{subsec:bulk and boundary constraints}). We expect the known terms \eqref{eq:bulk QbS commutator} in the bulk, and extra terms at the boundary. As both $\Qbc_+$ and $S_{+ \mu}(x)$ have bulk and boundary parts, there are four combinations from which new terms may arise: bulk-bulk, bulk-boundary, boundary-bulk, and boundary-boundary.

\subsubsection{Boundary contributions from bulk-bulk terms}

Let us now check how the changes introduced by the boundary affect the half-integrated algebra: the bulk-bulk term
\begin{equation}
\{\Qbc_{\ga,\bulk}, S_{\gb\mu,\bulk}\} = \{\Qbc_\ga, -\sqrt{2} i (\gamma_\mu \psib)_\gb \Wb'(\phib) + \sqrt{2} (\psi \gamma_\mu \gamma^\nu)_\gb \p_\nu \phib \}
\end{equation}
is affected by the changed relation \eqref{eq:change in d0phi bulk} in the term $\p_0\phib$, where we get an extra boundary term
\begin{equation} \label{eq: LG model bulk-bulk commutator}
\{\Qbc_{\ga,\bulk}, S_{\gb\mu,\bulk}(x)\} = \text{known bulk terms \eqref{eq:bulk QbS commutator}} +\gd(x^\perp) \underbrace{2 i (\psi \gamma_\mu \gamma^0)_\gb  ( \gamma^n\gamma^0\psib(x))_\ga}_{=:B_{\ga \gb, \mu}},
\end{equation}
which is also consistent with a similar result in \cite{DiPietro:2015zia}. The component most relevant to us is $B_{++}^{\ph+ \ph+ 0}$, as it appears in the integration of the part of the algebra that is unbroken. It can be rewritten to 
\begin{equation}
B_{++}^{\ph+ \ph+ 0} = 2i \psi_+ \psib_- =  i \psib \psi  -i \psib \gamma^\perp \psi = i \psib \psi  -i j^\perp,
\end{equation}
where we have inserted a bulk component of the supercurrent multiplet in the last equality, see appendix \ref{app:bulk components S multiplet}. It is noteworthy that this boundary term is neither real nor imaginary.

\subsubsection{Contributions from the boundary degrees of freedom}
Again using the commutation relations but not boundary conditions, we find
\begin{subequations} \label{eq:boundary commutators LG model}
\begin{align}
\{\Qbc_{+,\bdy}, S_{+,\mu,\bdy} (x) \} &=  -2 \gd_\mu^-\gd(x^\perp) (|J|^2 + |E|^2) , \\
\{\Qbc_{+,\bulk}, S_{+,\hmu,\bdy}(x) \} &= -2 i \gd(x^\perp) \gd_\hmu^- \psib_+(x)  \big(\Jb'(x)\etab(x) - i\Eb'(x)\eta(x)\big) , \\
\{\Qbc_{+,\bdy}, S_{+,\mu,\bulk} (x) \} &= 2 i \gd(x^\perp) \big(J'(x)\eta(x) + iE'(x)\etab(x)\big) (\psi(x)\gamma_\mu\gamma^0)_+ .
\end{align}
\end{subequations}
It is noteworthy that modulo boundary conditions, the third term cancels the boundary contribution of $\{\Qbc_{+,\bulk}, S_{+,\mu,\bulk} (x)\}$. This is expected since both extra terms have their origins in the changed relation \eqref{eq:change in d0phi full} which is identical to the original bulk relation modulo boundary conditions.

\subsubsection{Integrating the algebra}

As the integral of $S_{+}^{\ph+0}$ over a constant time slice yields the supercharge $\Qc_+$, an integration of the commutator $\int \{\Qbc_+, S_{+}^{\ph+0}\} = \{\Qbc_+, \Qc_+\}$ is a commutator which appears in the preserved supersymmetry algebra \eqref{eq:remaining zerotwo algebra}. We can thus check \eqref{eq: LG model bulk-bulk commutator} and \eqref{eq:boundary commutators LG model} by integrating $\{\Qbc_+, S_{+}^{\ph+0}\}$ and comparing the result to the known algebra. We will plug in the component expansions from appendix \ref{app:bulk components S multiplet}.

Let us first check the imaginary part, which is zero on the expected right-hand side of the equation. Interesting contributions come only from $\{\Qbc_{+,\bulk}, S_{+,\bulk}^{\ph+0}\}$, as all other contributions together are trivially real.
\begin{equation}
\Im \int_\gS \{\Qbc_+, {S_+}^0 \} =  \int_\gS 2 (i \p_+ j_0 - i \eta_{+ 0} \p^\rho j_\rho) + \int_\pgS \Im(B_{++}^{\ph+\ph+0}) = \int_\gS i \p_\perp j_\perp - \int_\pgS i j_\perp = 0.
\end{equation}
A similar computation in four dimensions was done in \cite{DiPietro:2015zia}.

For the real part we get (as expected from \eqref{eq:bulk QbS commutator})
\begin{equation}
\{\Qbc_+, S_{+}^{\ph+ 0} \}|_\bulk = -\gamma_{22}^\nu \big( 2 T_{\nu 0} + \tfrac{1}{4} \eps_{\nu 0 \rho} H^\rho\big) = - 4 (\hT^B)_+^{\ph+0},
\end{equation}
where $\hT^B$ is the bulk Noether (non-symmetric) energy momentum tensor \eqref{eq:bulk EM tensor nonsym}.
All boundary terms together (eqs.\ \eqref{eq: LG model bulk-bulk commutator} and \eqref{eq:boundary commutators LG model}) yield
\begin{equation} %
\{\Qbc_+, S_{+}^{\ph+ 0}\}|_\p = i \psib\psi + 2 |J|^2 + 2|E|^2 + 2 i\psib_+\big(\Jb'\etab - i\Eb'\eta\big) - 2i\big(J'\eta + iE'\etab\big)\psi_+ = - 4 (\hT^\p)_+^{\ph+ 0}
\end{equation}
using an explicit comparison to the non-symmetric boundary energy momentum tensor $\hT^\p$ \eqref{eq:bdy EM tensor nonsym no BC} which belongs to the bulk Noether energy momentum tensor $\hT^B$. Overall, we find that
\begin{equation}
\{\Qbc_+, \Qc_+\} = \int_\gS \{\Qbc_+, S_{+}^{\ph+ 0}\} = - 4 \int_\gS \big( (\hT^B)_+^{\ph+0} + \gd(x^\perp) (\hT^\p)_+^{\ph+ 0}\big) = - 4 P_+,
\end{equation}
which verifies the algebra. Notice that $P_+$ is independent of improvements, thus we may use improved versions of the energy-momentum tensor to compute the right-hand side of the equality. However, on the level of the half-integrated algebra, we see that the Noether ($\Sc$-frame) supercurrent generates the non-symmetric Noether energy momentum tensor, and both are sensitive to improvements.

Let us emphasize again that this argument works \emph{without} explicitly assuming boundary conditions and modifying the bulk fields in the presence of a boundary. Rather, we study the bulk theory without a boundary, then introduce it, and verify that the supersymmetry algebra is preserved by the equal-time commutators. Notice that without assuming any boundary conditions, the charges $\Qc_+$, $\Qbc_+$  are not conserved. However, no reference to the specific choice of boundary conditions was made in this argument (although in this simple model with one bulk chiral field and one boundary Fermi, \eqref{factorized neumann BC} is the only symmetric boundary condition compatible with stationarity).

\subsection{\texorpdfstring{The $\{\Qc,S\}$ commutator}{The \{Q,S\} commutator}} \label{subsec:Qc commutator}

In a similar way, we can also verify that $\{\Qc_+, S_{+ \mu}(x) \}$ integrates to the expected algebra $\{\Qc_+,\Qc_+\} = 0$. In the bulk, we get from \eqref{eq:bulk QS commutator}
\begin{equation}
\{\Qc_{+,\bulk}, S_{+ \mu,\bulk}(x) \} = i\eps_{\mu +\rho} (\gamma^+)_{++}\bar{Y}^\rho = - 8i\eps_{\mu +\rho} \p^\rho \Wb .
\end{equation}
From the commutation relations \eqref{eq:canonical quantization} we find that 
$\{\Qc_{\ga,\bulk}, S_{\gb \mu,\bdy} \}$ and $\{\Qc_{\ga,\bdy}, S_{\gb \mu,\bulk}\}$ are zero, but we do get a contribution from
\begin{equation}
\{\Qc_{\ga,\bdy}, S_{\gb \mu,\bdy}(x) \} = 4 i \gd_\mu^- \gd(x^\perp) \Jb(x) \Eb(x).
\end{equation}
Integrating the relation, we find
\begin{equation}
\{\Qc_+, \Qc_+\} = \int_\gS \{\Qc_+, S_+^{\ph+ 0}(x) \}
= 4i \int_\pgS \big( \Wb - \Jb(x) \Eb(x) \big),
\end{equation}
so if the factorization condition \eqref{factorization condition} is met, the algebra is preserved under the introduction of the boundary.

\section{Conclusions and Outlook} \label{sec:outlook}

In this paper we analyzed three-dimensional theories with $\mc N=2$ supersymmetry from various points of view: We constructed the symmetry currents by developing an extension of Noether's procedure, we considered the supermultiplets in the boundary situation and we commented on the $\Qb_+$-cohomology in a possible half-twisted theory. There are various ways to extend our results. First of all, while we focused on the case of $\N=2$ with specific boundary conditions, a similar analysis can be performed for other dimensions and other supersymmetries. Furthermore, our examples were limited to theories of chiral multiplets and it would be interesting to couple these systems to a gauge group.

We considered only flat backgrounds, and it would be very interesting to extend the discussion to  three-dimensional theories with boundaries in curved backgrounds. The partition function in such backgrounds has been studied recently in \cite{Yoshida:2014ssa,Jockers:2018sfl,Cheng:2018vpl}. Matrix factorizations have already been discussed on curved manifolds with boundary in \cite{Yoshida:2014ssa,Jockers:2018sfl}, see \cite{Aprile:2016gvn} for a discussion of A-type boundary conditions on curved manifolds. The formulation of boundary conditions must be compatible with the constraints imposed by the bulk supersymmetry on the background geometry. In particular, the manifolds have to admit transversely holomorphic foliations, see \cite{Closset:2012ru} for a discussion of the geometry compatible with supersymmetry. 

In theories without boundary,  supercurrent multiplets have been very useful to formulate theories in curved space. They also helped to understand the (in-)dependence of the partition function of part of the data \cite{Closset:2013vra}. It would be interesting to see whether boundary supermultiplets can help to draw similar conclusions in situations with boundaries.

In the example of Landau-Ginzburg models, we have exhibited in some detail the symmetries of models involving matrix factorizations. In the case of \emph{two-dimensional} Landau-Ginzburg models, matrix factorizations provided the key to fully solve the theories in situations with boundaries, in the sense that the full bulk and boundary spectrum and all correlation functions \cite{Kapustin:2003ga} were determined. It would be interesting to see to what extend these features have analogs in three dimensions. As the theory cannot be fully twisted, one would expect that a holomorphic dependence has to remain. 

Finally, one could generalize boundary conditions to defect gluing conditions, making potentially contact with \cite{Dimofte:2017tpi,Drukker:2017dgn}. In the Landau-Ginzburg example it is easy to see that defects between theories with different superpotentials involve factorizations of the difference of the two superpotentials on the two sides of the defect.

\acknowledgments
We thank Peter Mayr, Pantelis Fragkos, Tomáš Procházka, and Ismail Achmed-Zade for valuable discussions and input. 
The authors were supported by the Deutsche Forschungsgemeinschaft (DFG) grant ``Defects and non-perturbative operators in superconformal theories'' and Exzellenzcluster Universe.

\appendix
\addtocontents{toc}{\protect\setcounter{tocdepth}{1}}
\pdfbookmark[-1]{Appendices}{test}

\section{Conventions and notation} \label{app:conventions}
We mostly follow the notation and conventions of \cite{Dumitrescu:2011iu} which we recall for convenience.

\paragraph{Spacetime}
Our (half-)spacetime is given by
\begin{equation}
	M=\Set{(x^0,x^1,x^2) | x^1\leq 0},
\end{equation}
with mostly-plus metric $\eta_{\mu\nu}=\diag (-1,1,1)$. We most frequently use light-cone coordinates
\begin{equation}
	x^\pm=x^0\pm x^2, \quad x_\pm =\tfrac 1 2 (x_0\pm x_2),\quad  x^1=x_1=x^\perp,
\end{equation}
where the metric reads
\begin{equation}
	\eta_{\mu\nu}=
	\begin{pmatrix}
	0 & -\tfrac 1 2 & 0 \\
	-\tfrac 1 2 & 0 & 0 \\
	0 & 0 & 1
	\end{pmatrix},
	\quad
	\eta^{\mu\nu}=
	\begin{pmatrix}
	0 & -2 & 0 \\
	-2 & 0 & 0 \\
	0 & 0 & 1
	\end{pmatrix}.
\end{equation}
The Levi-Civita symbol is defined by $\eps_{012}=-1$, $\eps^{012}=1$. In light-cone coordinates it is $ \eps_{+-\perp}=-\tfrac 12, \;\eps^{+-\perp}=2$. It satisfies
\begin{equation}
	\begin{aligned}
		\eps_{\mu\nu\lambda}\eps^{\sigma\rho \lambda}&=\delta_\mu^{\ph\mu\rho}\delta_\nu^{\ph\nu\sigma}-\delta_\mu^{\ph\mu\sigma}\delta_\nu^{\ph\nu\rho},
		\\
		\eps_{\mu\rho\lambda}\eps^{\nu\rho\lambda}&=-2\delta_\mu^{\ph \mu \nu}.
	\end{aligned}
\end{equation}

\paragraph{Spinors}
Spinors in 3D are $SL(2,\R)$ fundamental representations, i.e.\ two component spinors $\psi_\alpha$, $\alpha\in\{1,2\}=\{-,+\}$. Indices are raised and lowered by $\eps_{\alpha\beta},\eps^{\alpha\beta}$, where $\eps_{12}=-1, \eps^{12}=1$ according to the rule
\begin{equation}
	\psi^\alpha=\eps^{\alpha\beta}\psi_\beta, \; \psi_\alpha=\eps_{\alpha\beta} \psi^\beta.
\end{equation}
Explicitly, we have
\begin{equation}
	\psi^\alpha=
	\begin{pmatrix}
		\psi^- \\\psi^+
	\end{pmatrix}
	=
	\begin{pmatrix}
		\psi_+ \\-\psi_-
	\end{pmatrix}
	, \quad
	\psi_\alpha=
	\begin{pmatrix}
		\psi_- \\ \psi_+
	\end{pmatrix}
	=
	\begin{pmatrix}
		-\psi^+ \\ \psi^-
	\end{pmatrix}.
\end{equation}
Indices that are contracted ``from top to bottom'' are omitted:
\begin{equation}
	\psi\chi\coloneqq \psi^\alpha\chi_\alpha=\psi^-\chi_-+\psi^+\chi_+.
\end{equation}
Note that $\psi\chi=\chi\psi$. 
Since Hermitian conjugation flips the order of spinors without flipping index position, we have that $\bar{\psi\chi}=-\psib\chib$. 

Some useful identities are given by
\begin{equation}
	\begin{aligned}
		\psi\psi&=2\psi^+\psi^-,
		\\\psi^\alpha\psi^\beta&=-\tfrac 1 2 (\psi\psi) \eps^{\alpha\beta},
		\\\psi_\alpha\psi_\beta &= \tfrac 1 2 (\psi\psi) \eps_{\alpha\beta}.
	\end{aligned}
\end{equation}

\paragraph{Clifford algebra}
We use the real gamma matrices 
\begin{equation}
	\gamma^\mu_{\alpha\beta}=(\gamma^0_{\alpha\beta},\gamma^1_{\alpha\beta},\gamma^2_{\alpha\beta})=(-\Unit , \sigma^1,\sigma^3).
\end{equation}
In light-cone coordinates these read explicitly
\begin{equation}
	\gamma^\mu_{\alpha\beta}=(\gamma^+_{\alpha\beta},\gamma^-_{\alpha\beta},\gamma^\perp_{\alpha\beta})=
	\Bigg(
	\begin{pmatrix}
	0 & 0 \\ 0 & -2
	\end{pmatrix}
	,
	\begin{pmatrix}
	-2 & 0 \\ 0 & 0
	\end{pmatrix}
	,
	\begin{pmatrix}
	0 & 1 \\ 1 & 0
	\end{pmatrix}
	\Bigg).
\end{equation}
They are symmetric $\gamma^\mu_{\alpha\beta}=\gamma^\mu_{\beta\alpha}$, real and satisfy the Clifford algebra
\begin{equation}
	(\gamma^\mu\gamma^\nu)_\alpha^{\ph\alpha\beta}=\eta^{\mu\nu}\delta_\alpha^{\ph\alpha\beta} +\eps^{\mu\nu\rho}(\gamma_\rho)_\alpha^{\ph\alpha\beta}.
\end{equation}
A useful list of identities follows from these:
\begin{equation}
	\begin{aligned}
		(\gamma^\mu)_{\alpha\beta} (\gamma_\mu)_{\gamma\delta}&=\eps_{\alpha\gamma}\eps_{\delta\beta}+\eps_{\alpha\delta}\eps_{\gamma\beta},
		\\ (\gamma^\mu\gamma^\rho\gamma_\mu)_{\alpha\beta}&=-(\gamma_\rho)_{\alpha\beta},
		\\A_\alpha^{\ph \alpha \beta}&=\tfrac 1 2 (\tr A)\delta_\alpha^{\ph \alpha \beta} +\tfrac 1 2 \tr(\gamma_\mu A) (\gamma^\mu)_\alpha^{\ph \alpha \beta}.
	\end{aligned}
\end{equation}
We may map vectors to bispinors and vice versa by 
\begin{equation}\label{bispinor relations}
	\begin{aligned}
	v_{\alpha\beta}=-2\gamma^\mu_{\alpha\beta}v_\mu,\quad v_\mu=\tfrac 1 4 \gamma^{\alpha\beta}_\mu v_{\alpha\beta}.
	\end{aligned}
\end{equation}
These imply in particular that 
\begin{equation}\label{explicit bispinor relations}	
	\begin{aligned}
	v_\pm&=\tfrac 14 v_{\pm\pm},
	\\v^\pm&=-\tfrac 12 v^{\pm\pm},
	\\ v^\perp&=v_\perp= -\tfrac 1 2 v_{+-}=-\tfrac 1 2 v^{-+}.
	\end{aligned}
\end{equation}

\paragraph{Integration}
We fix the integration order/constants by
\begin{equation}
	\int \d^2\theta \;\theta^2=1,\quad \int \d^2\thetab \;\thetab^2=-1 \quad \text{and}\quad\int \d \tp\d\tbp \tbp\tp =1 
\end{equation}
i.e.\ adjacent symbols cancel in $\zerotwo$-integration. These together imply
\begin{equation}
	\int \d^2\theta=\frac 1 2 \int \d \tm \d \tp, \quad \int \d^2\thetab =\frac 1 2 \int \d \tbp\d \tbm,
\end{equation}
and
\begin{equation}
	\int \d^4\theta = \frac 1 4\int \d^2\tp \d^2\tm,
\end{equation}
where we have defined
\begin{equation}
	\int \d^2 \theta^\pm=\int \d\theta^\pm \d \thetab^\pm.
\end{equation}

\paragraph{Supercharges and superderivatives}
The supercharges of 3D $\mc N=2$ supersymmetry in terms of $(x^\mu,\theta,\thetab)$-coordinates are
\begin{equation}\label{3d supercharges}
	\begin{aligned}
		Q_\alpha&=\pd{}{\theta^\alpha}+i(\gamma^\mu\thetab)_\alpha \p_\mu 
		=
		\begin{pmatrix}
			\pd{}{\tm}+2i\tbm \p_- -i \tbp \p_\perp\\
			\pd{}{\tp}+2i\tbp \p_+ -i \tbm \p_\perp
		\end{pmatrix},
		\\
		\Qb_\alpha&=-\pd{}{\thetab^\alpha}-i(\gamma^\mu\theta)_\alpha \p_\mu 
		=
		\begin{pmatrix}
			-\pd{}{\tbm}-2i\tm \p_- +i \tp \p_\perp\\
			-\pd{}{\tbp}-2i\tp \p_+ +i \tm \p_\perp
		\end{pmatrix}.
	\end{aligned}
\end{equation}
The covariant derivatives are
\begin{equation}\label{3d superderivatives}
	\begin{aligned}
		 D_\alpha&=\pd{}{\theta^\alpha}-i(\gamma^\mu\thetab)_\alpha \p_\mu
		 =
		 \begin{pmatrix}
			 \pd{}{\tm}-2i\tbm \p_- +i \tbp \p_\perp\\
			 \pd{}{\tp}-2i\tbp \p_+ +i \tbm \p_\perp
		 \end{pmatrix},
		\\ \Db_\alpha&=-\pd{}{\thetab^\alpha}+i(\gamma^\mu\theta)_\alpha \p_\mu
		=
		\begin{pmatrix}
			-\pd{}{\tbm}+2i\tm \p_- -i \tp \p_\perp\\
			-\pd{}{\tbp}+2i\tp \p_+ -i \tm \p_\perp
		\end{pmatrix}.
	\end{aligned}
\end{equation}
They satisfy the known algebra
\begin{equation}
	\begin{aligned}
		\{Q_\alpha,\Qb_\beta\}=2i \gamma^\mu_{\alpha\beta}\p_\mu, \quad \{D_\alpha,\Db_\beta\}=-2i \gamma^\mu_{\alpha\beta}\p_\mu.
	\end{aligned}
\end{equation}

\section{Decomposition of \texorpdfstring{\boldmath{3D $\mc N=2$}}{3D N=2}}

\subsection{Decomposition to \texorpdfstring{\boldmath 2D $\nzerotwo$}{2D N = (0,2)}}\label{app:branching to zerotwo}
We may constructively decompose 3D $\mc N=2$ superfields and operators in their $\zerotwo$-components. To do so, we use the \emph{branching coordinate} $\xi$.\footnote{$\xi^\perp$ is also called ``invariant coordinate'' in \cite{Drukker:2017xrb,Drukker:2017dgn}.} It is chosen such that in superspace with coordinates $(\xi^\mu, \gt^+, \gt^-)$, the representations of the preserved supercharge operators $Q_+$ and $\Qb_+$ commute with $\gt^-$ and $\tb^-$. Then $Q_+$ and $\Qb_+$ can be restricted to the sub-superspace without $\gt^-$. Another property of the branching coordinate is that the preserved supercharge operators do not contain or generate $P_\perp$.

If we want to preserve the $\zerotwo$-subalgebra generated by $Q_+,\Qb_+$, one can easily check that we need
\begin{equation}\label{branching coordinate}
	\begin{aligned}
		\xi^\mu=\big(x^+,x^-,x^\perp+i(\tp\tbm-\tm\tbp)\big).
	\end{aligned}
\end{equation}
Indeed, one can check that in terms of $\xi^\mu$ the operators \eqref{3d supercharges}, \eqref{3d superderivatives} take the following form
\begin{equation}\label{branched superoperators}
	\begin{aligned}
		Q_+&= \pd{}{\tp}+2i \tbp \pd{}{\xi^+}, &	  Q_-&=\pd{}{\tm}+2i\tbm \pd{}{\xi^-}-2i \tbp \pd{}{\xi^\perp},\\
		\Qb_+&= -\pd{}{\tbp}-2i \tp \pd{}{\xi^+}, &	  \Qb_-&=-\pd{}{\tbm}-2i\tm \pd{}{\xi^-}+2i \tp \pd{}{\xi^\perp},
		\\
		D_+&=\pd{}{\tp}-2i \tbp \pd{}{\xi^+}+2i\tbm\pd{}{\xi^\perp} ,	& D_- &=\pd{}{\tm}-2i\tbm\pd{}{\xi^-},\\
		\Db_+&=-\pd{}{\tbp}+2i \tp \pd{}{\xi^+}-2i\tm\pd{}{\xi^\perp}, 	& \Db_-&=-\pd{}{\tbm}+2i\tm\pd{}{\xi^-}.
	\end{aligned}
\end{equation}
In particular, the $Q_+, \Qb_+$ do not contain any $\p_\perp$ terms. The $\zerotwo$-covariant derivatives are defined by
\begin{equation}
	\begin{aligned}
		D_+^\zerotwo&=\pd{}{\tp}-2i \tbp \pd{}{\xi^+},\\
		\Db_+^\zerotwo&= -\pd{}{\tbp}+2i \tp \pd{}{\xi^+},
	\end{aligned}
\end{equation}
so we have that
\begin{equation}\label{D 3d vs D zerotwo}
	\begin{aligned}
		D_+&=D_+^\zerotwo+2i\tbm \p_\perp, \\
		\Db_+&=\Db_+^\zerotwo-2i\tm \p_\perp .
	\end{aligned}
\end{equation}
We often drop the $\zerotwo$-label when the covariant derivative type is clear from context. \\
We may now simply perform a Taylor expansion of an $\mc N=2$ superfield $\X$:
\begin{equation}
	\X(x,\theta,\thetab)=X^\zero(\xi,\tp,\tbp) +\tm X^\onea(\xi,\tp,\tbp) + X^\oneb(\xi,\tp,\tbp) \;\tbm+\tm\tbm X^\two(\xi,\tp,\tbp).
\end{equation}
It is clear that the $\zerotwo$-operators $\{Q_+,\Qb_+,D_+^\zerotwo,\Db_+^\zerotwo\}$ do not ``mix'' the coefficients of different orders in $\tm,\tbm$. In other words, the coefficients are exactly the $\zerotwo$-sub\-rep\-re\-sen\-ta\-tions of $\X(x,\theta,\thetab)$. 

As an example, let us decompose the 3D chiral field \eqref{chiral field} to its $\zerotwo$-submultiplets. 
In terms of $\xi$ we find that
\begin{equation}
	(y^+,y^-,y^\perp)=(\xi^+-2i\tp\tbm,\xi^--2i\tm\tbm, \xi^\perp+2i\tm\tbp).
\end{equation}
The expansion of $\Phi_{3D}$ gives then
\begin{equation}\label{branching of Phi_3D}
	\Phi_{3D}(x,\theta,\thetab)=\Phi(\xi, \tp,\tbp)-2i\tm\tbm\p_- \Phi(\xi, \tp,\tbp)+\sqrt 2 \tm \Psi (\xi, \tp,\tbp),
\end{equation}
where the chiral and Fermi multiplets are
\begin{equation}
	\begin{aligned}
		\Phi&=\phi+\sqrt 2 \tp\psi_+-2i \tp\tbp \p_+\phi, \\
		\Psi&=\psi_--\sqrt 2 \tp F - 2i \tp\tbp\p_+\psi_- 
		+\sqrt 2 i \tbp \p_\perp \phi -2i \tp\tbp \p_\perp \psi_+.
	\end{aligned}
\end{equation}
in agreement with \cite{Witten:1993yc}. These satisfy
\begin{equation}
	\Db_+\Phi=0,\quad \Db_+\Psi=-i\sqrt 2 \p_\perp\Phi.
\end{equation}
We can obtain the full expansion using \eqref{branching coordinate} on the right-hand side of \eqref{branching of Phi_3D}
\begin{equation}
	\Phi_{3D}=\Phi
	+\sqrt 2 \tm \Psi
	+i(\tp\tbm-\tm\tbp)\p_\perp \Phi
	-2i \tm\tbm\p_-\Phi  
	-\sqrt 2 i \tp\tm\tbm\p_\perp \Psi
	-\tp\tbp\tm\tbm\p_\perp^2 \Phi ,
\end{equation}
where now all (super-)functions depend on $x$.

\subsection{Decomposition to \texorpdfstring{\boldmath 2D $\mc N=\oneone$}{2D N=(1,1)}} \label{app:branching to oneone}
In the presence of a boundary we may also preserve a 2D $\mc N=\oneone$ subalgebra. For completeness, we also present decomposition of 3D $\mc N=2$ superfields and operators into this subalgebra. \\
We want to preserve the $\oneone$-subalgebra generated by\footnote{Note that in breaking from a 2D $(2,2)$ algebra to 2D $(1,1)$ we have some freedom of choice in ``picking phases'', i.e.\ we may choose
		\begin{equation*}
		\fQ_i\coloneqq\tfrac 1 2( e^{iv_i}Q_i +e^{-i v_i}\Qb_i),
		\end{equation*}
		for $i=1,2$ (see e.g.\ \cite{Hori:2000ic,Hori:2003ic}). However, in 3D $\mc N=2$ to $(1,1)$-subalgebra breaking, we further need to impose that the charges do not generate any $\p_\perp$-derivatives, i.e. that 
		\begin{equation*}
		\{\fQ_1,\fQ_2\}=\tfrac i 2 \Re(e^{i(v_1-v_2)})\p_\perp\overset{!}{=}0,
		\end{equation*}
		which fixes the phase to $v_1-v_2=\tfrac \pi 2 +k \pi$.}
\begin{equation}
\begin{aligned}
\fQ_-&\coloneqq\tfrac 1 2( Q_- +\Qb_-),\\
\fQ_+&\coloneqq\tfrac i 2( Q_+ -\Qb_+).
\end{aligned}
\end{equation}
These satisfy the $\oneone$-algebra:
\begin{equation}
\begin{aligned}
\{\fQ_-,\fQ_-\}&=-2i \p_-,\\
\{\fQ_+,\fQ_+\}&=-2i \p_+,\\
\{\fQ_-,\fQ_+\}&=0.
\end{aligned}
\end{equation}
Defining rotated, real Grassmann variables
\begin{equation}
\begin{aligned}
\vt^-&=-i(\tm-\tbm), &&& \tvt^- &=-(\tm+\tbm),\\
\vt^+&= -(\tp+\tbp), &&& \tvt^+&=-i(\tp-\tbp).
\end{aligned}
\end{equation}
and using the shifted ``branching'' coordinate
\begin{equation}
\begin{aligned}
	(\zeta^+,\zeta^-,\zeta^\perp)&=\big(x^+,x^-,x^\perp +\tfrac i 2 (\vtm\tvtp+\vtp\tvtm)\big)
	\\&=\big(y^+-\vtp\tvtp,y^-+\vtm\tvtm,y^\perp -\tfrac 1 2(\vtm\vtp -i\vtm\tvtp-i\vtp\tvtm-\tvtm\tvtp)\big),
\end{aligned}
\end{equation}
the supercharges  take the form
\begin{equation}
\begin{aligned}
\fQ_-&=-i \pd{}{\vtm}+\vtm \pd{}{\zeta^-},
\\\fQ_+&=-i\pd{}{\vtp}+\vtp \pd{}{\zeta^+}.
\end{aligned}
\end{equation}
We can then define also the $\oneone$-covariant derivatives
\begin{equation}
\begin{aligned}
\fD_-&=-i \pd{}{\vtm}-\vtm \pd{}{\zeta^-},
\\\fD_+&=-i\pd{}{\vtp}-\vtp \pd{}{\zeta^+}.
\end{aligned}
\end{equation}
which of course satisfy 
\begin{equation}
\begin{aligned}
\{\fD_\pm,\fD_\pm\}&=2i \p_\pm, 
\\ \{\fD_+,\fD_-\}&=0.
\end{aligned}
\end{equation}
Then, $\oneone$-irreducible multiplets are of the form
\begin{equation}
\Sigma=A+i\vtp B+i \vtm C +i\vtm\vtp D.
\end{equation}
Using the branching coordinate we find the decomposition
\begin{equation}
\Phi_{3D}(x,\theta,\thetab)=\vPhi+ \tvtp \fD_+\vPhi -\tvtm \fD_-\vPhi+\tvtm\tvtp(\fD_-\fD_+\vPhi-\p_\perp \vPhi)
\end{equation}
where on the right-hand side we have super(functions) of $(\zeta,\vt,\tvt)$ and we have defined the $\oneone$-multiplet 
\begin{equation}
\vPhi\coloneqq \phi 
-\tfrac 1 {\sqrt 2} \vtp \psi_+
+\tfrac i {\sqrt 2} \vtm \psi_-
+\tfrac 1 2 \vtm\vtp (\p_\perp \phi
+i  F).
\end{equation}
Around the same point, the full branching reads
\begin{equation}
\begin{aligned}
\Phi_{3D}&=
\vPhi
+ \tvtp \fD_+\vPhi 
-\tvtm \fD_-\vPhi
+\tfrac i 2 (\vtm\tvtp+\vtp\tvtm)\p_\perp \vPhi 
+\tvtm\tvtp(\fD_-\fD_+\vPhi
-\p_\perp \vPhi)
\\&\hspace{.4cm}
+  \tfrac i 2 \tvtm\tvtp\vtp\p_\perp\fD_+\vPhi
+  \tfrac i 2\tvtm\tvtp\vtm\p_\perp\fD_-\vPhi
+\tfrac 1 4 \vt^4\p_\perp^2 \vPhi,
\end{aligned}
\end{equation}
where $\vt^4\coloneqq \vtp\tvtp\vtm\tvtm$.\\
The $\oneone$-variations are induced by  $\dsym=i\veps_+\fQ_++i\veps_- \fQ_-$,  $\veps_\pm$  real spinors:
\begin{equation}
\begin{aligned}
\dsym \phi &= -\tfrac 1 {\sqrt 2} \veps_+ \psi_+ +\tfrac i {\sqrt 2} \veps_- \psi_-,
\\\dsym \psi_+&=\sqrt 2 i \veps_+\p_+\phi +\tfrac 1 {\sqrt 2} \veps_- (\p_\perp \phi +i F),
\\\dsym \psi_-&= -\tfrac i {\sqrt 2} \veps_+ (\p_\perp \phi +i F) -\sqrt 2 \veps_- \p_-\phi,
\\\dsym i F&=\sqrt 2 \veps_+ (\p_+\psi_-+\tfrac 1 2 \p_\perp \psi_+)
-\sqrt 2 i \veps_-(\p_-\psi_+
+\tfrac 1 2  \p_\perp \psi_-).
\end{aligned}
\end{equation}
The 3D $\mc N=2$ bulk Lagrangian of one chiral field $\Phi_{3D}$ is decomposed as follows
\begin{equation}
	\begin{aligned}
		\Lag&= \int \d^4 \theta \big(\Phib_{3D}\Phi_{3D}-(\theta\theta)(\thetab\thetab)\p_+\p_-(\Phib_{3D}\Phi_{3D})\big) +\int \d^2\theta W(\Phi_{3D})+\int \d^2\thetab \Wb(\Phib_{3D})
		\\&= \int \d^2 \vt \Big[
		2(\fD_-\vPhib \fD_+\vPhi-\fD_+\vPhib \fD_-\vPhi)
		+\vPhib\p_\perp \vPhi -\p_\perp \vPhib\vPhi
		-2i(W+\Wb)
		\\&\pheq
		+ \p_\perp \big[
		\tfrac i 2 \vtp(\fD_+\vPhib\vPhi-\vPhib\fD_+\vPhi)
		+\tfrac i 2 \vtm (\fD_-\vPhib \vPhi-\vPhib\fD_-\vPhi ) \\
		& \phantom{= + \p_\perp\big[} + \vtm\vtp\big(\tfrac 1 4\p_\perp (\vPhib\vPhi)+i (W -\Wb)\big)\big]
		\Big].
	\end{aligned}
\end{equation}
The equations of motion are in superspace
\begin{equation}
0=2\fD_-\fD_+\vPhi -\p_\perp \vPhi +i \Wb'(\vPhib),
\end{equation}
which in components are the usual bulk equations.

\section{Supercurrent multiplets in 3D}

\subsection{Decomposition to \texorpdfstring{\boldmath $\zerotwo$-multiplets}{(0,2)-multiplets}} \label{app: S decomposition to zerotwo}
We decompose the bulk multiplet $\Sc_{\alpha\beta}=-2\gamma^\mu_{\alpha\beta}\Sc_{\mu}$ using the branching coordinate $\xi$ according to
\begin{equation}\label{zerotwo expansion S}
	\Sc_{\alpha\beta}(x,\theta,\thetab)=
	\Sc_{\alpha\beta}^\zero(\xi,\tp,\tbp)
	+\tm \Sc_{\alpha\beta}^\one(\xi,\tp,\tbp)
	-\tbm\bar{\Sc_{\alpha\beta}^\one}(\xi,\tp,\tbp)
	+\tm\tbm \Sc_{\alpha\beta}^\two(\xi,\tp,\tbp).
\end{equation}
We obtain:
\begin{enumerate}
	\item The $+$-direction
	\begin{subequations}
	\label{+direction}
		\begin{align}
			\Sc^\zero_{++}&= 4j_+ -4i \tp (S_+)_+- 4i \tbp (\Sb_+)_+ -16 \tp\tbp T_{++},
			\\ 
			\begin{split}
			\Sc_{++}^\one&=
				-4i (S_+)_- -2\sqrt 2 \omegab_+
				+\tbp(4i  \p_\perp j_+ +4 K_{+\perp}+4i L_+) -4i \tp \Yb_+ \\	 
			&\pheq +8\tp\tbp \p_+ (S_+)_-,
			\end{split}
			\\
			\begin{split}\label{+direction two component}
				\Sc_{++}^\two&=
				-8 K_{+-}
				+8\tp \p_\perp (S_+)_-
				-8\tbp \p_\perp (\Sb_+)_-
				+8 \tp \p_-(S_+)_+
				-8\tbp \p_-(\Sb_+)_+
				\\&\hspace{.4cm}
				-4\sqrt 2 i \tp \p_+\omegab_-
				-4\sqrt 2 i \tbp \p_+\omega_-
				-4\sqrt 2 i \tp \p_\perp\omega_+
				-4\sqrt 2 i \tbp\p_\perp\omegab_+
				\\&\hspace{.4cm}
				-4\tp\tbp \p_\perp^2 j_+ 
				-8\tp \tbp \p_\perp L_+
				-4 \tp\tbp(-2\p_+\p_\nu j^\nu +\p^2 j_+).
			\end{split}
		\end{align}
	\end{subequations}
	\item The $-$-direction
	\begin{subequations}
	\label{-direction}
		\begin{align}
			\begin{split}
			\Sc_{--}^\zero&=4j_-
			-4i\tp(S_-)_+
			-4i\tbp(\Sb_-)_+
			+2\sqrt 2\tp\omegab_- \\ &\pheq
			-2\sqrt 2 \tbp\omega_-
			-8\tp \tbp K_{-+},
			\end{split}
			\\
			\begin{split}
			\Sc_{--}^\one&=
			-4i (S_-)_-
			-4i \tp \Yb_-
			+4\tbp (K_{-\perp}+i\p_\perp j_-+iL_-) \\
			&\pheq +8 \tp\tbp \p_+(S_-)_- +4\sqrt 2 i \tp\tbp\p_-\omegab_+,
			\end{split}
			\\ 
			\begin{split}\label{-direction two component}
				\Sc_{--}^\two&=
				-16 T_{--}
				+ 8 \tp \p_\perp (S_-)_-
				+8\tp \p_- (S_-)_+ 
				-8 \tbp \p_- (\Sb_-)_+ 
				-8\tbp \p_\perp(\Sb_-)_-
				\\&\hspace{.4cm} 
				+4\tp\tbp\p_\perp^2 j_-
				-8\tp\tbp \p_\perp L_-
				-4\tp\tbp (-2 \p_- \p^\nu j_\nu + \p^2 j_-).
			\end{split}
		\end{align}
	\end{subequations}
	\item The $\perp$-direction
	\begin{subequations}
	\label{perp-direction}
		\begin{align}
			\begin{split}
			\Sc_{-+}^\zero&=-2j_\perp
			+2i\tp(S_\perp)_+
			+2i \tbp(\Sb_\perp)_+
			+\sqrt 2\tp \omegab_+
							\\&\pheq
			-\sqrt 2\tbp \omega_+
			+4\tp \tbp K_{\perp +},
			\end{split}
			\\
			\begin{split}
				\Sc_{-+}^\one&=
				+2i (S_\perp)_-
				-\sqrt 2 \omegab_-
				-2\tbp(
				K_{\perp\perp}
				+i \p_\perp j_\perp
				+i L_\perp)
				+2i\tp \Yb_\perp
				\\&\hspace{.4cm}
				-4\tp\tbp \p_+(S_\perp)_-
				-2\sqrt 2i\tp\tbp\p_\perp \omegab_+
				-2\sqrt 2 i \tp \tbp\p_+\omegab_-,
			\end{split}
			\\ 
			\begin{split} \label{perp-direction two component}
				\Sc_{-+}^\two&=
				+4 K_{\perp-}
				-4 \tp \p_\perp (S_\perp)_-
				+4 \tbp \p_\perp (\Sb_\perp)_-
				-4\tp \p_- (S_\perp)_+  
				\\&\pheq
				+4\tbp \p_- (\Sb_\perp)_+
				+\sqrt 2 i\tp\p_-\omegab_+
				+\sqrt 2 i \tbp \p_-\omega_+
				+4\tp\tbp \p_\perp L_\perp 
				\\&\pheq
				+2\tp\tbp\p_\perp^2j_\perp
				-4\tp\tbp(\p_\perp \p^\nu j_\nu -\tfrac 1 2 \p^2 j_\perp ).
			\end{split}
		\end{align}
	\end{subequations}
\end{enumerate}
where 
\begin{equation}\label{defs of K and L}
\begin{aligned}
	K_{\mu\nu}&=2T_{\nu\mu}-\eta_{\mu\nu}A -\tfrac 1 4 \eps_{\mu\nu\rho}H^\rho &&=2T_{\nu\mu}-\eta_{\mu\nu}A -\tfrac 1 4 C_{\mu\nu},\\
	L_\mu&=\tfrac 1 4 \eps_{\mu\nu\rho}F^{\nu\rho} +\eps_{\mu\nu\rho}\p^\nu j^\rho&&=\tfrac 1 4 C_\mu +\eps_{\mu\nu\rho}\p^\nu j^\rho.
\end{aligned}
\end{equation}
where we have also defined the brane currents $C_{\mu\nu}=\eps_{\mu\nu\rho}H^\rho$ and $C_\mu=\eps_{\mu\nu\rho}F^{\nu\rho}$.
The decomposition for the $\Rc$-multiplet is found simply by setting the multiplet $\Y_\alpha\ni(\omega_\alpha, A, Y_\mu)$ to zero.

One may check that these decompositions are indeed $\zerotwo$-multiplets using the ($\zerotwo$-restriction of the) supersymmetry variations that follow.

\subsection{\texorpdfstring{\boldmath 3D $\N=2$}{3D N=2} supersymmetry variations of components}
Under a variation induced by $\dsym=\xi Q-\xib \Qb$, we may compute from \eqref{s,x,y expansions}
\begin{subequations}
	\begin{align}
		\dsym \lambda_\alpha&=i \xi_\alpha D 
		-i(\gamma^\mu\xi)_\alpha (H_\mu+\tfrac i 2 \eps_{\mu\nu\rho}F^{\nu\rho} ) +\tfrac i 2 \xib_\beta C,
		\\\dsym D&=\xi \gamma^\mu\p_\mu \lambdab-\xib\gamma^\mu\p_\mu\lambda,
		\\\dsym H_\mu&=- \xi \p_\mu \lambdab + \xib \p_\mu \lambda,
		\\\eps_{\mu\nu\rho} \dsym F^{\nu\rho}&=-2i\eps_{\mu\nu\rho}(\xi\gamma^\rho \p^\nu \lambdab+\xib \gamma^\rho \p^\nu \lambda),
		\\\dsym C&=0,\\
		\dsym \omega_\alpha&=\sqrt 2 \xi_\alpha B -\tfrac 1 {2\sqrt 2} \xib_\alpha C 
		-\sqrt 2 i (\gamma^\mu\xib)_\alpha Y_\mu,
		\\\dsym B&=- \sqrt 2i (\xib \gamma^\mu \p_\mu \omega),
		\\\dsym Y_\mu&= \sqrt 2  (\xi\p_\mu \omega), 
		\\\dsym j_\mu &=\tfrac 1 {\sqrt 2} (\xi\gamma_\mu \omegab)
		-\tfrac 1 {\sqrt 2} (\xib \gamma_\mu \omega)
		-i(\xi S_\mu)-i (\xib \Sb_\mu),\\
		\begin{split}
			\dsym (S_\mu)_\alpha&=\tfrac i 4 (\gamma_\mu \xi)_\alpha \Cb
			+\eps_{\mu\rho\nu}(\gamma^\nu\xi)_\alpha \Yb^\rho
						+ \xib_\alpha \big(\tfrac 1 4 \eps_{\mu\nu\rho}F^{\nu\rho}+\eps_{\mu\nu\rho}\p^\nu j^\rho\big)
			\\&\hspace{.4cm}
						-i(\gamma^\nu\xib)_\alpha (2T_{\nu\mu}-\tfrac 1 4 \eps_{\mu\nu\rho}H^\rho+i \p_\nu j_\mu- i\p_\rho j^\rho \eta_{\mu\nu}),
		\end{split}
		\\
		\begin{split}
			\dsym K_{\mu\nu}&=
			-\tfrac 1 2\big[
			2\eps_{\nu\rho\lambda}\xi\gamma^\lambda\p^\rho S_\mu
			-\sqrt 2 i\eps_{\nu\mu\rho}\xi \p^\rho \omegab
			-\tfrac i {\sqrt 2}\eta_{\mu\nu}\xi\gamma_\rho \p^\rho \omegab
			\\&\hspace{1cm}
			-2\eps_{\nu\rho\lambda}\xib\gamma^\lambda\p^\rho \Sb_\mu
			-\sqrt 2 i\eps_{\nu\mu\rho}\xib \p^\rho \omega
			-\tfrac i {\sqrt 2}\eta_{\mu\nu}\xib\gamma_\rho \p^\rho \omega
			\big],
		\end{split}
		\\
		\begin{split}
			\dsym T_{\mu\nu}&=
			-\tfrac i {4 \sqrt 2}  \eta_{\mu\nu}(\xi\gamma_\rho \p^\rho\omegab)
			-\tfrac 1 2 \xi \gamma_{(\nu}\gamma_\rho \p^\rho S_{\mu)}
			+\tfrac 1 2 \xi \p_{(\nu}S_{\mu)}
			\\&\hspace{.4cm}
			-\tfrac i {4 \sqrt 2}  \eta_{\mu\nu} (\xib \gamma_\rho \p^\rho\omega)
			-\tfrac 1 2 \xib \gamma_{(\nu}\gamma_\rho \p^\rho \Sb_{\mu)}
			+\tfrac 1 2 \xib \p_{(\nu}\Sb_{\mu)}.
		\end{split}
	\end{align}
\end{subequations}

\subsection{Decomposition of bulk constraints}
Using the branching coordinate $\xi$ and the expansions
\begin{equation}\label{zerotwo expansions chi and Y}
	\begin{aligned}
		\chi_\alpha^B&=\chi_\alpha^{B\zero} +\tm \chi_\alpha^{B\onea} +\tbm \chi_\alpha^{B\oneb} +\tm\tbm \chi_\alpha^{B\two},
		\\\Y_\alpha^B&=\Y_\alpha^{B\zero} +\tm \Y_\alpha^{B\onea} +\tbm \Y_\alpha^{B\oneb} +\tm\tbm \Y_\alpha^{B\two}.
	\end{aligned}
\end{equation}
we may rewrite the constraints \eqref{constraints 3d n=2} for the bulk $\Sc$-multiplet to the following collection of equations:
\\
From $\Db_- \chi_\alpha = \tfrac 1 2 \eps_{-\alpha} C$ (recall $\eps_{-+}=-1$ otherwise zero):
\begin{subequations}
	\label{branching of S - chi1}
	\begin{align}
	\chi^{B\oneb}_\alpha &= \tfrac 1 2 \eps_{-\alpha} C, 
	\\ 
	\chi^{B\two}_\alpha &= -2i \p_- \chi^{B\zero}_\alpha.
	\end{align}
\end{subequations}
From $\Db_+ \chi_\alpha = \tfrac 1 2 \eps_{+\alpha} C$:
\begin{subequations}
	\label{branching of S - chi2}
	\begin{align}
	\Db_+  \chi^{B\zero}_\alpha &= \tfrac 1 2 \eps_{+\alpha} C, 
	\\ 
	\Db_+  \chi^{B\onea}_\alpha &= 2i\p_\perp \chi^{B\zero}_\alpha, 
	\\ 
	\Db_+  \chi^{B\two}_\alpha &= 0.
	\end{align}
\end{subequations}
From $\Im D^\alpha \chi_\alpha = 0$:
\begin{subequations}\label{branching of S - chi3}
	\begin{align}
	\Im(D_+\chi_-^{B\zero}-\chi_+^{B\onea})&=0,
	\\
	\Db_+\bar{\chi_-^{B\onea}}+\chi_+^{B\two} 
	-2i\p_-\chi_+^{B\zero }
	-2i\p_\perp \chi_-^{B\zero}&=0,
	\\
	\Im(D_+\chi_-^{B\two}-2i\p_-\chi_+^{B\onea}-2i\p_\perp\chi_-^{B\onea})&=0.
	\end{align}
\end{subequations}
From $D_\alpha \Y_\beta +D_\beta \Y_\alpha=0$ we obtain 
\begin{subequations}
	\label{branching of S - Y1}
	\begin{align}
	D_+ \Y_-^{B\zero} +\Y_{+}^{B\onea}&=0,
	\\ D_+ \Y_{-}^{B\onea}&=0,
	\\ D_+ \Y_{-}^{B\oneb} &=
	2i \p_\perp\Y_-^{B\zero}
	+ \Y_+^{B\two}
	-2i\p_-\Y_+^{B\zero},
	\\D_+ \Y_+^{B\zero}&=0,
	\\D_+ \Y_{+}^{B\onea}&=0,
	\\D_+ \Y_{+}^{B\oneb}&=2i \p_\perp\Y_+^{B\zero}, 
	\\D_+ \Y_+^{B\two}&=-2i\p_\perp\Y_{+}^{B\oneb},
	\\\Y_{-}^{B\onea}&=0,
	\\ \Y_-^{B\two}&=2i\p_-\Y_-^{B\zero}.
	\end{align}
\end{subequations}
From $\Db^\alpha \Y_\alpha+C=0$ we obtain 
\begin{subequations}
	\label{branching of S - Y2}
	\begin{align}
	\Db_+ \Y_-^{B\zero} +\Y_{+}^{B\oneb}+C&=0,
	\\ \Db_+ \Y_{-}^{B\onea}
	+2i \p_\perp\Y_-^{B\zero}
	+ \Y_+^{B\two}
	+2i\p_-\Y_+^{B\zero}&=0,
	\\ \Db_+ \Y_{-}^{B\oneb}&=0,
	\\ \Db_+ \Y_-^{B\two} -2i\p_\perp  \Y_{-}^{B\oneb} -2i\p_-\Y_{+}^{B\oneb} &=0.
	\end{align}
\end{subequations}
And the relation $\Db^\beta \Sc_{\alpha \beta}=\chi_\alpha+\Y_\alpha$:
\begin{subequations}
	\label{branching of S - S}
	\begin{align}
	\chi^{B\zero}_\alpha +\Y^{B\zero}_\alpha &= \Db_+ \Sc^{B\zero}_{\alpha-} - \bar\Sc^\one_{\alpha+}, 
	\\
	\chi^{B\onea}_\alpha+ \Y^{B\onea}_\alpha &= -\Db_+ \Sc^\one_{\alpha-} - \Sc^{B\two}_{\alpha+} - 2i\p_\perp\Sc^{B\zero}_{\alpha-} - 2i\p_-\Sc^{B\zero}_{\alpha+}, 
	\\
	\tfrac 1 2 \eps_{-\alpha} C +\Y_\alpha^{B\oneb}&=\Db_+ \bar\Sc^\one_{\alpha-}, 
	\\
	\chi^{B\two}_\alpha+ \Y^{B\two}_\alpha &= \Db_+ \Sc^{B\two}_{\alpha-} + 2i\p_\perp\bar\Sc^\one_{\alpha-} + 2i\p_-\bar\Sc^\one_{\alpha+}.
	\end{align}
\end{subequations}

\subsection{Decompositions of bulk improvements}
Using the branching coordinate and the expansions \eqref{zerotwo expansion S} and \eqref{zerotwo expansions chi and Y} we may compute the improvements of decomposed multiplets:

\begin{subequations}\label{bulk improv decomposed S-zero}
	\begin{align}
	\Sc_{++}^{B\zero}&\mapsto \Sc_{++}^{B\zero}+[D_+,\Db_+]U^{B\zero},
	\\\Sc_{--}^{B\zero}&\mapsto \Sc_{--}^{B\zero}-2U^{B\two},
	\\\Sc_{+-}^{B\zero}&\mapsto \Sc_{+-}^{B\zero}+D_+\bar{U^{B\one}}-\Db_+U^{B\one}.
	\end{align}
\end{subequations}

\begin{subequations}
	\begin{align}
	\Sc_{++}^{B\one}&\mapsto \Sc_{++}^{B\one}
	+[D_+,\Db_+]U^{B\one} +4i \p_\perp D_+ U^{B\zero},
	\\\Sc_{--}^{B\one}&\mapsto \Sc_{--}^{B\one} -4i\p_-U^{B\one},
	\\\Sc_{+-}^{B\one}&\mapsto \Sc_{+-}^{B\one}
	- D_+ U^{B\two}
	+2i \p_\perp U^{B\one}
	-2i\p_-D_+U^{B\zero}.
	\end{align}
\end{subequations}

\begin{subequations}
	\begin{align}
	\begin{split}
	\Sc_{++}^{B\two}&\mapsto \Sc_{++}^{B\two}
	+[D_+,\Db_+] U^{B\two}
	+4i\p_\perp D_+ \bar{U^{B\one}}
	\\&\hspace{1.2cm}
	+4i\p_\perp \Db_+ U^{B\one}
	-8 \p_\perp^2U^{B\zero},
	\end{split}
	\\\Sc_{--}^{B\two}&\mapsto \Sc_{--}^{B\two} -8\p_-^2 U^{B\zero},
	\\\Sc_{+-}^{B\two}&\mapsto \Sc_{+-}^{B\two}
	-2i\p_-D_+\bar{U^{B\one}}
	-2i \p_-\Db_+U^{B\one}
	+8\p_-\p_\perp U^{B\zero}.
	\end{align}
\end{subequations}

\begin{subequations}
	\begin{align}
	\chi_+^{B\zero}&\mapsto \chi_+^{B\zero} 
	+2\Db_+D_+\bar{U^{B\one}},
	+4i\p_\perp\Db_+ U^{B\zero},
	\\\chi_-^{B\zero}&\mapsto \chi_-^{B\zero}
	+2\Db_+U^{B\two}
	-4i\p_-\Db_+U^{B\zero},
	\\
	\begin{split}
	\chi_+^{B\onea}&\mapsto \chi_+^{B\onea}
	+ 2\Db_+ D_+U^{B\two}
	-4i \p_\perp\Db_+ U^{B\one}
	-4i \p_\perp D_+\bar{U^{B\one}}
	\\&\hspace{1.3cm}
	+8\p_\perp^2 U^{B\zero}
	+4i\p_-\Db_+D_+U^{B\zero},
	\end{split}
	\\\chi_-^{B\onea}&\mapsto \chi_-^{B\onea}
	+4i( 2\p_-\Db_+U^{B\one}
	- \p_\perp U^{B\two}
	+2i \p_\perp \p_- U^{B\zero}),
	\\\chi_\alpha^{B\oneb}&\mapsto \chi_\alpha^{B\oneb},
	\\\chi_+^{B\two}&\mapsto \chi_+^{B\two}
	-4i\p_-(\Db_+D_+\bar{U^{B\one}}
	+2i\p_\perp\Db_+ U^{B\zero}),
	\\\chi_-^{B\two}&\mapsto \chi_-^{B\two}
	-4i( \p_-\Db_+U^{B\two}
	-2i\p_-^2\Db_+U^{B\zero}).
	\end{align}
\end{subequations}

\begin{subequations}\label{bulk improv decomposed Y}
	\begin{align}
	\Y_+^{B\zero}&\mapsto \Y_+^{B\zero}
	-D_+\Db_+\bar{U^{B\one}},
	\\\Y_-^{B\zero}&\mapsto \Y_-^{B\zero}
	+\Db_+U^{B\two}
	+2i \p_\perp\bar{U^{B\one}}		
	+2i\p_-\Db_+U^{B\zero},
	\\\Y_+^{B\onea}&\mapsto \Y_+^{B\onea}
	- D_+\Db_+U^{B\two}
	-2i \p_\perp D_+\bar{U^{B\one}}
	-2i\p_-D_+\Db_+U^{B\zero},
	\\\Y_-^{B\onea}&\mapsto \Y_-^{B\onea},
	\\\Y_+^{B\oneb}&\mapsto \Y_+^{B\oneb}
	-2i \p_\perp\Db_+\bar{U^{B\one}},
	\\\Y_-^{B\oneb}&\mapsto \Y_-^{B\oneb}
	+4i\p_-\Db_+\bar{U^{B\one}},
	\\
	\begin{split}
	\Y_+^{B\two}&\mapsto \Y_+^{B\two}
	+2i\p_-D_+\Db_+\bar{U^{B\one}}
	-2i\p_\perp\Db_+U^{B\two}
	\\&\hspace{1.2cm}
	+4  \p_\perp^2\bar{U^{B\one}}
	+4 \p_\perp\p_-\Db_+U^{B\zero},
	\end{split}
	\\\Y_-^{B\two}&\mapsto \Y_-^{B\two}
	+2i\p_- \Db_+U^{B\two}
	-4\p_- \p_\perp\bar{U^{B\one}}		
	-4\p_-^2\Db_+U^{B\zero}.
	\end{align}
\end{subequations}

\subsection{Explicit bulk components of \texorpdfstring{\boldmath $\Sc_\mu$}{S} for LG model} \label{app:bulk components S multiplet}
We may compute the components of the supercurrent multiplets for the Landau-Ginzburg model where 
\begin{equation}
	\begin{aligned}
	 	\Sc_{\alpha\beta}&=D_\alpha \Phi_{3D} \Db_{\beta} \Phib_{3D}+D_\beta \Phi_{3D} \Db_{\alpha} \Phib_{3D},
	 	\\\chi_\alpha&=-\tfrac 1 2 \Db^2 \D_\beta(\Phib_{3D}\Phi_{3D}),
	 	\\\Y_\alpha&=-\Db^2\Phib_{3D} D_\alpha \Phi_{3D}.
	\end{aligned}
\end{equation}
according to the expansions \eqref{s,x,y expansions}. In other words, we are in the $\Sc$-frame (cf.\ page \pageref{introduction of 'improvement frame'}) and we obtain:
\begin{subequations}\label{explicit components in LG}
	\begin{align}
		R\text{-``current''}:&\footnotemark&j^\mu&=(\psib\gamma^\mu\psi),\\
		\text{supercurrent}:&&S_{\mu\alpha}&=\sqrt 2 (\gamma^\nu \gamma_\mu  \psi)_\alpha \p_\nu \phib 
		-\sqrt 2 i(\gamma_\mu \psib)_\alpha \Wb',\\
		\text{lowest in }\chi_\alpha:&&\lambda_\alpha&=  2 \sqrt 2 (\gamma^\mu\psib)_\alpha \p_\mu \phi 
		+2 \sqrt 2 i  W' \psi_\alpha, \\
		\text{lowest in }\Y_\alpha: &&\omega_\alpha&=4 W' \psi_\alpha,\\
			\begin{split}
			\text{EM tensor}:&
			\\&
			\end{split} &
			\begin{split} T_{\nu\rho}&=
			(\p_\rho\phi \p_\nu \phib +\p_\nu \phi \p_\rho \phib)
			\!-\!\eta_{\nu\rho} (|\p\phi|^2\!+ |W'|^2)
				\\
				&\hspace{.4cm}
			+\tfrac i 2 (\p_{(\rho} \psib \gamma_{\nu)} \psi)
			-\tfrac i 2(\psib\gamma_{(\nu} \p_{\rho)} \psi),
			\end{split}
		\\		
		\text{irrelevant auxilliary}:&& A&=-4|W'|^2+i\p_\mu \psib \gamma^\mu \psi -i\psib\gamma^\mu \p_\mu \psi, \\
		\{Q,S\} \text{ 1-brane  charge}:&&Y_\mu&= 4 \p_\mu W,\\
		\{\Qb, S\} \text{ 1-brane  charge}:&&H^\mu &= -2i \p^\mu (\psib \psi),\\
		\{\Qb, S\} \text{ 0-brane charge}:&&\eps_{\rho\mu\nu}F^{\mu\nu} &=-4 \eps_{\rho\mu\nu}\p^\mu j^\nu -8i \eps_{\rho\mu\nu} \p^\mu \phi \p^\nu \phib .
	\end{align}
\end{subequations}
\footnotetext{This particular $R$-``current'' is not conserved for most superpotentials. If we improve the multiplet to an $\Rc$-multiplet, this component of the multiplet is the conserved $R$-current.}
Note that all (Hodge duals to) brane currents are \emph{exact} forms. This is to be expected, since we are working on a trivial space, and it only shows local triviality in general backgrounds. For example, if $W$ is not a properly defined function, then $Y^\mu$ is not identically exact.  

\section{Delta distributions on the boundary}\label{app:deltadistributions}

The commutators of quantized operators involve some subtleties with respect to delta distributions and boundaries. We define
\begin{equation}
	\cpint{(-\infty, 0]} f(x) \gd(x) \coloneqq f(0),
\end{equation}
with the reasoning that we want the entire boundary to be part of the theory, thus anything on the boundary is fully part of the system, and that if the $\gd$-distribution is to be understood as a limit of functions $g_n(x) \to \gd(x)$ which fulfill $\int_{(-\infty, 0]} g_n (x) = 1$, then the above will follow automatically.
This results in one important subtlety: switching from $\tdel{y} \gd(x-y)$ to $-\tdel{x} \gd(x-y)$ introduces a boundary term, specifically
\begin{equation}
\cpint{(-\infty, 0]} \d y \: \cpint{(-\infty, 0]} \d x  f(x) g(y) \big( \tdel{x} \gd(x-y) +  \tdel{y} \gd(x-y) \big) = f(0) g(0).
\end{equation}
This, in turn, implies
\begin{align}
-\cpint{(-\infty, 0]} \d y  f(x) g(y) \tdel{y} \gd(x-y) &= -f(0)g(0) + \cpint{(-\infty, 0]} \d y  f(y) g'(y), \\
\cpint{(-\infty, 0]} \d y  f(x) g(y) \tdel{x} \gd(x-y) &= \cpint{(-\infty, 0]} \d y  f(y) g'(y),
\end{align}
as both boundary terms cancel in the second equation.

\providecommand{\href}[2]{#2}\begingroup\raggedright\endgroup
\bibliographystyle{JHEP}

\end{document}